\documentclass[twoside,11pt]{article}
\pdfoutput=1

\usepackage{amsmath}
\usepackage{arXiv2e}
\usepackage{caption}
\usepackage{subcaption}
\RequirePackage[]{natbib}

\usepackage{graphics}
\usepackage{epstopdf}
\usepackage{pgfplots}
\usepackage{url}
\usepackage{multirow}
\usepackage[percent]{overpic}
\pgfplotsset{compat=default,no markers}
\pgfkeys{/pgf/number format/.cd,
 fixed,fixed zerofill,
 set thousands separator={\,}
}
\mathchardef\Intop= "1352
\mathchardef\Sum= "1350
\mathchardef\Prod="1351
\DeclareMathOperator*{\argmin}{arg\,min}

\arXivheading{VOLXXX}{2016}{xxx-xxx}{1/16}{xx/xx}{Jonathan M. Friedman}

\ShortHeadings{Accurate Estimates for Proportions}{Friedman}
\firstpageno{1}

\begin{document}

\title{ Binomial and Multinomial Proportions:  Accurate Estimation and Reliable Assessment of Accuracy} 


\author{\name Jonathan M. Friedman \email jmfriedman7@alum.mit.edu \\
       \addr Canbas Co., Ltd.\\
       2-2-1 Otemachi\\
       Numazu, Shizuoka 410-0801, Japan
       }

\editor{xxx xxx xxx}

\maketitle

\begin{abstract} 
Misestimates of $\sigma_{P_o}$, the \emph{uncertainty} in $P_o$ from a 2-state Bayes equation used for binary classification, apparently
arose from $\hat{\sigma}_{p_i}$, the uncertainty in underlying 
pdfs estimated from experimental $b$-bin histograms. 
To address this, several Bayesian estimator pairs $(\hat{p}_i, \hat{\sigma}_{p_i})$ were compared for
agreement between nominal confidence level ($\xi$) and calculated coverage values ($C$).
Large $\xi$-to-$C$ inconsistency for large $b$ and $ p_i \gg \frac{1}{b}$ arises for all multinomial estimators since priors downweight low likelihood, high $p_i$ values.
To improve $\xi$-to-$C$ matching, $(\xi-C)^2$ was minimized against $\alpha_0$ in a more general prior pdf ($\mathcal{B}[\alpha_0,(b-1)\alpha_0;x]$) to obtain $(\hat{p_i})_{\xi\leftrightarrow C}$.
This improved matching for $b=2$, but for $b>2$, $\xi$-to-$C$ matching by $(\hat{p_i})_{\xi\leftrightarrow C}$ required an effective value "$b=2$" and renormalization, and this reduced $\hat{p}_i$-to-$p_i$ matching.
Better $\hat{p}_i$-to-$p_i$ matching came from the original multinomial estimators, a new discrete-domain estimator $\hat{p}(n_i,N)$, or an earlier \emph{joint} estimator, $(\hat{p_i})_{\bowtie}$ that 
co-adjusted all estimates $p_i$ for James-Stein shrinkage to a mean vector.
Best simultaneous $\xi$-to-$C$ and $\hat{p}_i$-to-$p_i$ matching came by \emph{de-noising} initial estimates of underlying pdfs. 
For $b=100$, $N<12800$, de-noised $\hat{p}$ needed $\approx 10\times$ fewer observations to achieve $\hat{p}_i$-to-$p_i$ matching equivalent to that found for $\hat{p}(n_i,N)$, $(\hat{p_i})_{\bowtie}$ or the original multinomial $\hat{p}_i$.
De-noising each different type of initial estimate yielded similarly high accuracy in Monte-Carlo tests.
\end{abstract}


\section{Introduction} 

Experimental estimates of probability density function (pdf) values and of the uncertainty in these values were required for a computational classification model to estimate drug effectiveness \citep{friedmanBayesICA}.
These estimates of pdfs were based on the frequency, $p_i$, of finding $n_i$ occurrences out of $N$ total experimentally observed values sorted into bin $i$ of a $b$-bin histogram according to a general multinomial distribution. 
Large inconsistencies were found for the confidence intervals estimated using earlier estimators.
Here we characterize these inconsistencies and describe some methods that circumvent them.

Initially, computational tests of Bayesian classification of drug sensitivity using underlying pdfs based on earlier bin-by-bin estimators for $ p_i $ and ${{\sigma}^2_{p_i}} $ \citep{brownStatSci,brownAnnals,laplace,wald2}
led to calculated confidence levels that were sometimes at odds with the observed accuracy of classification.
Whereas it was expected that predictions from the Bayesian model for drug sensitivity would be sometimes be incorrect, 
it was expected that analysis of the propagation of uncertainty would allow one to tell when sets of predictions might be unreliable.
In fact, initial estimates of uncertainty for different types of data did not correlate very well with the accuracy of predictions.
The fraction of \emph{incorrect} predictions was much higher than expected from calculated confidence levels for one class of test set measurements,
while for learning set and another class of test set measurements the fraction of \emph{correct} predictions was 
often higher than expected, suggesting problems. 

\subsection{Overview and Organization}
Here we examine suspected errors in the \emph{uncertainty} in $\hat{p_i}$ values and test possible corrections.
Histogram-based estimators $\hat{p}_i$ and $\hat{\sigma}_{p_i}$ designed to estimate pdfs of \emph{unknown} functional form from observed data are tested 
by examining bin-by-bin agreement for estimates of arbitrary underlying pdfs of \emph{known} functional form based on random samplings from these known underlying pdfs $p_i$.
Whereas bin-by-bin $\hat{ p_i }$-to-$p_{i}$ agreement is important, here as earlier \citep{thatcher}, 
it was expected that histogram estimates $\hat{p_i}$ would vary unavoidably due to small sample size and sample-to-sample variability for some experiments. 
Accurate estimates $\hat{p}_i$ were ultimately desired, but to avoid misclassification in cases with insufficient experimental information, 
it was equally important to detect indeterminable cases by having reliable estimates of $\hat{\sigma}_{p_i}$.
When combining component $\hat{p}_i$ to form a composite Bayesian $P_o$ for classification, if one simultaneously combined accurate component $\hat{p}_i$ and $\hat{\sigma}_{p_i}$,
to get $\hat{\sigma}_{P_o}$, then one would be able to reject cases for which $|P_o - P_{crit}|<Z\hat{\sigma}_{P_o}$.

Uncertainty in $\hat{p_i}$ depends on the number of experimental observations $N$ and the size of $\hat{p}_i$. 
It can be expressed as an estimated confidence interval, $\hat{\delta_i} \equiv (\hat{\delta}^-_i, \hat{\delta}^+_i)$, about $\hat{p_i}$ to a stated nominal level of confidence, $\xi$.\footnote{
Here, confidence level $\xi$ is used since the symbol for tolerance, $\alpha$, is identical to the symbol that is generally used as an argument to the beta pdf $\mathcal{B}[\alpha,\beta; x]$.
For equal-tailed confidence intervals, $\alpha = \frac{1-\xi}{2}$.
}
Many estimators $\hat{p}_i$ and associated estimators for confidence intervals $(\hat{\delta}^-_i, \hat{\delta}^+_i)$ are available. 
Whereas much earlier emphasis had been placed on choosing $\Hat{p}_i$ $\hat{\delta_i}$ pairs with narrow, theoretically consistent intervals,
poor estimates of uncertainty in $P_o$ for Bayes classification using propagation of uncertainty suggested that nominal $\xi$ values for the the usual choices of estimators might not sufficiently reflect the actual reliability.
Instead, it might be preferable to choose the $\hat{p_i}$ $\hat{\delta_i}$ estimator pair based on
empirical agreement between nominal confidence, $\xi$, and coverage $C$, 
the fraction of $N$-observation trials for which $p_{i,true} \in  (\hat{\delta}^-_i, \hat{\delta}^+_i)$ \citep{brownStatSci,brownAnnals,newcombe}.

After introducing earlier estimators $\hat{p}_i$ and $\hat{\delta}_i$ (Sections 1.2-1.4), the first strategy taken was to compare $\xi$-to-$C$ matching for different estimators (Sections 1.5-1.7)
to see if choosing the estimator set with the most consistently high degree of $\xi$-to-$C$ matching (or low degree of mis-matching) was sufficient to improve the error modeling.
Continued problems with Bayes classification using the best of four initially tested $\hat{p_i}$ $\hat{\delta_i}$ pairs, 
led to a closer examination (Section 1.7.2) that at first indicated problems for all initially examined $\hat{p_i}$ $\hat{\delta_i}$ pairs.

To avoid these problems, since earlier non-informative Bayes priors for deriving $\hat{p_i}$ and $\hat{\delta}_i$ were particular cases of the continuous beta pdf $\mathcal{B}[\alpha,\beta;x]$ with different values for $\alpha$ and $\beta$, 
we examined the effect of
minimizing $(\xi-C)^2$ with respect to an arbitrary parameter $\alpha_0$ in a more general non-informative beta pdf $\mathcal{B}[\alpha_0,(b-1)\alpha_0;x]$ prior (Section 2.1).
Multinomial estimators from this prior follow from general equations in \citet{jaynesCh18}.
In Sections 2.2-2.6, values of $\langle (\xi-C)^2 \rangle$ were minimized with respect to $\alpha_0$ for fixed arbitrary ranges $p_i=p^* \in (\psi_0,\psi_1) $ and constant $b$, $\xi$, and $N$. 

The prior pdf $\mathcal{B}[\alpha_0,(b-1)\alpha_0;x]$ using optimal $\alpha_0 = \alpha_0(N, \xi, b, \psi_0<p^*<\psi_1)$ is essentially a probability matching prior \citep{dattaMukerjee} for  a discrete lattice of outcomes \citep{rousseau2000,rousseau2002}.
Instead of rescaling $|\hat{\delta}^+_i-\hat{\delta}^-_i|$, $\alpha_0$ in the prior is adjusted for $\xi$-to-$C$ matching.
For $b>2$, poor $\xi$-to-$C$ matching was found with numerically optimized $(\hat{p_i})_{\xi\leftrightarrow C}$ and $(\hat{\delta_i})_{\xi\leftrightarrow C}$ (Section 2.4.1).
Within most limited local ranges of $p^* \in (\psi_0,\psi_1)$, optimal $\alpha_0 \approx 0$, and 
$\sigma_{\alpha_0} \gg \alpha_0$.
However, when $b=2$, improved $\xi$-to-$C$ matching, much larger optimal values for $\alpha_0$, and smaller relative variation, $\sigma_{\alpha_0} \sim O(0.01-0.10\times\alpha_0)$ were found (Section 2.4.2) over most $(\psi_0,\psi_1)$.  
This allowed $\xi$-to-$C$ to match well when a single value for $\alpha_0$ was used for the entire range $p^* \in (\psi_0,\psi_1) = (0,1)$.
Further improvements to $\alpha_0 = \left[\alpha_0(N, \xi)\right]_{b=2}$ allowed accurate Bayes classification despite poorer initial $\hat{p}_i$-to-$p_i$ matching. 
As a further option, discrete Bayes posterior functions with domain limited to $\mathbb{Z}$ (integers) and corresponding priors were derived to define binomial $\hat{p}(n,N)$ and $\hat{\delta}(n,N)$ or multinomial $\hat{p_i}(n_i,N)$ and $\hat{\delta_i}(n_i,N)$ (Appendix B).  
The binomial form for the discrete estimator exhibits reasonable $\xi$-to-$C$ matching. The multinomial form
maintains $\hat{p}_i$-to-$p_{i}$ matching but exhibits $\xi$-to-$C$ mismatches similar to those for earlier multinomial $\hat{p}_i$ and $\hat{\delta}_i$.

Attempts to improve initially optimized $\alpha_0(N,\xi)$ values at each $N$ for the continuous $\mathcal{B}[\alpha_0,\alpha_0;x]$ priors (Section 2.6) led to a de-noising procedure for $\alpha_0(N,\xi)$ (Appendix A) 
that could also be used directly (Section 3) to de-noise the initial estimate $\hat{p}_{i,0}$
 of an underlying pdf to get an improved estimate $\widehat{p_{i,smooth}}$.
Since $\sigma_{p_i}$ was important, an empirical method was also derived (Section 3.4-3.5) to estimate $\widehat{\sigma_{p_{i,smooth,0}}}$ from individual smoothed-histograms, $\widehat{p_{i,smooth}}$, 
by assuming a parametric form $\hat{\sigma}_{p_i} = \sigma_{est}(\widehat{p_{i,smooth}},N,A_0, B_0 )$. 
First, parameter values $\bar{A_0}$ and $\bar{B_0}$ to estimate $\sigma_{p_i}$ from a single histogram were established for the case $b$=$100$,
 by fitting the parametric function $\sigma_{est}(\widehat{p_{i,smooth}},N,A_0,B_0 )$ simultaneously to sets of single histograms derived from random samples
generated by 6 arbitrarily chosen underlying test pdfs.
These initial single-run Monte-Carlo (MC) estimates $\hat{\sigma}_{p_i}=\sigma_{est}(\widehat{p_{i,smooth}},N,\bar{A_0},\bar{B_0} )$
were then corrected to account for observed run-to-run \emph{variation} in each single-run estimate $\sigma_{est}$
by using an empirical relationship between single-run estimates $\sigma_{est}$ and $\sigma_{p_i,MC}$, the actual long-run MC variance in each bin-by-bin $\hat{p}_i$ estimated from many $N$-observation histograms.  
Since run-to-run variance in the ratio $\rho_i=\frac{\sigma_{i,est,opt}}{\sigma_{p_i,MC}}$ was relatively insensitive to the underlying pdf and to $|p_i|$, a correction factor could be estimated between single-run and long run estimates of $\sigma_{p_i}$.
Including this correction avoids statistically likely underestimates by single-run estimates from parametric $\sigma_{est}$ that arise from run-to-run variance in these single-run parametric estimates.
Since run-to-run variance $\sigma^2_{\rho_i}$ 
was fairly constant from bin-to-bin (varying $p_i$), varied systematically as a fairly smooth function of $N$, and was fairly independent of the tested underlying pdfs,
values of $\sigma^2_{\rho_i}$ 
for a given value $N$ could be retrieved from a fit curve without repeating MC.
Retrieving this value and presuming: 
$$ \left( \frac{\widehat{\sigma_{p_{i,smooth,0}}}}{\widehat{\sigma_{p_i,MC}}}\right) \sim \mathcal{N}[{\mu_{\frac{\sigma_{est}}{\sigma_{p_i,MC}}}(N)},{\sigma^2_{\frac{\sigma_{est}}{\sigma_{p_i,MC}}}(N)}]\text{~;~~}\mu_{\frac{\sigma_{est}}{\sigma_{p_i,MC}}}(N)\approx 1 $$ 
\noindent where ${\sigma^2_{\sigma_{est}/\sigma_{p_i,MC}}(N)}$ is effectively invariant over bins and trial pdfs, 
allowed initial parametric estimates $\widehat{\sigma_{p_{i,smooth,0}}} = \sigma_{est}$ to be correctively rescaled to $\widehat{\sigma_{p_{i,smooth}}}$ by the limiting low value of the ratio $\rho_i$ at a given tolerance $\xi_{\frac{\sigma_{est}}{\sigma_{p_i,MC}}}$.
This probability-matching adjustment to account for run-to-run variation in $\hat{\sigma}_{p_i}$ raises the final estimated value $\hat{\sigma}_{p_i}$ above the initial single-run parametric estimate $\sigma_{est}$  
 to avoid occasional, statistically expected, large underestimates of $\sigma_{p_i}$ by $\sigma_{est}$.
 Such occasional large underestimates of $\sigma_{p_i}$, and in turn of $\sigma_{P_o}$ for binary Bayes classification, were a likely cause of the falsely large values for $Z$ or Student-$t$ for individual predictions.
The originally faulty error analysis for Bayes classification arose from a disconnect between the concept of nominal confidence $\xi_{p_i}$ and the required level of confidence $\xi_{p_i}$.
Getting the average value for $C$ to match $\xi$ means that about half the time $C$ is less than $\xi$ by an unspecified, often large amount.
What was actually required was for $C > \xi$ "essentially always".
Scaling the initial $\sigma_{est}$ based on $\sigma^2_{\sigma_{est}/\sigma_{p_i,MC}}(N)$, using the lower limiting value of the ratio $\sigma_{est}/\sigma_{p_i,MC}(N)$ at the $0.01$ tolerance level,
increases $\sigma_{est}$ from the initial parametric estimate and leads, 0.99 of the time, to minimal $C$ that is \emph{at least as good as} the desired confidence level.
Most of the time, rescaled $\sigma_{est}$ is too large, but this way the situation with $C < \xi$ that increases the frequency of Bayes classification errors can be much more completely avoided.
This scaling brings $\xi$ in line with confidence intervals often expected by experimentalists, but is inconsistent with standard definitions of $\xi$ based on agreement with \emph{average} $C$ (essentially $\xi_{\frac{\sigma_{est}}{\sigma_{p_i,MC}}} = 0.5$).

Basing estimators $\hat{p}_i$  and  $\hat{\sigma}_{p_i}$ on \emph{multinomial} Uniform priors was originally thought to cause the poor error estimates seen for $P_o$ in Bayes classification, 
because large mismatches were observed between $C$ and nominal $\xi_{p_i}$ in MC tests.
Upon deriving forms for the discrete prior and posterior (Appendix B), 
these earlier large $\xi$-to-$C$ mismatches were seen to be attributable to statistically unlikely restrictions imposed on other histogram bins by the statistically unlikely high trial values for $p_i$ in one bin in the anomalous MC tests. 
A more likely cause for poor estimates of error in $P_o$ was the failure to account for the uncertainty in the estimate of $\hat{\sigma}_{p_i}$ itself 
(or more generally in each interval limit $\hat{\delta}^+$ or $\hat{\delta}^-$ itself). 

The improved accuracy in $\hat{p}_i$ from de-noising initial histogram estimates was next characterized for different trial pdf's at $N \le 12800$ (Section 4).
De-noising histograms reduces the $N$ required to achieve the same MC average signal-to-noise (S/N) by a factor of about 10. (Here S/N is a measure of $\hat{p}$-to-$p$ matching, with "noise" based on the error $\sqrt{(\hat{p_i}-p_{i,true})^2}$.)
De-noised histograms with 4 to 25 times fewer observations\footnote{  
Improvement factors varied from test pdf to test pdf and for different values of $N$ (see \ref{tab:P0Tables}): usually above 10 for $N < 800$, increasing at lower $N$ to as high as 25, but as low as 3.6 for the highest examined $N$. }
achieve S/N comparable to histograms based on earlier unsmoothed \emph{multinomial} estimators including the \emph{joint} estimator $(\widehat{p_i})_{\bowtie}$ from Rgbp in R \citep{stein,morrisLysy,Rgbp}.
For comparable S/N when using the joint estimator $(\widehat{p_i})_{\bowtie}$ as a \emph{starting} point for smoothing, a slight change was required in the procedure after smoothing to adjust the pdf's scale and baseline. 
Applying the same adjustments to smoothed pdfs from all other initial estimators also improved the S/N of these further.
At present, the MC average of S/N for estimated pdfs derived by de-noising starting estimates $(\widehat{p_i})_{\xi\leftrightarrow C}$, the discrete $\hat{p_i}(n_i,N)$, 
the multinomial Rule-of-Succession estimator $\hat{p_i}(n_i,N)$,
or the joint estimator $(\widehat{p_i})_{\bowtie}$ are more comparable,
 with each initial $\hat{p}_i$ exhibiting slight advantages for different trial pdfs and $N$-values.
Differences in MC average S/N among the \emph{smoothed} estimates from different starting points are fairly small.
Larger increases in S/N are found on going from 
unsmoothed $b=2$ $(\widehat{p_i})_{\xi\leftrightarrow C}$ to either unsmoothed $b>2$ joint $(\widehat{p_i})_{\bowtie}$, 
continuous multinomial $\hat{p}_i$, or discrete multinomial $\hat{p_i}(n_i,N)$ but only at low $N$.
Consistently larger increases in S/N at all $N$ are found on going from unsmoothed to smoothed estimators.
Aside from improved $\widehat{p_i}$-to-$p_i$ matching, parametric single-run estimates $\widehat{\sigma_{p_{i,smooth}}}$, even after upward rescaling,
 are lower than initial unsmoothed estimates $\hat{\sigma}_{p_i}$, and have sufficiently consistent $\xi$-to-$C$ matching.  

Given this overview and organizational layout, the details of the experimental numerical investigation into improved estimates for $\hat{p_i}$ and $\hat{\delta_i}$ are described below.
We start by describing earlier Bayes estimators $\hat{p_i}$ and $\hat{\delta_i}$ used for initial $\xi$-to-$C$ comparison.

\subsection{Uniform Density as the Bayes Prior} 

To understand better the initially observed misestimation of uncertainty for Bayes classification,
expressions for estimators of multinomial $\hat{ p_i }$ and ${\hat{\sigma}_{p_i}}^2 $ were re-derived,
by assuming that any value of $p_i$ was equally likely at the outset (a Uniform Bayes prior pdf, $\mathcal{U}[0,1;x]$).
\begin{gather}
\mu_{p_i} = \hat { p_i }  = \frac { n _i + 1 } { N + b } \\
 {\hat{\sigma}_{p_i}}^2 = {\langle {p_i}^2 \rangle - \langle { p_i } \rangle^2} = {\frac{\hat{p_i} \left(1-\hat{ p_i } \right)}{N+1+b}} 
\end{gather}
Similar, more general equations for these estimators were found to have been derived by \citet{jaynesCh18}, 
who recognized Eq. 1 \& 2 to be a generalization of Laplace's Rule of Succession from the binomial ($b$=$2$) to the multinomial case \citep{jaynes}.

Re-examination of the literature, showed that repeated use of $\mathcal{U}[0,1;x]$ as the Bayes prior pdf is in fact common for estimating $\hat{ p}_i $ in histograms, and the extra '+1' associated with each bin is referred to as a 'pseudocount'.  
However the expression for $\sigma^2_{p_i}$ was not encountered or used as often.
As Jaynes has argued \citep{jaynesCh18}, consideration of $\sigma_{p_i}$ diminishes some objections to the use of the Rule of Succession.
For example, an experiment with 0 observations to determine a binary outcome ($b=2$) gives $\hat{ p_i } = 0.5000$, which was thought to be conceptually unreasonable.
However, consideration of the estimated variance leads to, $\hat{\sigma}_{p_i} = 0.29$, so that at
$\xi = 0.95$, the expected range of values is $ p_i  = 0.5000 \pm 0.57 $, suggesting that $p_i$ might possibly well have any value from 0 to 1 at this confidence level. 
This is clearly consistent with the state of knowledge specified by 0 observations.

\subsection{The Jeffreys-Bayes Prior} 

Use of $\mathcal{U}[0,1;x]$ as the Bayes prior pdf for estimates $\hat{p}_i$ and $\hat{\sigma}_{p_i}$ seemed reasonable at the outset since all values of $p_i$ in $(0,1)$ seemed equally possible in the absence of information.
On further examining the literature though, it was found that even for the case of the simplest histogram, one with 2 bins dictated by a binomial distribution,
the proper choice of Bayes prior pdf was a point of debate.
For more general multinomial histograms, bins with pseudocount values other than 1, including non-integer values, are often used. 
The criterion for choosing the optimal number of pseudocounts in each bin is also unsettled.

Some of the contention arises from the argument that an experiment that distinguishes between the values $p_a=1.0\times 10^{-23}$ and $p_b=0.0001$ establishes more \emph{meaningful} information than one
that distinguishes between the values  $p_a=0.5000$ and $p_b=0.5001$ would, even though $p_b-p_a \approx 1\times10^{-4}$ in both instances.  
That is, differences in $log(p)$ are more relevant than differences in $p$ and there is a higher density of possible \emph{meaningful} values for $log(p)$ and $log(1-p)$ as one approaches the limiting values of $p$=$0$ or $p$=$1$.
Since more new meaningful information lies close to the values $p=0$ and $p=1$, consideration of the density of information content, as quantified by Fisher information (Jeffreys) \citep{brownStatSci,brownAnnals,jeffreys}, 
would dictate use of
beta pdf $\mathcal{B}[\frac{1}{2},\frac{1}{2};x]$ instead of $\mathcal{U}[0,1;x] = \mathcal{B}[{1},{1};x]$ as the Bayes prior pdf.
Use of $\mathcal{B}[\frac{1}{2},\frac{1}{2};x]$ for binomial $\hat{p}$ adds enhanced weight, symmetrically about $p=\frac{1}{2}$, to outcomes near $0$ and $1$.
For the $b$-bin multinomial case, the expressions for the expected values of the mean and variance with the Jeffreys-Bayes prior are:
\begin{gather}
\mu_{p_i} = \hat { p_i } = \frac { n _i + \frac{1}{2} } { N + \frac{1}{2}\text{~}b } \\ 
  {\hat{\sigma}_{p_i}}^2 = {\langle {p_i}^2 \rangle - \langle { p_i } \rangle^2 } = { \frac{\hat{ p_i } \left(1-\hat{ p_i } \right)}{N+1+\frac{1}{2}\text{~} b}  } 
\end{gather}

\noindent These expressions for the Jeffreys estimates are identical to those for the generalized Rule of Succession, except that a pseudocount of $\frac{1}{2}$ instead of $1$ is used in each bin.
Note that an alternative argument for estimating information content using Shannon entropy \citep{shannonA,shannonB} instead of Fisher information as the information measure 
leads back to the Rule of Succession and the Uniform Bayes prior \citep{dimitrov,jaynesMaxEnt1,jaynesMaxent2}.\footnote{
Presuming a higher relative importance of logarithmic differences can lead to the paradoxical conclusion that there is less importance in learning about a change in the probability of death from 0.0010 to 0.9999, a factor of $\approx 10^3$ than
in learning about a change in the risk of death from $10^{-12}$ to $10^{-3}$, a factor of $10^9$. 
Information gain by $p \log{p}$ measures is nearly equal in both instances.
Emphasizing regions of $p$ with high logarithmic differences makes sense though if all events are of very low or very high probability.
}
Also note that Jeffreys's original derivation \citep{jeffreys} was not based on information theory, 
but on an invariance argument that holds only when Bayes prior and posterior densities are continuously differentiable over all possible outcome values for the estimates, $\hat{ p_{i}} $.
Such assumptions of continuity of the derivative do not hold strictly when proportions are estimated from counts of discrete events as is typically done.  
In such 'counting' experiments with integer results,  the resulting estimates $\hat{ p_i }$ may assume only a finite set of discrete outcome values for any given experiment \citep{rousseau2000,rousseau2002,wangBinom}.

\subsection{Posterior pdfs to Get Limits for the Confidence Intervals} 

\subsubsection {Posteriors for the Approximately Continuous Priors}

For estimators from each class of approximate continuous Bayes prior pdf (Jeffreys or Uniform), two ways were examined to estimate confidence interval limits $(\widehat{\delta_i})_\xi = (\hat{\delta}^-_i, \hat{\delta}^+_i)$,
each based on a different approximate continuous Bayes posterior pdf. 
The first possible posterior was the standard normal (Gaussian) pdf, $\mathcal{N}[\widehat{p_i},\widehat{\sigma^2_{p_i}}; \theta ]$.  
Use of this class of Bayes posterior can be justified by the law of large numbers since parameters defining the confidence intervals are expected to be valid in the limit of an infinite number of $N$-observation trials. 
Conveniently, one may use $\hat{p}_i$ and ${\hat{\sigma}_{p_i}}^2$ (Eq. 1-4) as parameters, based on the estimators from either prior above.
For such estimates from a single $N$-observation trial though, \emph{this form for the posterior is only approximate};
for finite numbers of $N$-observation trials, some values of $p_{i,true}$ close to $0$ or $1$ cause $\mathcal{N}[\widehat{p_i},\widehat{\sigma^2_{p_i}}; \theta ]$ 
to have a substantial amount of probability density for $\theta$ lying outside the range of possible values for $p_i$.

Instead of the clearly approximate form $\mathcal{N}[\widehat{p_i},\widehat{\sigma^2_{p_i}}; \theta]$, one may use a Beta-pdf, $\mathcal{B}[\alpha,\beta; \theta]$, the "exact" form for the Bayes posterior, as advocated in \citep{brownStatSci,brownAnnals}. 
Since $supp(\mathcal{B}[\alpha,\beta; \theta]) = [0,1]$ or $(0,1)$ no density lies outside the range of possible values for $p_i$, but intervals from $\mathcal{B}[\alpha,\beta; \theta]$ are generally not symmetric about $\widehat{p_i}$.  
To estimate confidence intervals ($\hat{\delta}$)$_\xi$ using $\mathcal{B}[\alpha,\beta; \theta]$ as the posterior, presumed continuity of $\theta$ and integration allow parameters $\alpha$ and $\beta$ 
to be updated by adding the number of 'successes' and 'failures' counted in an experiment to initial values, $\alpha_0$ and $\beta_0$, from the Bayes prior pdf:
\begin{equation}
f(\theta) \text{~} d\theta = \mathcal{B}[n_i + \alpha_0,~ N-n_i + \beta_0;~ \theta] \text{~} d\theta = K~\left( \theta \right)^{n_i+\alpha_0-1} \text{~} (1- \theta )^{N-n_i+\beta_0-1} \text{~} d\theta 
\end{equation}
$K$ is a normalization constant.  
For the Jeffreys-Bayes prior, $\alpha_0= \frac{1}{2}$, $\beta_0= (b-1) \frac{1}{2}$ and for the Uniform Bayes prior, $\alpha_0=1$, $\beta_0={b-1}$.
This is seen by comparing expressions for expected values $\mu_{p}$ and $\sigma_{p}$ (Eq.1-4) with expressions for $\mu(\alpha,\beta)$ and $\sigma(\alpha,\beta)$ for a generalized beta pdf, $\mathcal{B}[\alpha,\beta; \theta]$.
By Eq. 5, other than adding $\alpha_0$ and $\beta_0$, the experimental counts are used directly to form parameters $\alpha$ and $\beta$ for the $\mathcal{B}[\alpha,\beta; \theta]$ posterior.
For either continuous posterior, $f(\theta)$, confidence interval limits $\delta^{-}$ and $\delta^{+}$ at a confidence value $\xi$ derive from:
\begin {gather}
\Intop_{-\infty}^{\delta^{-}} f(\theta) d\theta = \frac{1-\xi}{2} \text{~and~} \Intop_{\delta^{+}}^{\infty} f(\theta) d\theta = \frac{1-\xi}{2} 
\end {gather}
\noindent for equal-tailed confidence intervals.
When $\mathcal{N}[\mu_{p_i},~\sigma^2_{p_i};~\theta]$ is chosen as $f(\theta)$ using $\mu_{p_i}$ and $\sigma_{p_i}$ from Eq.1-4,
then good agreement between $C$ and $\xi$ is obtained by choosing $\hat{\delta}^{-} = max(0,\delta^{-})$ and  $\hat{\delta}^{+} = min(1,\delta^{+})$ \citep{agrestiCoull}.
For Bayes classification, sufficient $\xi$-to-$C$ matching required only a few decimal places of accuracy in $\hat{\delta}^{+}$ and $\hat{\delta}^{-}$, but sufficiently consistent run-to-run accuracy was initially absent in our Bayes classification test case.

One should note that whereas $\mathcal{B}[\alpha,\beta; \theta]$ is considered an exact posterior, $P(\hat{p_i}$$=$$\theta|X$$=$${n_i})$, 
its derivation and typical application to derive expressions for $\hat{p}_i(n_i,N)$ and $\hat{\delta}_i(n_i,N)$ 
presume that the experimental value $\theta$ varies approximately continuously for purposes of differentiation and integration.
Application of $\mathcal{N}[\widehat{p_i},\widehat{\sigma^2_{p_i}}; \theta ]$ as the posterior also presumes such continuity.
Continuity of outcomes is not met in experiments to determine proportions based on counting small discrete numbers of events \citep{rousseau2000,rousseau2002,wangBinom}.
In essence, this presumption of continuity where it is lacking causes the "exact" $\mathcal{B}[\alpha,\beta; \theta]$ posterior to be another approximation.
$\mathcal{B}[\alpha,\beta; \theta]$ can be an exact posterior, but only when its argument $\theta$ may assume all values in a continuous range ($\theta\in\mathbb{R}$). 
Otherwise integrals to derive $\hat{p}_i(n_i,N)$ and $\hat{\delta}_i(n_i,N)$ presume that $\theta$ may have been defined by observed fractional occurrences. 
For an $N$ observation \emph{binomial} trial, the range of estimator $\hat{p}(n,N)$ consists of only $N+1$ possible outcome values, $\Theta_N = \theta_j \in \{\theta_0, \theta_1, \theta_2 \dots \theta_N\}$.
For an $N$ observation, $b$-bin multinomial trial having a $b$-dimensional observation vector $\textbf{n} = (n_1, n_2, n_3, \dots n_{b})^T $ of integer outcome values,
the much larger number of possible values for $\hat{p}_i(\textbf{n},N)$ from $P(\hat{p_i}$$=$$\theta|\textbf{n})$ is equal to the number of combinations of $b$ integers that sum to $N$, $\binom{N+b-1}{b-1}$. 

The continuity approximation generally works fairly well, particularly for large $N$ in the binomial case, but as others have noted \citep{rousseau2000,rousseau2002,wangBinom} and in tests here,
the relative accuracy of this approximation differs slightly for different presumed forms for the continuous posterior. 
Sections 1.6-1.7 compare the relative accuracy found for different continuous approximations to the posterior.
An alternative form for estimators $\hat{p_i}$ based on discrete forms for the Bayes prior and posterior pdfs is derived in Appendix B.

\subsection {Calculation of Coverage}

\begin{figure}
\centering
\scalebox{0.85} {
\centering
\includegraphics[width=15 cm]{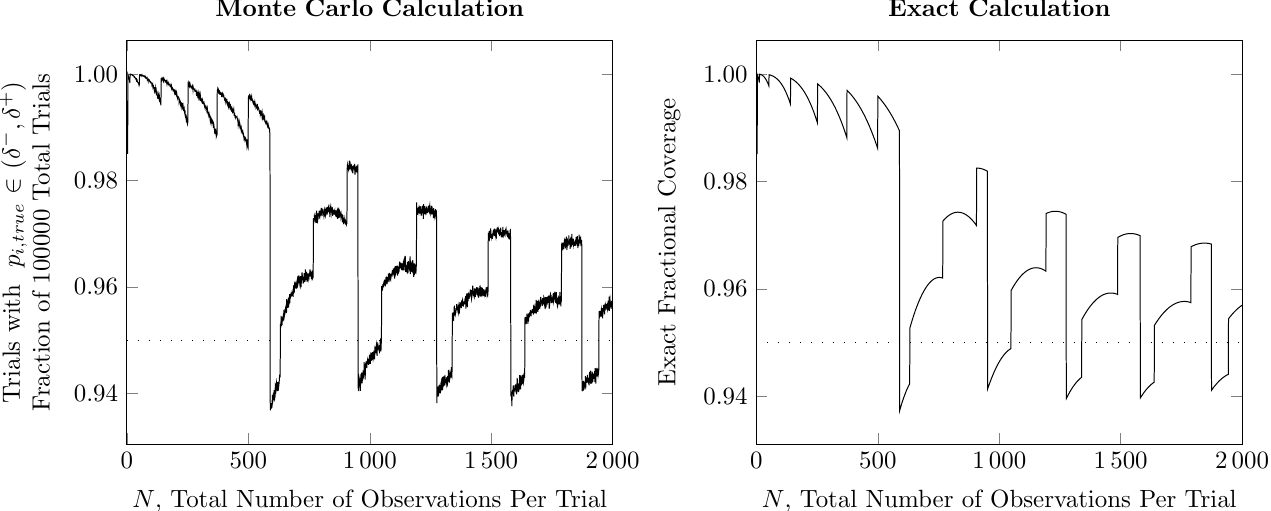}
}
                \caption{ 
                        A comparison of MC random binomial variates versus exact \citep{wangBinom} methods to calculate coverage, $C(N)$.
                        $C(N)$ is shown for fixed $p_{true}=0.005$ using $\hat{p}$ and $\hat{\sigma}_p$ based on $\mathcal{U}[0,1;x] = \mathcal{B}[1,1;x]$ priors (Eq.1-2) to get $\hat{\delta}_\xi$ based on $\mathcal{N}[\hat{p}, \hat{\sigma_p};x]$ posteriors (Eq.6).
                        Compare with Fig. 2 of \citet{brownStatSci}.  
                        $C$, the fraction of trials with $p_{true}$ $\in$ ($\hat{p}$-$1.96\hat{\sigma_p}$, $\hat{p}$+$1.96\hat{\sigma_p}$), is expected to match $\xi=0.95$. 
                        Negative deviations (values of $C < 0.95$) remain fairly close to 0.95 for all values of $N$ from $1$ to $2000$. 
                        Transitions of $C(N)$ about $\xi=0.95$ occur at values of $N$ that are 2 less than found earlier for $\hat{\delta}_\xi$  from $\mathcal{B}[\frac{1}{2}, \frac{1}{2}; x]$ prior $\times \hspace{2pt} \mathcal{B}[\alpha,\beta; x]$ posterior \citep{brownStatSci},   
                        but $C$ remains near or above 95\%. \label{fig:MCvsWang}}
\end{figure}

Uncertainties about which known estimators $\hat{p_i}$ and $\hat{\delta_i}$ were sufficiently accurate for Bayes classification led us to compare $\xi$-to-$C$ matching for some of the choices. 
Similar, fairly recent tests \citep{brownStatSci,brownAnnals} compared binomial $\hat{p}$ and $\hat{\delta}$ based on the continuous $\mathcal{B}[\frac{1}{2},\frac{1}{2};\theta]$ Jeffreys-Bayes prior with earlier asymptotic estimators \citep{agrestiCoull,laplace,wald2,wilson}, 
but $\hat{p}$ and $\hat{\delta}$ from the continuous $\mathcal{U}[0,1;x]$-Bayes prior were omitted.   
More recent comparisons including this prior for $\mathcal{B}[\alpha,\beta; \theta]$ posterior are in \citep{newcombe}.
For more complete comparison including the  $\mathcal{N}[\widehat{p_i},\widehat{\sigma^2_{p_i}}; \theta ]$ posterior and to test computational methods, continuous Bayes estimators $\hat{p}$ and $\hat{\delta}$ were first compared for the simpler binomial case.
Initially, 
$C$ was calculated from random binomial variates ($b=2$) using estimated confidence interval limits $(\widehat{\delta_i})_\xi$ from either the Jeffreys or Uniform Bayes prior.
Later, use was made of a discrete summation \citep{wangBinom}, from which $C$ is calculated exactly, given the lower and upper limits of the confidence intervals $(\delta^{-}_{i|n_i},\delta^{+}_{i|n_i})$ for each of the $N+1$ possible outcomes $n_i\in\{0, \dots N\}$ for a given sample size $N$ (Fig \ref{fig:MCvsWang}).  

From \citet{brownStatSci,brownAnnals}, it was clear that $(\widehat{\delta_i})_\xi$ based on the Jeffreys-Bayes prior were preferred over those from earlier \emph{asymptotically valid} estimators, such as the Wald estimator \citep{wald2,laplace}
for ($N < 100$).
However, for many values of $p_{true}$, at least 20-50 experiments were required 
to avoid particularly deviant $N$ with low $C$ values ($ \xi-C > 1-\xi$;  $\xi \gg C$) that can lead to erroneous Bayes classification.
For most values of $p_{true}$, deviant $N$ are absent for $N>50$, but for small values of $p_{true}$ deviant $N$ are present at larger values of $N$ ($N>100$, $N>1000$).
It may not be practical to repeat difficult, time sensitive, or expensive experiments the required 20-50 times to assure that the interval ($\hat{\delta}$)$_\xi$ 
is actually reliable at nominal confidence level $\xi$, as judged by calculated $C$.
More relevantly for a 100-bin multinomial case used in Bayesian classification, small-valued $p_i$ 
typically occur throughout the entire histogram since $b$ is large, and these
require even larger $N$ to avoid deviant $N$.
To know when the final predictions from statistical modeling were reliable, it was important 
to have confidence intervals ($\hat{\delta}$)$_\xi$ for which $C$ reliably reflected the nominal confidence $\xi$
in addition to accurate estimates of the observation frequency $p_i$.

Considering application to Bayes classification, even with $\xi$-to-$C$ consistency for most of the many required estimates $\widehat{p_i}$, if there are chance large inconsistencies in enough instances, this could cause incorrect final classification outcomes.
For commonly encountered values of $N$ and $p_{i,true}$, $C$ and $\xi$ often differed by significant amounts at the first or second decimal place.
Overestimates of $|\delta^+ - \delta^-|$, avoid false classification, but increase the number of cases deemed to be indeterminable by decreasing the overall confidence. 
Underestimates of $|\hat{\delta}^+ - \hat{\delta}^-|$ are more serious, leading to $\xi \gg C$, undue confidence, and therefore misclassification.
Whereas reducing measurement uncertainty $\sigma_{p_i}$, and hence  $|\delta^+ - \delta^-|$, for a given $N$ is the ultimate goal of experimental design, this is only useful if the estimates $\widehat{\sigma_{p_i}}$ are themselves accurate.
Given the need for only two to three-place accuracy in $\delta^-$ and $\delta^+$, 
and the simplification from using well-known results about propagation of uncertainty that a $\mathcal{N}[\mu, \sigma^2;  \theta]$ posterior permits, 
the degree of $\xi$-to-$C$ matching was compared for ($\hat{\delta}$)$_{\xi=0.95}$ from both the $\mathcal{N}[\mu_{p_i}, \sigma^2_{p_i}; \theta]$ and the "exact" $\mathcal{B}[\alpha,\beta;x]$ posterior pdf.

\subsection {Initial Comparisons for the Binomial Case ($b=2$)}

\subsubsection{The $\mathcal{N}[\mu_{p_i}, \sigma^2_{p_i}; \theta]$ Bayes Posterior and $\xi$-to-$C$ Consistency }

\begin{figure}
\centering
\scalebox{0.85} {
\centering
\includegraphics[width=15 cm]{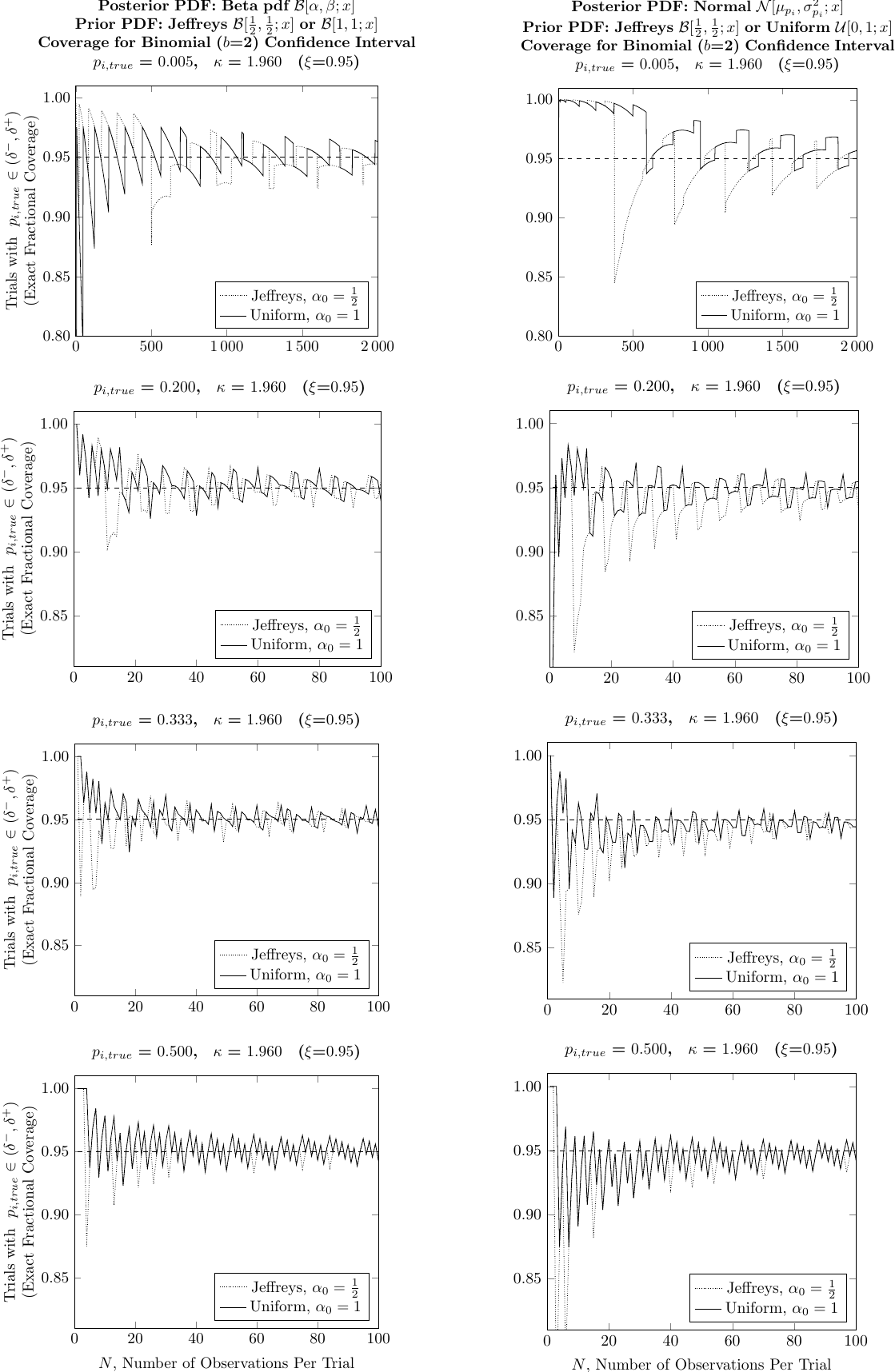}
}
                \caption{ Binomial Case ($b=2$):  Comparison of $C(N)$ at fixed $p_{i,true}$ for 95\% confidence intervals based on either the $\mathcal{B}(\alpha,\beta;x)$ beta-pdf (left, Eq.5, $\beta_0 = \alpha_0$ ) or the $\mathcal{N}(\hat{\mu}_p,\hat{\sigma}_p)$ normal Bayes posteriors (right), 
                         and either the Jeffreys ($\alpha_0 = \frac{1}{2}$) or Uniform ($\alpha_0 = 1$) Bayes Priors. 
                         Oscillation amplitudes about $\xi=0.95$ decrease with increasing $N$ outside these ranges of maximal difference.
                        \label{fig:varyNju}}
\end{figure}

When intervals ($\hat{\delta}_i$)$_{\xi=0.95}$ in the $b=2$ binomial case are determined from Eq. 6 using $\mathcal{N}[\mu_{p_i}, \sigma^2_{p_i}; \theta]$ as Bayes posterior $f(\theta)$, 
calculated $C$ values are practically identical when either the uniform $\mathcal{U}[0,1;x]$ ($= \mathcal{B}[ 1, 1 ; \theta]$; $\mu_{p_i}$, $\sigma_{p_i}$ from Eq.1-2) 
or for the Jeffreys-Bayes  $\mathcal{B}[\frac{1}{2},\frac{1}{2} ; \theta]$ prior pdfs ($\mu_{p_i}$, $\sigma_{p_i}$ from Eq.3-4) are used,
over most values of $N$ and $p$.
However, in regions for which $\xi$-to-$C$ consistency differs most between $\hat{p}$ and $\hat{\sigma}$ from these two priors (Fig \ref{fig:varyNju}),  
$\mathcal{U}[0,1;x]$ is generally the choice that leads to fewer error-prone situations for which $ \xi-C > 1-\xi$.

\begin{figure}
\centering
\scalebox{0.85} {
\centering
\includegraphics[width=15 cm]{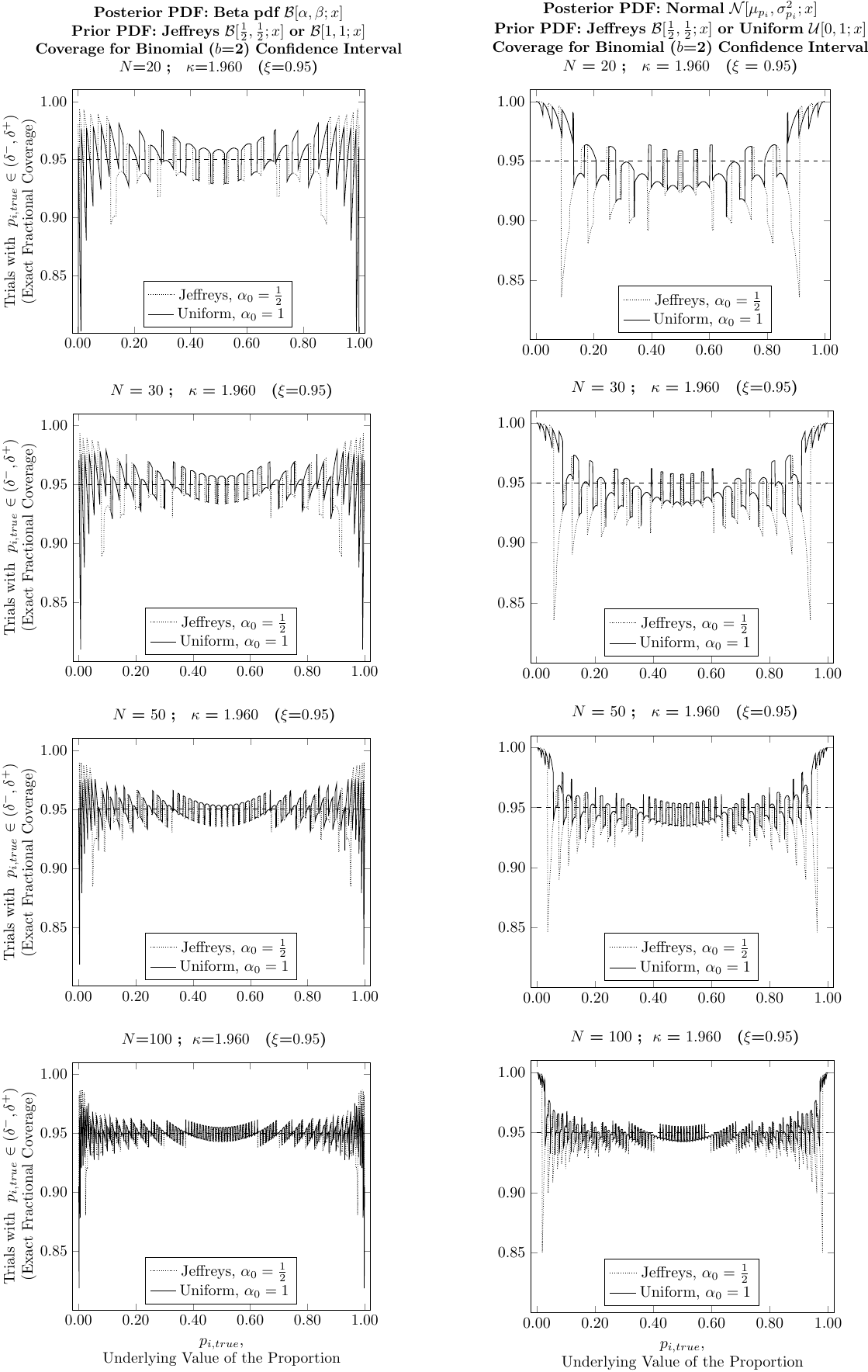}
}
                \caption{ Binomial Case ($b=2$):  Comparison of $C(p_{i,true})$ at fixed $N$ for 95\% confidence intervals based on either the  $\mathcal{B}(\alpha,\beta;x)$ beta-pdf (left, Eq.5, $\beta_0 = \alpha_0$ ) or  $\mathcal{N}(\hat{\mu}_p,\hat{\sigma}_p)$ normal Bayes posteriors (right), 
                        and either the Jeffreys ($\alpha_0 = \frac{1}{2}$) or Uniform ($\alpha_0 = 1$) Bayes Priors.
                        \label{fig:varyPju}}
\end{figure}

\begin{figure}
\centering
\scalebox{0.85} {
\centering
\includegraphics[width=15 cm]{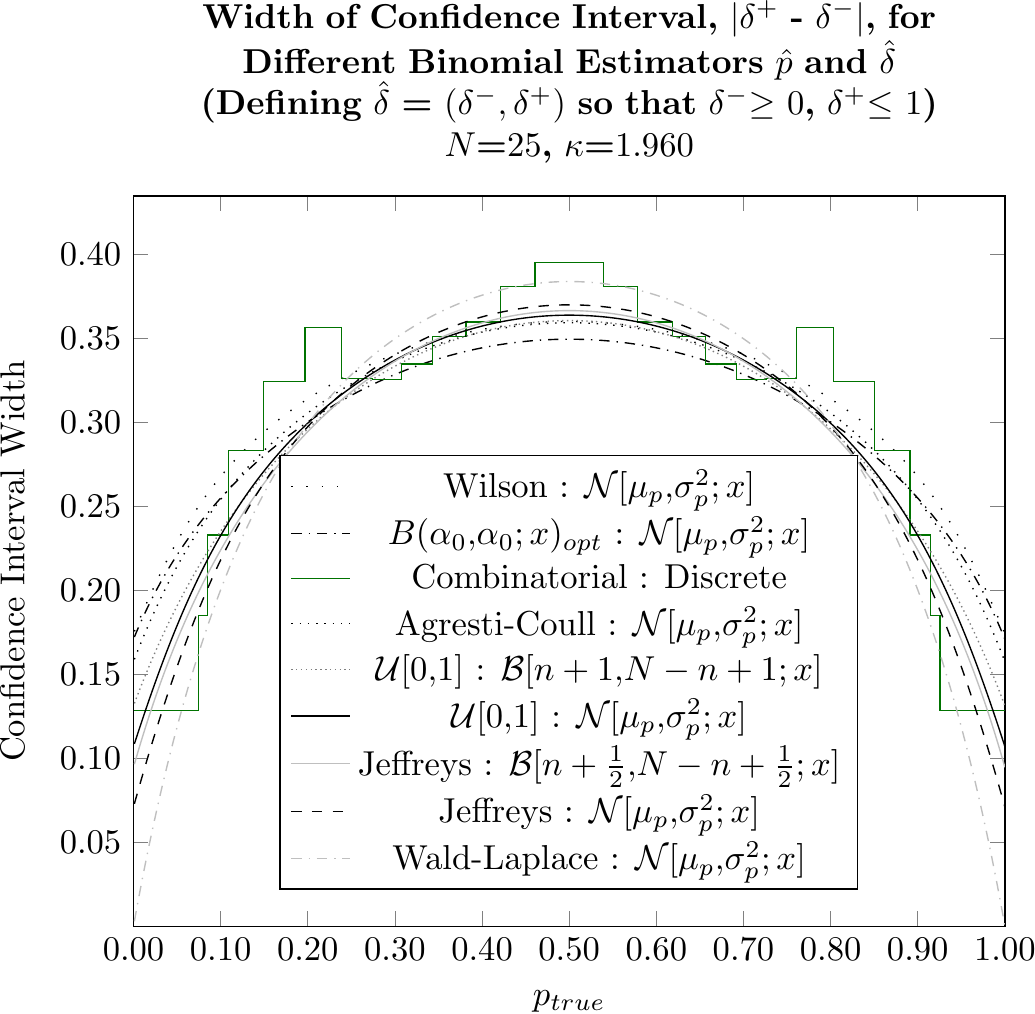}
}
                \caption{ Comparison of $|\delta^+-\delta^-|$ at $\xi=0.95$ for different estimators of intervals $\delta$ for binomial $p$ \citep{agrestiCoull,laplace,wald2,wilson} for the particular case $N=25$. 
Bayes posterior pdfs include the continuous  $\mathcal{B}[n+\alpha_0,N-n+\alpha_0;x]$ or $\mathcal{N}[\mu_p, \sigma^2_p; x]$ posteriors, or the discrete posterior of Appendix B.
Use of $\mathcal{N}[\mu_p, \sigma^2_p; x]$ requires estimates $\mu_p$ and $\sigma^2_p$ from Eq.1-4.
In the legend, plots are identified as "Prior (or Estimator Type) : Posterior".
Compare with Fig. 8 of \citep{brownStatSci}.  \label{fig:intervalLengths}
             }
\end{figure}

Use of the $\mathcal{U}[0,1;x]$ prior with the $\mathcal{N}[\mu_{p_i}, \sigma^2_{p_i}; \theta]$ posterior avoids such error-prone mismatches between $C$ and $\xi$ for all but the very lowest values of $N$ \emph{total} observations. 
For small $p_{true}$ values, the combination of Jeffreys $\mathcal{B}[\frac{1}{2},\frac{1}{2} ; \theta]$ prior with the $\mathcal{N}[\mu_{p_i}, \sigma^2_{p_i}; \theta]$ posterior exhibits error-prone mismatches ($C \ll \xi$) at some moderately large values of $N$, 
approaching several 100 to several 1000.
These values for $p_{true}$ and $N$ are in a range typically encountered in the histogram analysis used for classification.
When large $| C - \xi|$, is found for the $\mathcal{U}[0,1;x]$ prior $\times~\mathcal{N}[\mu_{p_i}, \sigma^2_{p_i}; \theta]$ posterior combination, typically for some smaller values of $p_{true}$, 
it is always in the direction for which $C > \xi$, usually $C \approx 1$, avoiding errors leading to false Bayes classification outcomes (Fig \ref{fig:varyNju}).
When maximally large $| C - \xi|$ are found from the the Jeffreys $\mathcal{B}[\frac{1}{2},\frac{1}{2} ; \theta]$ prior  $\times~\mathcal{N}[\mu_{p_i}, \sigma^2_{p_i}; \theta]$ combination, $| C - \xi|$ is typically larger, with  $C < \xi$, extending closer to error-prone regions of low $C$ values.  
In plots at constant $N$ and varying $p_{i,true}$ [Fig \ref{fig:varyPju}, $N$ values as in \citet{brownStatSci,brownAnnals}], estimates ($\hat{\delta}_i$)$_{\xi=0.95}$ from $\mathcal{U}[0,1;x]$ lead to increased  $| C - \xi|$ at small values of $p_{i,true}$,
but again $C > \xi$ always and error-prone situations for which $ \xi-C > 1-\xi$ are avoided.
The Jeffreys prior has some error-prone situations for which $C \ll \xi$.
Interval lengths for ($\mathcal{U}[0,1;x]$ prior $\times \mathcal{N}[\mu_{p_i}, \sigma^2_{p_i}; \theta]$ posterior) are comparable to those for earlier estimators (Fig \ref{fig:intervalLengths}).

\subsubsection{The $\mathcal{B}[\alpha,\beta;x]$ Bayes Posterior and $\xi$-to-$C$ Consistency }

When $\mathcal{B}[\alpha,\beta;x]$ beta-pdf Bayes posteriors are used 
to derive ($\hat{\delta}_i$)$_{\xi=0.95}$ from Eq. 5-6, 
$\xi$-to-$C$ matching is fairly consistent when either the
$\mathcal{U}[0,1;x]$ ($=\mathcal{B}[1,1 ; \theta]$ ) or the Jeffreys $\mathcal{B}[\frac{1}{2},\frac{1}{2} ; \theta]$ priors are used, 
but there are marked points of inconsistency for both.  
With the Jeffreys prior ($\mathcal{B}[\frac{1}{2},\frac{1}{2} ; \theta]$ prior $\times\hspace{2pt}\mathcal{B}[\alpha,\beta;x]$ posterior), 
$C < \xi$ at low to intermediate values of $p_{i,true}$, but usually only slightly, for several small values for $N$ (Fig \ref{fig:varyNju},\ref{fig:varyPju}).
In general, $C < \xi$ fairly often, but with $|C - \xi|$ of limited magnitude.  
For $p_{true}$ in $(0,1)$, the overall $\langle (\xi-C)^2 \rangle$ for the other estimator pair based on
$\mathcal{U}[0,1;x]$ prior $\times\hspace{2pt}\mathcal{B}[\alpha,\beta;x]$ posterior, 
is smaller than for Jeffreys $\mathcal{B}[\frac{1}{2},\frac{1}{2} ; \theta]$ prior $\times\hspace{2pt}\mathcal{B}[\alpha,\beta;x]$ posterior,
but $|\xi-C|$ exhibits large fluctuations near $p_{true}=0$ and $p_{true}=1$.
Points in these relevant ranges for $p_{true}$ exhibit $C \ll \xi$, conducive to catastrophic errors in Bayes classification.
The overall $\langle (\xi-C)^2\rangle$ value for Jeffreys $\mathcal{B}[\frac{1}{2},\frac{1}{2} ; \theta]$ prior $\times \hspace{2pt} \mathcal{B}[\alpha,\beta;x]$ posterior, 
although larger than for $\mathcal{U}[0,1;x]$ prior $\times\hspace{2pt}\mathcal{B}[\alpha,\beta;x]$ posterior,
is still smaller than it was for Jeffreys $\mathcal{B}[\frac{1}{2},\frac{1}{2} ; \theta]$ prior $\times \hspace{2pt} \mathcal{N}[\mu_{p_i}, \sigma^2_{p_i}; \theta]$ posterior.

\subsubsection{Comparing the Best Two Continuous Prior-Posterior Combinations}
When comparing $\mathcal{U}[0,1;x]$ prior $\times~\mathcal{N}[\mu_{p_i}, \sigma^2_{p_i}; \theta]$ posterior versus
 $\mathcal{B}[\frac{1}{2},\frac{1}{2} ; \theta]$ prior $\times~\mathcal{B}[\alpha,\beta;x]$ posterior,
the two "best" Bayes prior-posterior combinations leading to the fewest large $\xi$-to-$C$ inconsistencies,
interval lengths $|\delta^+ - \delta^-|$ for these two were often practically identical (Fig \ref{fig:intervalLengths}), and the average
 $\langle (\xi-C)^2 \rangle$ was often slightly lower for the $\mathcal{B}[\frac{1}{2},\frac{1}{2} ; \theta]$ prior $\times~\mathcal{B}[\alpha,\beta;x]$ posterior combination.
However, despite better \emph{average} $\xi$-to-$C$ consistency with $\mathcal{B}[\frac{1}{2},\frac{1}{2} ; \theta]$ prior $\times~\mathcal{B}[\alpha,\beta;x]$ posterior,
this consistency is occasionally interrupted by unpredictable coincidental situations with increased $\xi$-to-$C$ inconsistency, larger maximal errors with $C < \xi$ or $C \ll \xi$, 
particularly in the relevant range of small $p_{true}$ and large to moderate $N$.
Such large, albeit infrequent, inconsistencies for unpredictable combinations of $N$ and $p_{true}$ values in the range relevant to histogram analysis reduce the reliability of classification. 
Since $\xi$-to-$C$ mismatch with $C < \xi$ led to more problems for Bayes classification, 
since the $\mathcal{U}[0,1;x]$ prior $\times~\mathcal{N}[\mu_{p_i}, \sigma^2_{p_i}; \theta]$ posterior combination had comparable $\langle (\xi-C)^2 \rangle$, 
and since it simplifies analysis of propagation of uncertainty, the $\mathcal{U}[0,1;x]$ prior $\times~\mathcal{N}[\mu_{p_i}, \sigma^2_{p_i}; \theta]$ posterior combination was the initial choice for trying Bayes classification.
First though, a few multinomial $\xi$-to-$C$ comparisons were examined for some relevant small values of $p_{i,true}$.

\subsection {Initial Comparisons for the Multinomial Case ($b>2$)}

\subsubsection{Trends for $\xi$-to-$C$ Matching Apparently Similar for $b>2$, Low $p_{i,true}$}
\begin{figure}
\centering
\scalebox{0.85} {
\centering
\includegraphics[width=15 cm]{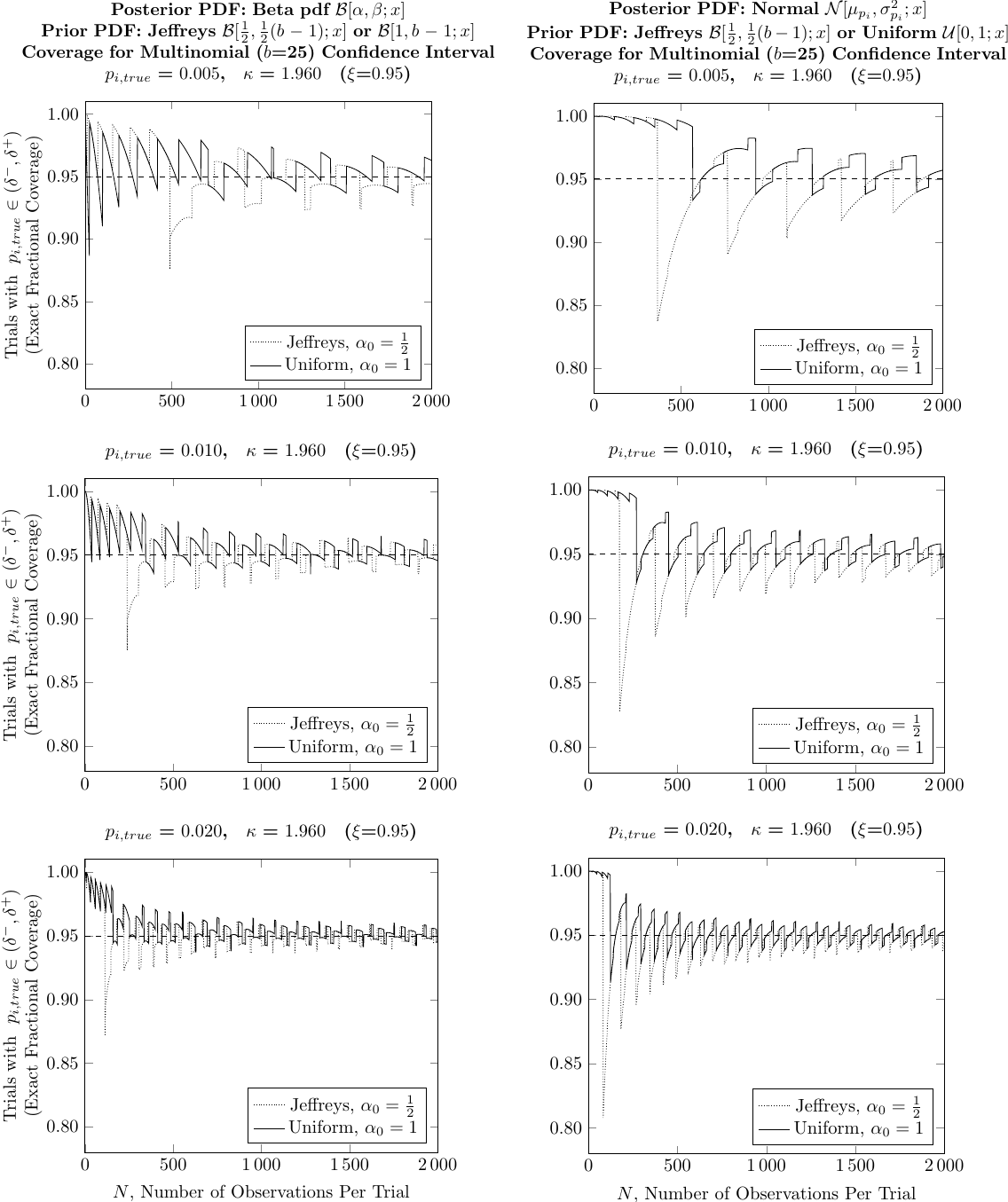}
}
                \caption{The 25-Bin Multinomial Case:  Comparison of coverage for 95\% confidence intervals for either the beta- (left, Eq.5, $\beta_0 = 24\alpha_0$ ) or normal  (right, $\sigma$ from Eq.2 or Eq.4, $b=25$) Bayes posterior pdfs, and either the Jeffreys ($\alpha_0 = \frac{1}{2}$) or Uniform ($\alpha_0 = 1$) Bayes Prior pdf at small fixed values of $p_{i,true}$ for the 25 bin multinomial case.  
                        \label{fig:multiLowPb25}}
\end{figure}

Similar conclusions about prior-posterior choice were reached for multinomial intervals, 
using $b>2$ in Eq. 1-4 for parameters in the $\mathcal{N}[\mu_{p_i}, \sigma^2_{p_i}; \theta]$ posterior 
or $\beta_0 = (b-1)\alpha_0$, $\alpha_0\in\{\frac{1}{2},1\}$ in Eq. 5 for the $\mathcal{B}[\alpha,\beta;x]$ posterior. 
Initially, only values of $p_{i,true} < 0.05$ were examined for $b>2$ because these values occurred often for large $b$ and they sometimes led to larger $|C -\xi|$ for relevant-sized $N$ in the earlier $b=2$ case.
As had been seen for $b=2$, 
intervals ($\hat{\delta}_i$)$_{\xi=0.95}$ from the $\mathcal{U}[0,1;x]$ prior $\times~\mathcal{N}[\mu_{p_i}, \sigma^2_{p_i}; \theta]$ posterior combination led more consistently to $C\approx\xi$ and $C \ge \xi$
than ($\hat{\delta}_i$)$_{\xi=0.95}$ from the Jeffreys $\mathcal{B}[\frac{1}{2},\frac{b-1}{2} ; \theta]$ prior $\times~\mathcal{B}[\alpha,\beta;x]$ posterior combination
for histogram of moderate $b$, $b \in \{6,10,25\}$ (Fig \ref{fig:multiLowPb25}). 

%

Initial applications of estimates of $\hat{p}_i$ and $\hat{\delta}_i$ from the $\mathcal{U}[0,1;x]$ prior $\times~\mathcal{N}[\mu_{p_i}, \sigma^2_{p_i}; \theta]$ posterior combination to Bayes classification for known test cases provided fairly consistent results.
However, analyses of more demanding experimental test data led to possible inconsistencies, 
with some data sets having samples incorrectly classified yet estimated to have high certainty ($ |P_o - P_{crit}| \gg 4\sigma_{P_o} $)
and other data sets having a high fraction of correctly classified samples based on $sgn(P_o-P_{crit})$, but unexpectedly consistently large values $\hat{\sigma}_{P_o}$. 
Estimators of $\hat{p}_i$ and $\hat{\delta}_i$ were thus reexamined to try to rationalize these apparent anomalies.

\subsubsection{Large Inconsistency at High $p_{i,true}$ for Both Priors}
\begin{figure}
\centering
\scalebox{0.85} {
\centering
\includegraphics[width=15 cm]{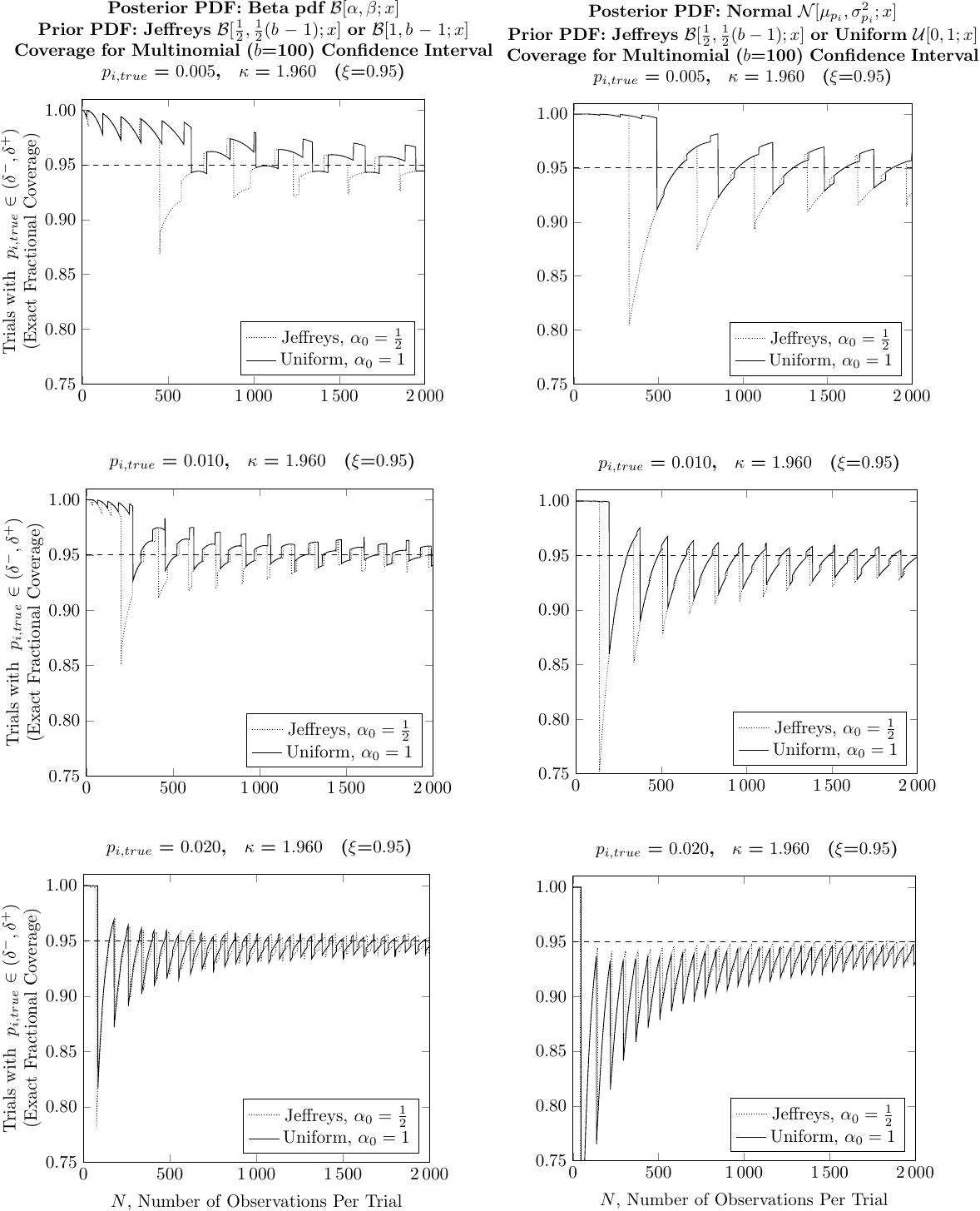}
}
                \caption{ The 100-Bin Multinomial Case (small $p_{i,true}$):  Comparison of $C$ for ($\hat{\delta}_i$)$_{\xi=0.95}$, 95\% confidence intervals, for either the beta-  (left, Eq.5, $\beta_0 = 99\alpha_0$ ) or normal  (right, $\sigma$ from Eq.2 or Eq.4, $b=100$)  Bayes posterior pdfs, and either the Jeffreys ($\alpha_0 = \frac{1}{2}$) or Uniform ($\alpha_0 = 1$) Bayes Prior pdf at \emph{small} fixed values of $p_{i,true}$ for the 100-bin multinomial case.  
                        \label{fig:varyNlowPb100}}
\end{figure}

\begin{figure}
\centering
\scalebox{0.85} {
\centering
\includegraphics[width=15 cm]{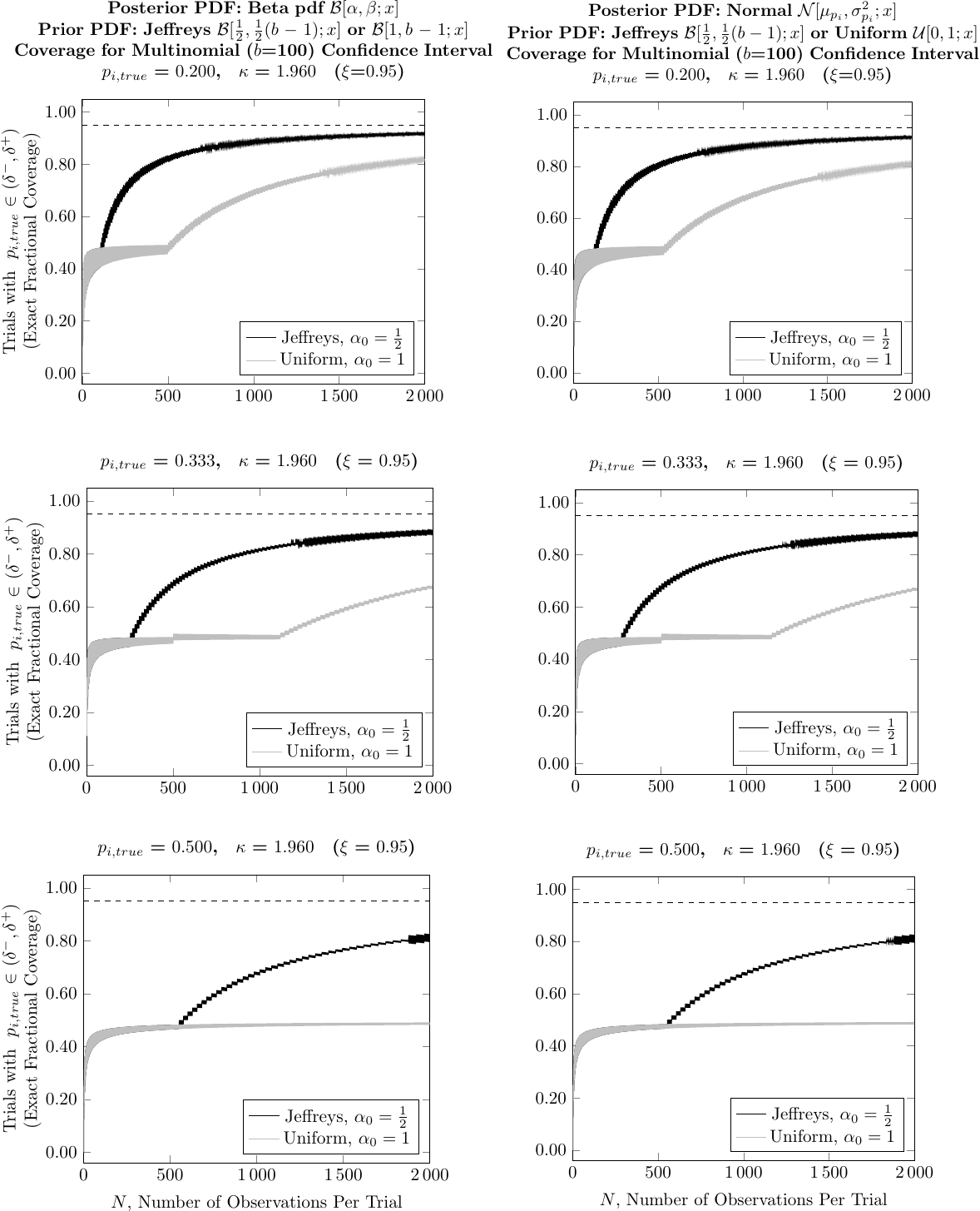}
}
                \caption{The 100-Bin Multinomial Case (large $p_{i,true}$):  Comparison of $C$ for 95\% confidence intervals for either the $\mathcal{B}[\alpha,\beta;x]$ -  (left, Eq.5, $\beta_0 = 99\alpha_0$ ) or $\mathcal{N}[p_i,\sigma_{p_i};x]$ (right, $\sigma$ from Eq.2 or Eq.4, $b=100$)  Bayes posterior pdfs, and either the Jeffreys ($\alpha_0 = \frac{1}{2}$) or Uniform ($\alpha_0 = 1$) Bayes Prior pdf at \emph{larger} fixed values of $p_{i,true}$ for the 100-bin multinomial case.  
                        \label{fig:varyNhighPb100}}
\end{figure}

Further $\xi$-to-$C$ comparison over histograms with more bins and a wider range of presumed $p_{i,true}$ revealed large $|\xi-C|$ with $C\ll \xi$ 
for estimates from \emph{both} initially considered pairs of continuous Bayes priors and posteriors (Fig \ref{fig:varyNlowPb100},\ref{fig:varyNhighPb100}).  
For $b>2$, the more significant cases of large $|\xi-C|$ occurred for \emph{larger} values of $p_{i,true}$ closer to $p_{i,true}=0.500$ that had been well-behaved in the binomial, $b=2$ case. 
Initial indications of these problems had begun to appear at small values of $N$ even at moderately low values of $p_{i,true}$ for the 100-bin case (Fig \ref{fig:varyNlowPb100}), but initially the significance of these had not been fully appreciated.
At $\xi=0.95$ with 'larger' values of $p_{i,true}$ ($p_{i,true} > 0.100$), values of $C$ fell to between $0.4$ and $0.6$ for sample sizes $N \sim {10}^2$ and remained well below $0.95$ for $N \sim {10}^3$ (Fig \ref{fig:varyNhighPb100}).

Whereas different Aristotelian arguments for using either the Jeffreys or Uniform Bayes priors based on information content each appear to have merit,
numerous other information measures are available \citep[see \emph{e.g.}][]{basseville2011,reidWilliamson2011,kumar2005}, each with their own merit as a basis for alternative Bayes prior pdfs.
Therefore, it appears that basing Bayes prior selection solely on information theory, is equivalent to replacing an arbitrary choice of a Bayes prior pdf with an arbitrary choice of information measure.
Furthermore both Jaynes's \citep{jaynesCh18} and Jeffreys's \citep{brownStatSci,brownAnnals,jeffreys} information-based criteria to select optimal Bayes priors presume priors and posteriors to be continuously differentiable with respect to the measurement value.
This makes any conclusions only approximate in the absence of continuity, when possible $\theta$ values are limited by discrete integer counts of observed occurrences.
Fine scale comparative predictions would then require confirmation by direct calculation.

For practical purposes though, statistical classification of drug sensitivity
required accurate estimators $\hat{p_i}$ and $\hat{\delta_i}$, and aside from (i) $\hat{p}_i$ matching $p_{i,true}$, sufficient accuracy demanded that
(ii) $| \delta^+_i - \delta^-_i |$ be sufficiently short and
(iii) the nominal confidence $\xi$ used to define ($\hat{\delta}_i$)$_{\xi}$ be matched or exceeded by the calculated coverage $C$ [ $C \approx \xi$ or at worst $C > \xi-(1-\xi) $; \{$C> 2\xi - 1$\} ] with no large errors at relevant $N$ and $p_{true}$ values.
Thus it was of interest to see if alternate choices of Bayes priors could provide estimators with improved $\xi$-to-$C$ matching relative to the Jeffreys or the Uniform Bayes prior. 
In the first attempt at such priors, use was made of a more general form for a continuous multinomial prior.

\section {Optimization of a Parametric Prior for $\xi$-to-$C$ Matching}
\subsection {A Generalized Parametric Multinomial Prior}
For multinomial distributions, a generalized Bayes prior
corresponding to the "multidimensional" $\mathcal{B}[\alpha_0,\beta_0; x]$ is a Dirichlet pdf: $\mathcal{B}[\alpha_1, \alpha_2, \dots ; x]$ where each bin has its own associated value for $\alpha$.
When a Dirichlet pdf is used as a Bayes prior though, most often a 'non-informative' Dirichlet Bayes prior is chosen \citep{jaynes,jaynesCh12,uninformative} for which all $\alpha_i$ share a common value $\alpha_0$.
For a general Dirichlet-Bayes prior pdf, the expressions for $\hat{p}_i$ and $\hat{\sigma}^2_{p_i}$ for use with the $\mathcal{N}[\hat{p}_i,\hat{\sigma}^2_{p_i};\theta]$  posterior are: 

$$
\mu_{p_i} = \hat { p_i } = \frac { n _i + \alpha_i } { N + \alpha_i + \Sum^b_{\substack{j=1\\ \hspace{-27 pt} j\neq i }} \alpha_j } $$. 
$$
 \widehat{{{\sigma}_{p_i}}^2} = {\langle {p_i}^2 \rangle - \langle { p_i } \rangle^2} = { \frac{\langle p_i \rangle \left(1-\langle p_i \rangle \right)}{N+1+\alpha_i + \Sum^b_{\substack{j=1\\ \hspace{-27 pt} j\neq i }} \alpha_j } } $$

\noindent and for the 'non-informative' case, when considering the parameters for a single bin $i$, this is equivalent to $\mathcal{B}[\alpha_0,\beta_0; x]$ with $\alpha_0=\alpha_i$ and $\beta_0 = \Sum^b_{\substack{j=1\\ \hspace{-27 pt} j\neq i }} \alpha_j = (b-1) \alpha_0   $:

\begin{gather}
\mu_{p_i} = \hat { p_i } =  \frac { n _i + \alpha_0 } { N + b\alpha_0 } \\ 
 \widehat{{{\sigma}_{p_i}}^2} = {\langle {p_i}^2 \rangle - \langle { p_i } \rangle^2} = { \frac{\langle p_i \rangle \left(1-\langle p_i \rangle \right)}{N+1+ b\alpha_0 } } 
\end{gather}

\noindent The earlier expressions (Eq.1-4) for either the Jeffreys or the Uniform-Bayes priors are clearly special cases of this general expression when all of the values for $\alpha_i$ here are either $\frac{1}{2}$ or $1$ respectively.
The  $\mathcal{B}[\alpha_0,\alpha_0;x]$ pdf is a more general special case for which $b=2$.

\subsection {Bayesian Priors with $\alpha_0$ as an Adjustable Hyperparameter}

The generality of the form  $\mathcal{B}[\alpha_0,(b-1)\alpha_0; x]$ as a non-informative prior for multinomial distributions based on the Dirichlet pdf is particularly convenient for numerical optimization of $\xi$-to-$C$ matching.
Only a single value $\alpha_0$ needs to be determined by $\xi$-to-$C$ optimization of estimators $\hat{p}_i$ and $(\hat{\delta}_i$)$_\xi$, 
but the optimal value for $\alpha_0$ in these estimators is expected to change with $N$, $b$, $\xi$, and perhaps $n_i$.
This numerical optimization of a parametric prior is less direct than an earlier approach for $\xi$-to-$C$ probability matching for discrete estimation \citep{rousseau2000,rousseau2002} 
that entailed determining factors for direct rescaling of the confidence interval to achieve $\xi$-to-$C$ matching.
That method was advocated for use in retrospective statistical analyses but was cautioned for use in predictive ones.
Although improved $\hat{p}_i$ and $\hat{\sigma}_{p_i}$ were needed for predictive Bayesian classification of drug sensitivity, 
part of the reason for pursuing optimization of $\alpha_0$ for $\xi$-to-$C$ matching was simply the curiosity about how
close the generally presumed values $\alpha_0=\frac{1}{2}$ (Jeffreys) and $\alpha_0=1$ (Uniform) in the prior were to the actual values that provided optimal for $\xi$-to-$C$ matching.
The severe $\xi$-to-$C$ mismatches that had been observed for the 100-bin multinomial example with $\alpha_0\in\{\frac{1}{2},1\}$ suggested possibly greater limitations about this class estimators that needed to be better understood.

The bulk of Section 2 entails finding the parameters $\alpha_0$ and $\beta_0$ in these generalized pdfs that optimize matching between $C$ and $\xi$ and characterizing the resulting $\hat{p_i}$ and $\hat{\delta_i}$.
A numerical de-noising method originally developed to improve optimization of $\alpha_0$ was also found, on its own, dramatically to improve the accuracy of estimates $\hat{p_i}$ and $\hat{\sigma}_{p_i}$ beyond that which can be obtained by simple estimators (Sec. 3).

\subsection{Optimizing $\alpha_0$ in $\mathcal{B}[\alpha_0,(b-1)\alpha_0;x]$ Prior: Subranges $p_{ik}$ of $(0,1)$ at Each $N$ }
Computational optimization of $\xi$-to-$C$ matching (minimization of $(\xi-C)^2$) with respect to $\alpha_0$ for was performed at each point on a grid of fixed values for $N$
and with trial $p_i$ values for optimization initially \emph{selected from fixed ranges} $p_{ik} \in (\psi_k,\psi_{k+1})$.
Since empirical functional relationships were not initially found between the optimal values for $\alpha_0$ on this ($N,k$)-grid, pre-optimized values of $\alpha_0$ could be read from a stored table.
It had been presumed that $\alpha_0$ might vary between different subranges $p_{ik}$ of (0,1), but that within each $p_{ik}$,
 $\alpha_0(p_{ik})$ would need to be sufficiently invariant, $\sigma^2_{\alpha_0(p_{ik})} \ll \mu^2_{\alpha_0(p_{ik})}$, for consistency of the prior using $\alpha_0(p_{ik})$.
While apparently restrictive, this is less stringent than the presumption made for either the Jeffreys or Uniform priors that $\alpha_0$ is invariant over the entire range $p_{i} \in (0,1)$.
To use tabulated ($\xi\leftrightarrow C$)-optimized $\alpha_0$ values to estimate $\hat{p}_{\xi\leftrightarrow C}$, values of $\alpha_0$ were selected from the table by a recursion using intermediate best estimates of values for $\alpha_0$ and $p_{ik}$.

During optimization of $\alpha_0$,  
exact calculation of $C$ by discrete summation \citep{wangBinom} obviated the need to perform many random trials.  
This procedure requires an exhaustive list of confidence interval limits $\left(\delta^{(-)}_i(\alpha_0), \delta^{(+)}_i(\alpha_0)\right), i\in\{0,1, \dots N\}$ for all possible discrete outcomes of a binomial or multinomial experiment with a fixed sample size of $N$ observations,
and does not specify how the set of confidence interval limits should be chosen.
Optimization of $\alpha_0$ is fraught with technical difficulties as might be discerned from the irregular abrupt variation in the earlier plots of $C(N)$ or $C(p_{true})$ for the binomial or multinomial estimators (Fig. \ref{fig:varyNju}, \ref{fig:varyPju}, \ref{fig:multiLowPb25}, \& \ref{fig:varyNlowPb100}).
Such jaggedness arises since the outcome space for experiments to estimate proportions is limited to integer counts ($n_i: \mathbb{Z}$).
It can cause numerical optimization of $C(\alpha_0)$ at fixed $N$ and limited $p_{true}$ to become trapped either at local minima of $(\xi-C)^2$ or at nondescript positions in flat regions of invariant $C$.

Despite such computational complications, 
optimization could be achieved by a Newton-Raphson procedure by treating Wang's discrete summation for $C$ as an approximately continuous function and using numerical estimates for $\frac{\partial{C}}{\partial{\alpha_0}}$ based on this presumption of smoothness and continuity.
In actuality, $C(\alpha_0)$ is jagged since the summation limits for $C$ vary and abrupt jumps occur as new terms enter or leave this summation.
Instabilities when minimizing $(\xi-C)^2$ are mitigated by that fact that 
only a single optimal parameter $\alpha_0$ is needed for each ($N,k$)-grid point.
One may easily identify and accommodate situations for which the numerical derivative becomes unreasonably too small or too large.  

Such modification of the discrete numerical derivatives of the presumed smooth form for the true coverage function, $C(\alpha_0)$, 
often prevents convergence to a fixed final optimal value for $\alpha_0$, instead leading to endless cycling in the vicinity of the optimum.  
To adjust for this, several near-optimal paired values for $\alpha_0$ and $\left(\xi-C(\alpha_0)\right)^2$ from intermediate cycles are stored for reference.
 Then, upon detecting such cycling, a direct search algorithm is invoked in the region of cycling, 
typically a fairly narrow range, to find nearby values of $\alpha_0$ that reduce $\left(\xi-C(\alpha_0)\right)^2$ .
Fortunately, the change in $C(\alpha_0)$ with changing $\alpha_0$ values is fairly monotonic, with few changes in the \emph{sign} of the discrete derivative.
This apparently provides values for $\alpha_0$ fairly close to the minimal values of $\langle(\xi-C)^2\rangle$ for $p_{i,true}\in(\psi_k,\psi_{k+1})$. 

\subsection{Observed $\xi$-to-$C$ Matching When Optimized $\alpha_0 = \alpha_0(N,b,p_{ik})$}
For \emph{multinomial} estimators (fixed $b>2$ in Eq.7-8) and the $\mathcal{N}[\mu_p,\sigma^2_p; x]$ posterior, optimized values of $\alpha_0$ varied significantly \emph{within} isolated ranges $p_{ik}\in(\psi_k,\psi_{k+1})$ at different $N$, contrary to the initial assumptions.
For the 100-bin multinomial prior pdf ($b=100$, $\mathcal{B}[\alpha_0, 99\alpha_0; x]$), 
minimal $\langle (\xi-C)^2 \rangle$ was found when $\alpha_0$ became essentially $0$ for most zones $p_{i,true} \in (\psi_k,\psi_{k+1})$, 
but $\alpha_0$ had significant nonzero values for zones of $p_{i,true}$ values close to either $0$ or $1$.
Testing showed that $C$ at individual $p_{i,true}$ within each range generally did not match $\xi$ particularly well using intervals from estimators $\hat{\delta}_n$ based on the near-zero $\alpha_0$ values.
On examining ever smaller optimization zones near $0$ or $1$, by repeated halving of the initial ranges, the optimal values for $\alpha_0$ eventually achieved a less jagged functional form, 
but still failed to provide $\hat{\delta}_n$ giving accurate $C$.

\begin{figure}
\centering
\scalebox{0.75} {
\centering
\includegraphics[width=15 cm]{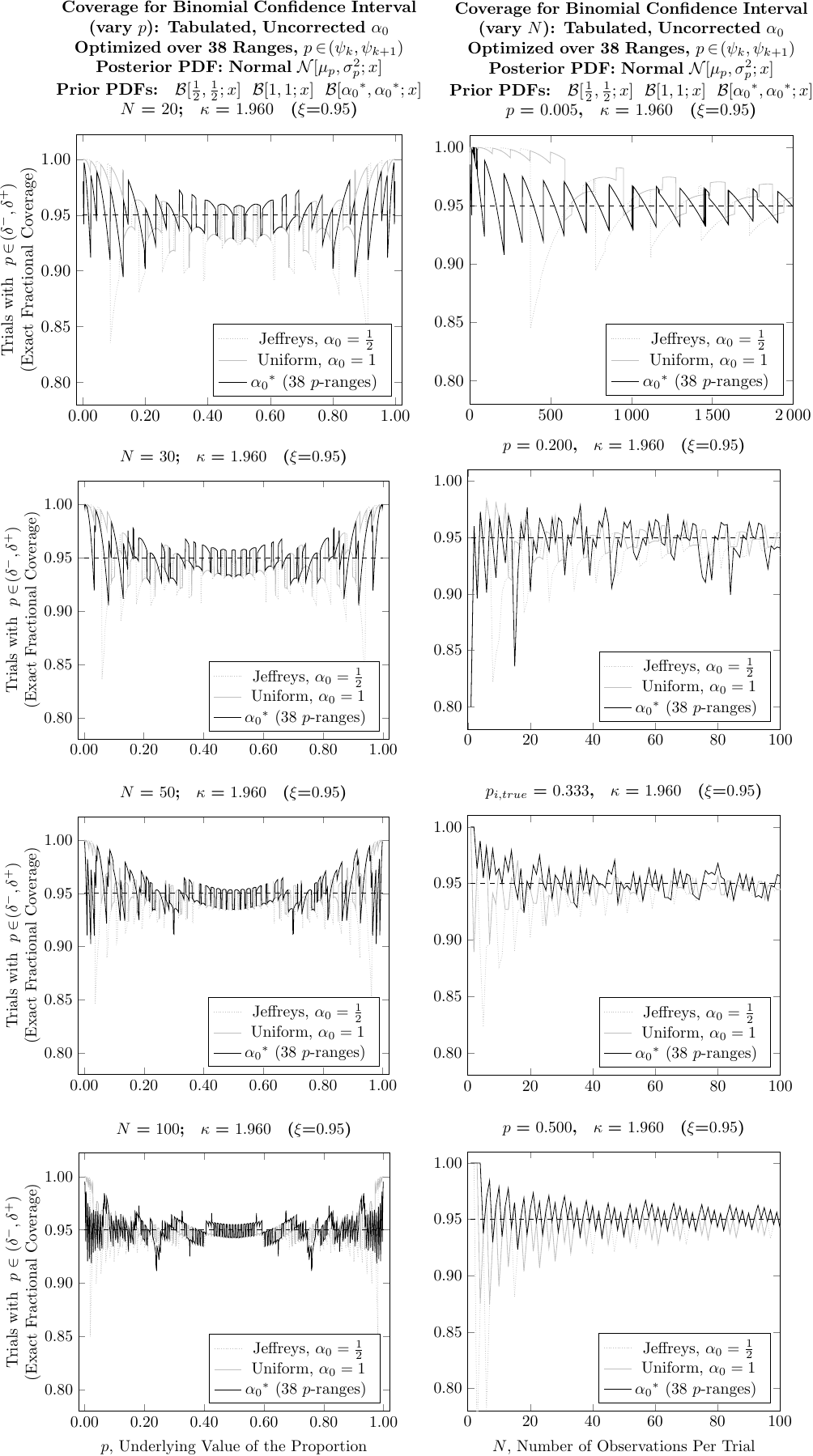}
}
                \caption{ Comparison of $\xi$-to-$C$ for binomial $\mathcal{B}[\alpha_0, \alpha_0; x]$\hspace{2pt}prior $\times~\mathcal{N}[\mu_p,\sigma^2_p; x]$\hspace{2pt}posterior 
                          for $\alpha_0$\hspace{1pt}$\in$\hspace{1pt}$\{\frac{1}{2}, 1, {\alpha_0}^*\}$ (Eq.7-8, for $\mu_p$, $\sigma^2_p$).
                          Here ${\alpha_0}^*$ are tabulated values optimized for $C$ to match $\xi$=$0.95$ with separate optimizations in each of 38 ranges $p\in(\psi_k,\psi_{k+1})$ at each $N$.
                        \label{fig:38pRange}}
\end{figure}

In contrast to those from the \emph{multinomial} ($b=100$) Dirichlet-Bayes prior,
values of $\alpha_0$ optimized for the \emph{binomial} estimators $\hat{p}$ and $\hat{\delta}$, based on $\mathcal{B}[\alpha_0, \alpha_0; x]$ as the Bayes prior (Eq.7-8, $b$=2), retained significant non-zero values for all ranges of $p_{true}$.  
For such $\alpha_0$ using intervals $\hat{\delta}$ based on the posterior $\mathcal{N}[\mu_p,\sigma^2_p; x]$, 
there were initially particular regions of spikes and dips in optimized $\alpha_0$ value over 38 examined sub-ranges $p_{i,true}\in(\psi_k,\psi_{k+1})$ at each $N$, but only for some particular values of $N$.
By-and-large though, optimized values for $\alpha_0$ remained above a set limit and with $\alpha_0 \gg \sigma_{\alpha_0}$  within each value of $N$ across all 38 tested sub-ranges $(\psi_k,\psi_{k+1})$ comprising (0,1).
Also, $\xi$-to-$C$ matching 
was qualitatively improved relative to using un-optimized values
$\alpha_0=\frac{1}{2}$ or $\alpha_0=1$ for the Jeffreys- or Uniform-Bayes priors (Fig \ref{fig:38pRange}).  
It was recognized though that spikes and dips in $C(N)_{p_i}$ and $C(p_i)_N$ might be signalling that the initial numerically optimized $C(\alpha_0)$ values were trapped at local minima of $\langle(\xi-C)^2\rangle$ for some $N$ and ranges $p_{true}\in(\psi_k,\psi_{k+1})$.

Numerical optimization of $\alpha_0$ over each sub-range $p_{i,true}\in(\psi_k,\psi_{k+1})$ entailed monitoring $\langle (\xi-C)^2\rangle$ over 1000 evenly spaced test values of $p_{i,true}$ in the sub-range.
On examining the statistics of $C$ values over the 1000 test values of $p_{i,true}$ in each range, 
a fairly large variance in $C$, $\sigma^2_C = \langle C^2\rangle - \langle C\rangle^2$, \emph{within} each zone was found using estimators $\hat{p}_i$ and $\hat{\delta}_i$ from the non-informative Dirichlet Bayes prior pdf ($\mathcal{B}[\alpha_0,99\alpha_0;x]$) for the $b=100$ multinomial case,
but a fairly small $\sigma^2_C$ was found in each zone for the $b=2$ binomial case ($\mathcal{B}[\alpha_0,\alpha_0;x]$).
Apparently, for the 100-bin case, the large $\sigma^2_C$ combined with a heuristic restraint to keep $\alpha_0>0$ 
led the computational optimization to select $\alpha_0$ values as close to zero as would be allowed.
Convergence of $\alpha_0$ to significant non-zero values occurred only for $p_{i,true} \approx 0$ or $p_{i,true}\approx 1$ (up to values approaching $\frac{1}{b}$ or $1-\frac{1}{b}$). 
This difference in behavior of $\sigma^2_C$ values suggests that the initial assumption of the existence of a unique, invariant optimal value for $\alpha_0$ 
might be invalid when $\min\{p_{i,true},(1-p_{i,true})\} \gg \frac {1}{b}$.

\subsection{Problems with and Modified Treatment for Multinomial Priors}

\subsubsection{Problems with Non-informative Multinomial Priors for Large $b$}
Such difficulties for large $b$ brought to question how 'non-informative' the Dirichlet Bayes priors with equal $\alpha_0$ values actually were and hence the practice of using pseudocounts to estimate proportions for the multinomial case.  
It seemed possible that the absence of a single optimal value for $\alpha_0$ that avoids large local $\sigma_C$ also might explain why the earlier large $\xi$-to-$C$ inconsistency occurred
(Fig \ref{fig:varyNlowPb100},\ref{fig:varyNhighPb100}) when using estimators derived from either the Jeffreys- or Uniform-Bayes priors when $p_i \approx 0.5$ in the 100-bin multinomial case.
To rationalize this problem, the difference in the results for $b=2$ and $b=100$ initially focused attention on the $b$ terms in the denominators of the expressions (Eq.1-4,7-8) for $\hat{p}_i$ and $\hat{\sigma}^2_{p_i}$ derived from the initially examined Bayes priors.

Considering again the experiment with $0$ observations ($N$$=$$0$), earlier considerations of $\hat{\sigma}_{p}$ together with $\hat{p}$ left any value possible for $p$ at the $95\%$ confidence level for the binomial case.
In the $b=100$ multinomial case though, $\hat{p}_i(N$=$0) =\frac{1}{b}$ 
instead of $\frac{1}{2}$.  
This, combined with a decreased value for $\hat{\sigma}_{p_i}$---\hspace{2pt}due to a large value for $b$ in the denominators of Eq. 2,4,8\hspace{2pt}---\hspace{2pt}causes many estimates $\hat{p}_i$ for $p_i \gg \frac{1}{b}$ to lie outside of the $95\%$ confidence range.
This can lead to particularly poor $\xi$-to-$C$ matching for large $p_i$.
Hence, when using 'non-informative' priors, there is a tacit underlying assumption that the histogram will not be dominated by the partitioning of observations to a small fraction of the available histogram bins.
In other words, 'non-informative' priors appear to bias estimates for $p_i$ toward the value $\frac{1}{b}$, particularly for large values of $b$ and small to moderate values of $N$.
This provides rationalization for the $b$$=$$100$ case when $\mathcal{B}[\alpha_0, \beta_0; x]$ is used as a Bayes prior,
but a more cogent explanation and rationale for ignoring this bias was later found on comparing $\mathcal{B}[\alpha_0, \beta_0; x]$ to the form for discrete multinomial priors $\bar{\pi}_j$ (Appendix B).

\subsubsection{Substitution of Binomial for Multinomial Priors as Work-Around}
As a makeshift alternative for multinomial estimation, the multinomial Dirichlet Bayes prior $\mathcal{B}[\alpha_0, (b-1)\alpha_0; x]$ was replaced by an optimized binomial Bayes prior, $\mathcal{B}[\alpha_0,\alpha_0;x]$, ($b$=$b_{\text{eff}}$=$2$).  
This optimized binomial prior was used in turn for each bin $i$, with a binomial 'success' event considered to be an occurrence in bin $i$ and a 'failure' event corresponding to an occurrence in any other bin of the multinomial histogram.
A binomial prior is less "informative" than the 'non-informative' multinomial Dirichlet Bayes prior in that no knowledge is required whatsoever about what is going on in bins other than the one for which the proportion and its variance are being estimated.
This leads to other inaccuracies compared to the continuous multinomial priors or discrete multinomial priors $\bar{\pi}$. 

Although $\hat{\delta}_i$ from the binomial $\mathcal{B}[\alpha_0,\alpha_0;x]$ prior appeared to improve
$\xi$-to-$C$ matching, 
the price paid was the loss of the constraint between values of $p_i$ to keep $\Sum p_i = 1$. 
This normalization was instead applied later in a separate scaling step to improve estimates. 
A second problem is that the bias to $\frac{1}{b}$ is replaced by a bias to $\frac{1}{2}$ that is strongly apparent for small $p_i$, leading to $\hat{p}$-to-$p$ inconsistency, but this too was eventually correctable.  

\subsection{Optimizing $\alpha_0$ in $\mathcal{B}[\alpha_0,\alpha_0;x]$ Prior (b=2): Full Range $p\in(0,1)$ at Each $N$ }

Given that (i) the general \emph{binomial} $\mathcal{B}[\alpha_0,\alpha_0;x]$ prior seemed more reliable and applicable to estimating both binomial and general multinomial proportion values than the multinomial Dirichlet-Bayes prior, 
and that (ii) for any $N$, $\alpha_0$ for this general form could be optimized for $\xi$-to-$C$ matching to a single unique $\alpha_0(N)$ value that should be applicable for all values of $p_{i} \in (0,1)$ at that $N$, 
it seemed worthwhile to have values for such optimal $\alpha_0$ for many values of $N$. 
The earlier tests of $\xi$-to-$C$ matching appeared to be successful for $\hat{\delta}$ derived from $\mathcal{B}[\alpha_0,\alpha_0;x]$ priors with $\alpha_0$ optimized separately in local zones $p_{(k)}\in(\psi_k,\psi_{k+1})$ at each $N$.
However, residual differences between optimized values for $\alpha_0$ in some adjacent zones $p_{(k)}\in(\psi_k,\psi_{k+1})$ at each $N$ suggested that some zone-by-zone optimizations might be stuck in local minima.
Use of a single optimization range $(\psi_k,\psi_{k+1}) = (0,1)$ 
to determine a single optimal $\alpha_0$ at each $N$ might lead to a readily recognizable functional form $\alpha_0(N)$ for the optimal values that might aid in recognizing and correcting such anomalous $\alpha_0$ values.  

\subsubsection{Optimizing Binomial $\alpha_0(N)$: $\mathcal{B}[\alpha_0,\alpha_0;x]$ Prior $\times~\mathcal{N}[\hat{p},\hat{\sigma}^2_p;x]$ Posterior}
\begin{figure}
\centering
\scalebox{0.55} {
\centering
\includegraphics[width=15 cm]{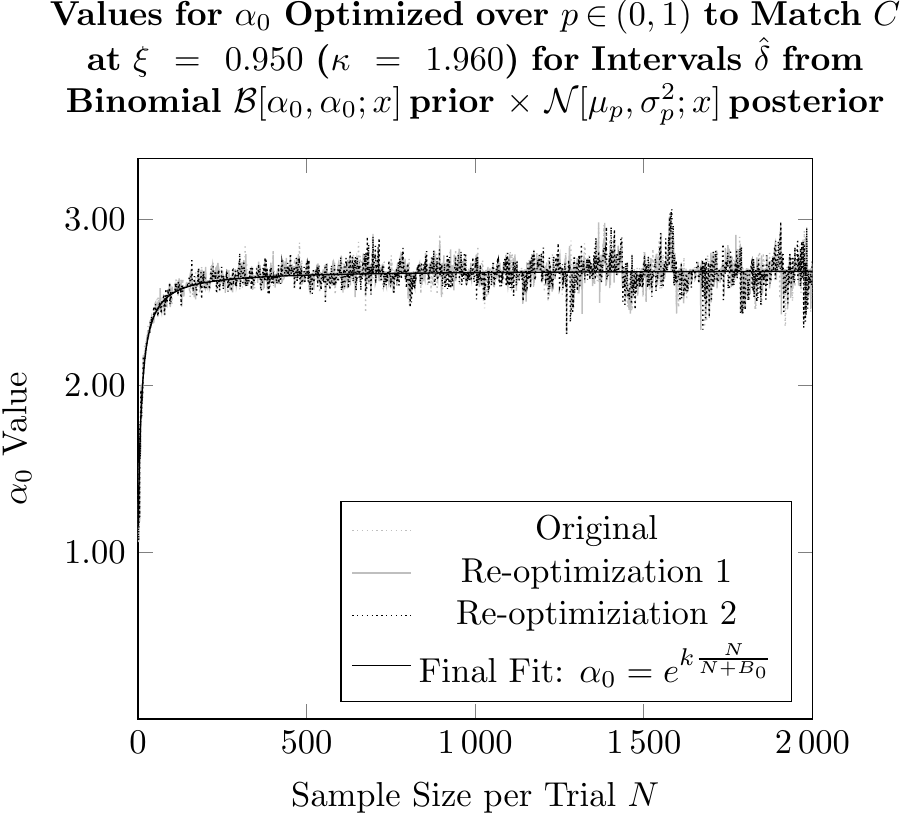}
}
                \caption{ Optimization of $\alpha_0$ over $p$\hspace{2pt}$\in$\hspace{1pt}$(0,1)$ for $\xi$-to-$C$ matching ($\xi=0.950$) at each $N$.
                          Idealized points from initial fits $\alpha_0(N) \approx e^{k\frac{N}{N+B_0}}$ were starting points for further rounds.  
                          After 3 rounds of $\xi$-to-$C$ optimization at each $\xi$, the values for $k$ and $B_0$ were:
                          ($\xi$=$0.95$: $k$=$0.991$, $B_0$=$5.863$);  ($\xi$=$0.975$: $k$=$1.186$, $B_0$=$5.138$);  ($\xi$=$0.990$: $k$=$1.392$, $B_0$=$5.219$).  
                        \label{fig:alphaVsNexponent}}
\end{figure}

For the binomial prior $\mathcal{B}[\alpha_0,\alpha_0;x]$, initial optimizations (Sec 2.3) of $\alpha_0$ over different restricted ranges $p\in(\psi_k,\psi_{k+1})$ led to a median value $\alpha_0 \approx 1.7$ for $N \in \{1,2\dots 200\}$.
Thus $\alpha_0$ was varied, starting from $\alpha_0$=$1.7$, to minimize $(\xi-C)^2$ over the full range $p \in (0,1)$ for fixed confidence value $\xi=0.95$,
with $\hat{\delta}$ defined by $\mathcal{N}[\hat{p},\hat{\sigma}^2_p;x]$ through Eq. 6.
Separate optimization was performed at each $N \in \{1,2\dots 2000\}$.  
Plotting optimized $\alpha_0(N)$ against $N$, 
the resulting curve had the form of a very noisy exponential curve that approached a value of $\alpha_0=1.0$ as $N$ approached 0 and a value $\alpha_0=2.7$ for large $N$ (Fig \ref{fig:alphaVsNexponent}).

The noise in this initial curve had been anticipated.
It was expected that there be would some $N$ for which the optimal $C(\alpha_0)$ value would be constant for a contiguous range of values for $\alpha_0$. 
Noise might arise for initial plots of $\alpha_0$ versus $N$ if the computational procedure only indicated the \emph{initial random entry point} into such a zone. 
Furthermore, for some sub-ranges of $\alpha_0$ within such a zone of constant $C$, round-off error might cause a slightly lower computed value for $(\xi-C)^2$, causing optimization to become stuck in such a sub-range. 
Variation with $N$ of the positions of the sub-ranges arising from round-off error might add additional spurious noise.
Finally, it is also possible for noise to result from true entrapment at a local minimum in $(\xi-C)^2$ for some values of $N$ and $\alpha_0$.

On seeing the near-exponential form for the curve it was decided to re-optimize $\alpha_0$ as before, but starting from points on the exponential fit.  
Values of $\alpha_0$ re-optimized starting from this fit were displaced toward it,  
occasionally much closer to the fit curve of starting points than to the direct results of the first optimization. 
A third optimization of $\alpha_0$ starting from the points on a second exponential fit 
led to a new optimal $\alpha_0(N)$ set with even less scatter, 
but still with several recurring zones of deviation from the fit.

\begin{figure}
\centering
\scalebox{0.75} {
\centering
\includegraphics[width=15 cm]{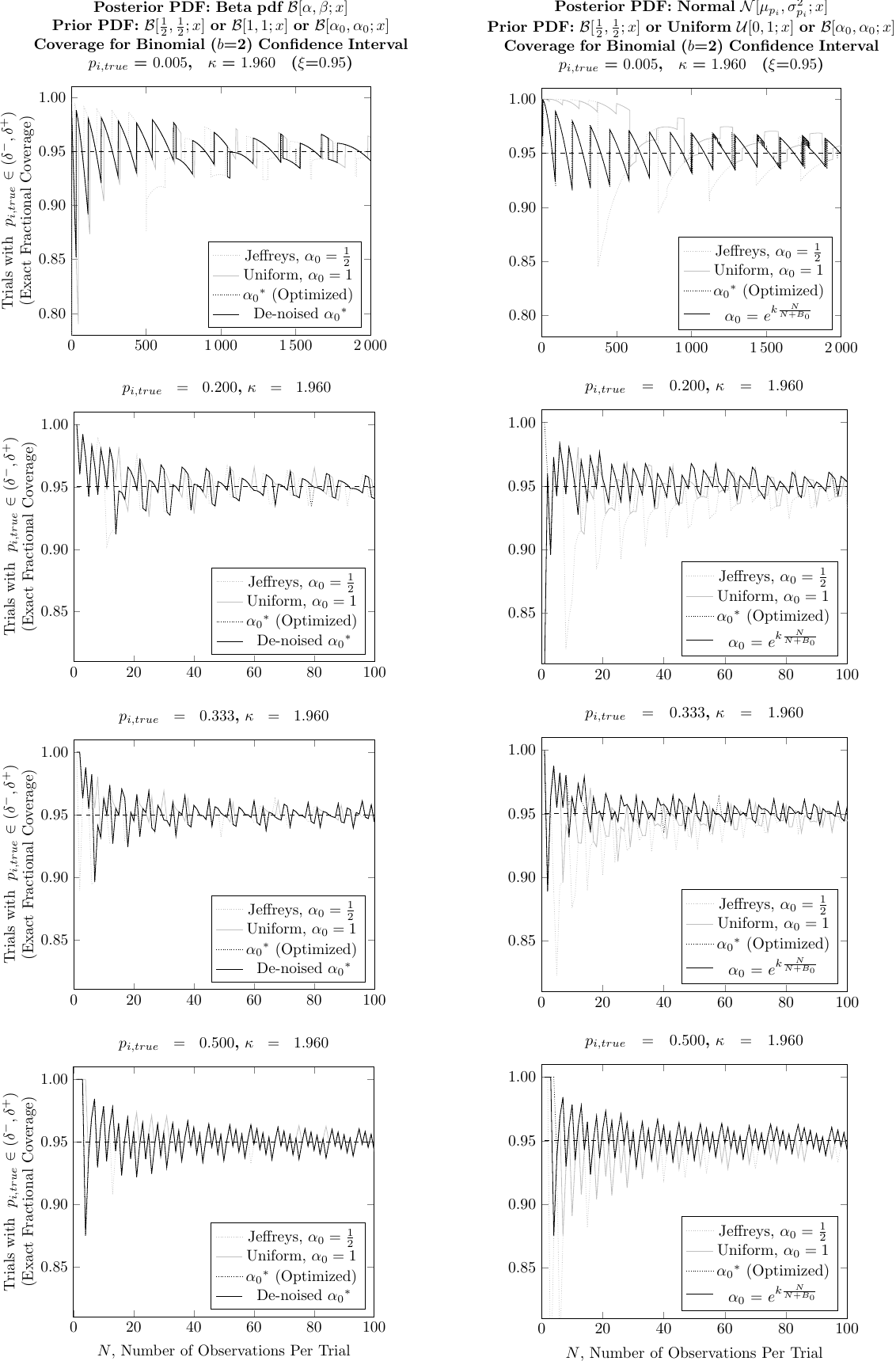}
}
                \caption{ Coverage plot $C(N)$ using binomial Bayes priors with $\alpha_0$ optimized for $\xi$-to-$C$ ($\xi$=$0.950$) matching once over the full range $p_{i,true}\in(0,1)$ at each $N$:  Comparison of $C(N)$ for several fixed values of $p_{i,true}$.  
                          Separate optimizations of $\alpha_0(N)$ in the prior pdf were done for either $\mathcal{B}(\alpha,\beta;x)$ beta-pdf Bayes Posterior (left) or for $\mathcal{N}(\hat{\mu}_p,\hat{\sigma}_p;x)$ Gaussian Bayes Posterior (right), 
                          to provide optimal $\xi$-to-$C$ matching for evenly spaced test $p_{i,true}\in(0,1)$.
                          Tests compare tabulated unsmoothed and smoothed \emph{optimal} $\alpha_0(N)$ values (Beta posterior) or tabulated unsmoothed and fit (smoothed) \emph{optimal} $\alpha_0(N)$ values (Gaussian posterior). 
                        \label{fig:adaptiveVaryN}}
\end{figure}

\begin{figure}
\centering
\scalebox{0.75} {
\centering
\includegraphics[width=15 cm]{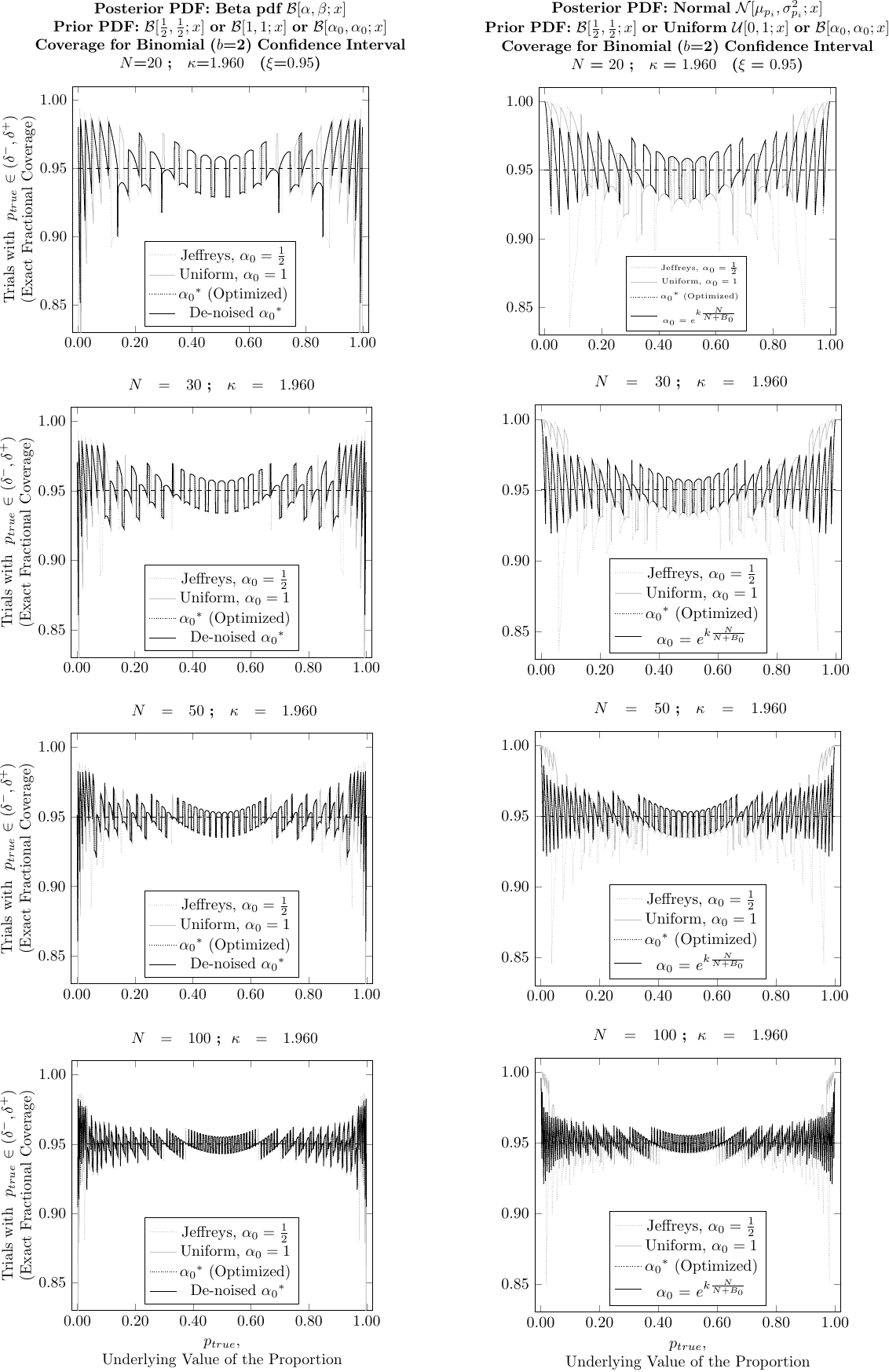}
}
                \caption{ Coverage plot $C(p_{i,true})$ using binomial Bayes priors with $\alpha_0$ optimized for $\xi$-to-$C$ ($\xi$=$0.950$) matching once over the full range $p_{i,true}\in(0,1)$ at each $N$:  Comparison of $C(p_{i,true})$ for several fixed values of $N$.  
                          Separate optimizations of $\alpha_0(N)$ in the prior pdf were done for either $\mathcal{B}(\alpha,\beta;x)$ beta-pdf Bayes Posterior (left) or for $\mathcal{N}(\hat{\mu}_p,\hat{\sigma}_p;x)$ Gaussian Bayes Posterior (right), 
                          to provide optimal $\xi$-to-$C$ matching for evenly spaced test $p_{i,true}\in(0,1)$.
                          Tests compare tabulated unsmoothed and smoothed \emph{optimal} $\alpha_0(N)$ values (Beta posterior) or tabulated unsmoothed and fit (smoothed) \emph{optimal} $\alpha_0(N)$ values (Gaussian posterior). 
                        \label{fig:adaptiveVaryP}}
\end{figure}

Values for $\alpha_0$ that had been optimized for $\xi$-to-$C$ matching at $\xi$$=$$0.95$ ($\alpha^{[\xi=0.95]}_0$) led to improved coverage. 
This was observable in plots of $C(N)$ at fixed $p_{true}$ (Fig \ref{fig:adaptiveVaryN}, rhs) or $C(p_{true})$ at fixed $N$ (Fig \ref{fig:adaptiveVaryP}, rhs).
These plots of $C$ versus $N$ or $p_{true}$ oscillated nearly \emph{symmetrically} about nominal confidence $\xi=0.95$.
Also noteworthy was that when $p\approx 0$ or $p\approx 1$, estimated intervals ($\hat{\delta}_{|n\approx 0}$)$_{\xi=0.95}$ and ($\hat{\delta}_{|n\approx N}$)$_{\xi=0.95}$
based on $\mathcal{B}[\alpha_0,\alpha_0;x]^{[\xi=0.95]}$ prior $\times~\mathcal{N}[\hat{p},\hat{\sigma}^2_p;x]$ posterior 
encroached considerably less into the regions beyond $p < 0$ and $ p  > 1$ than had earlier intervals using $\alpha_0$$=$$1$ or $\alpha_0$$=$$\frac{1}{2}$ to get $\hat{p}$ and $\hat{\sigma}^2_p$ for the $\mathcal{N}[\hat{p},\hat{\sigma}^2_p;x]$ posterior. 

When $\xi$-to-$C$ agreement was examined for $\xi = \xi^{'} \neq 0.95$, but using $\alpha^{[\xi=0.95]}_0$ to derive ($\hat{\delta}$)$_{\xi=\xi^{'}}$ by Eq. 6,
the $C$ still matched the alternate nominal confidence levels $\xi^{'}$ fairly well, but oscillations in $C(N)$ and $C(p_{true})$ were no longer symmetric about the new $\xi^{'}$, tending slightly toward underconfidence ($C > \xi$).
To understand this, $\alpha_0$ was re-optimized for $\xi$-to-$C$ matching using $\xi=\xi^{'}\neq 0.95$ implicitly in the optimization target.
Initially for the $\xi$$=$$0.95$ target, the apparent limiting values of $1.00$ and $2.7$ ($\approx e^{k\frac{N}{N+B_0}} $) led us to hold the value of $k$ fixed at $1.00$ and use a single parameter $B_0$ to describe the exponential curve.
For $\xi$$=$$\xi^{'}$$>$$0.95$, the plots of $\alpha_0(N)$ were again exponentially shaped, however the limiting value at high $N$ differed from the value 2.7, varying with the different choices for the target $\xi$
and forcing us to use the more general two parameter fit to describe the exponential curve.
When $C(N)_{p=p_0}$ and $C(p_{true})_{N=N_0}$ curves were calculated 
using $\alpha^{[\xi^{'}]}_0$ specifically optimized at the newly targeted confidence levels $\xi=\xi^{'}$, 
 $\xi$-to-$C$ matching was still observed as $N$ or $p_{true}$ varied, but the oscillations about the new values $\xi^{'}$ became more symmetric.

To estimate an unknown pdf using a histogram with $b=100$, if the unknown underlying pdf had been $\sim\mathcal{U}[0,1;x]$, then $\langle p_{i,true} \rangle $ would be $0.01$.  
Accurate estimates $\hat{p}_i$ were necessary for values as low as $0.005$ or even below. 
For $p_{i,true}$ values this low, the earlier forms for binomial estimators often led to woefully inconsistent agreement between $C$ and $\xi$ unless $N$ was fairly high (Fig \ref{fig:varyNju}, \ref{fig:varyPju}).
Of the four earlier prior-posterior combinations $\mathcal{U}[0,1,;x]$ prior $\times~\mathcal{N}[\mu_{p_i},\sigma_{p_i};x]$ posterior and the Jeffreys $\mathcal{B}[\frac{1}{2},\frac{1}{2};x]$ prior  $\times~\mathcal{B}[\alpha,\beta;x]$ posterior 
combination had the best $\xi$-to-$C$ matching while avoiding low $C$ values with $|\xi - C| > 1 - \xi$; $ C < 2\xi - 1$.
Matching of $\xi$-to-$C$ for
the newly optimized binomial $\mathcal{B}[\alpha_0,\alpha_0;x]$ prior $\times~\mathcal{N}[\mu_{p_i},\sigma_{p_i};x]$ posterior appeared to be closer than had been found for either of these two combinations
(Fig \ref{fig:adaptiveVaryN}-\ref{fig:adaptiveVaryP}, rhs).  
Fairly decent $\xi$-to-$C$ matching was observed over the entire range of $N$,  
with maximal $|\xi-C|$ at low values of $N$ for which generally $|\xi-C|_{max} < 1-\xi$.

Initial tests of $\hat{\delta}$ derived from the $\mathcal{N}[\mu_p,\sigma^2_p;x]$ \emph{posterior} pdf using  priors with $\alpha_0$ optimized over limited ranges $p\in(\psi_k,\psi_{k+1})$ (Fig \ref{fig:38pRange}) or over $p\in(0,1)$ (Fig \ref{fig:adaptiveVaryN}-\ref{fig:adaptiveVaryP}, rhs)
indicated improved $\xi$-to-$C$ consistency.
However, it was anticipated that $\hat{\delta}$ based on priors with $\alpha_0$ optimized specifically for 
the "exact" Bayes posterior function $\mathcal{B}[\alpha, \beta; x]$ might provide even more accurate coverage values.

\subsubsection{Optimizing Binomial $\alpha_0(N)$: $\mathcal{B}[\alpha_0,\alpha_0;x]$ Prior $\times~\mathcal{B}[\alpha, \beta; x]$ Posterior}

\begin{figure}
\centering
\scalebox{0.55} {
\centering
\includegraphics[width=15 cm]{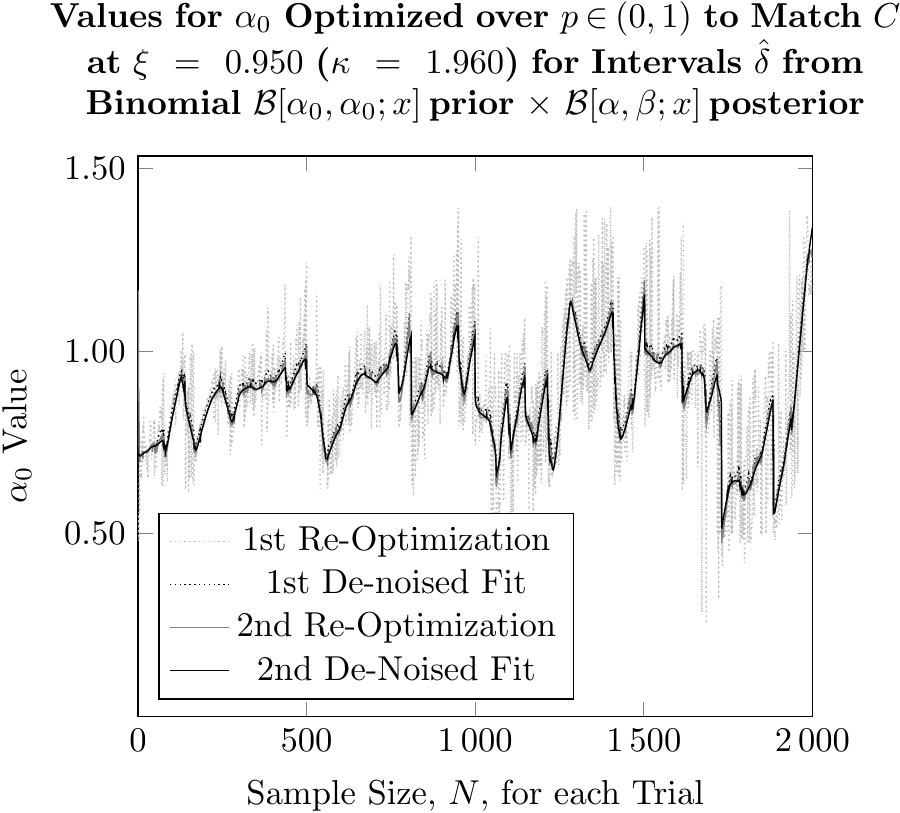}
}
                \caption{Optimization of $\alpha_0$ over $p$\hspace{2pt}$\in$\hspace{1pt}$(0,1)$ for $\xi$-to-$C$ matching ($\xi=0.950$) at each $N$ using $\hat{\delta}$ based on $\mathcal{B}[\alpha_0,\alpha_0;x]$\hspace{2pt}prior $\times~\mathcal{B}[\alpha,\beta;x]$\hspace{2pt}posterior.
                          A noisy $\alpha_0(N)$ curve (not shown) from optimizing $\xi$-to-$C$ matching over 4 very limited contiguous ranges spread over $N$ gave a poor fit to $\alpha_0(N) \approx e^{k\frac{N}{N+B_0}}$,
                          but idealized points were used to start the first full round of optimization for all $N$.  
                          Curves of $\alpha_0(N)$ for each full round of optimization were de-noised by a 'local-linearity' approximation (Appendix A) to derive starting points for further rounds.
                        \label{fig:alphaVsNhorrid}}
\end{figure}

Separate optimizations for $\xi$-to-$C$ consistency were performed to obtain optimal $\alpha_0$ values for use with $\mathcal{B}[\alpha,\beta;x]=\mathcal{B}[n+\alpha_0,N-n+\alpha_0;x]$ as the Bayes \emph{posterior} pdf.
Optimizations at 4 separate small groups of contiguous $N$ values below $N=1000$ starting from the constant value $\alpha_0=1.70$, were used to try to jump start to an approximate exponential form that could provide a starting estimate of $\alpha_0$ at each $N$.
It was immediately found that use of $\mathcal{B}[\alpha,\beta;x]$ as Bayes posterior required consistently lower values for $\alpha_0$ in the Bayes prior than before, 
so the starting value in the initial optimization was changed to $\alpha_0=1$ to avoid systematic entrapment at local minima in $(\xi-C)^2$.
Re-optimizing $\xi$-to-$C$ matching over the same small fraction of $N$ values after this switch led to an exponential approximation to get initial $\alpha_0$ for all $N$ values. 
Using these initial values, and performing $\xi$-to-$C$ optimization of $\alpha_0$ at $\xi=0.95$ for each value of $N$ from $1$ to $2000$, 
one obtained a 'horrid', noisy curve of optimal $\alpha_0$ versus $N$ for which there was no obviously simple functional form (Fig \ref{fig:alphaVsNhorrid}).
Interestingly for the $\mathcal{B}[\alpha,\beta;x]$ posterior, optimal $\alpha_0$ 
differed from $\alpha_0$ for the Jeffreys prior ($\alpha_0$=$0.5$) or the Uniform prior ($\alpha_0$=$1.0$), but was usually somewhere in between.

Given the earlier experiences with optimizing $\alpha_0$ for use with $\mathcal{N}[\mu_p,\sigma^2_p;x]$ as the Bayes posterior, 
it was anticipated that there might be scatter in the initially optimized values for $\alpha_0$,
and that these values 
might be regularized, using $\alpha_0$ values from nearby $N$ to define improved starting points for iterative re-optimization of $\alpha_0$.  
Since the functional form for such regularization of $\alpha_0(N)$ was not as obvious as for the earlier near-exponential form though, it was decided to use an assumption of 'approximate local linearity' to effect the de-noising.
A new procedure was developed for this (Appendix A) and computer source code is provided in the supplemental material.
Note that the de-noising method used here appears to be less liable to loss of sharp features than other similar recent de-noising methods \citep{boseAhuja,fleishman,loess}.

\subsubsection{Re-optimization of $\alpha_0$ from Idealized Fits to $\alpha_0(N)$}

Re-optimization of $\xi$-to-$C$ matching for the $\mathcal{B}[\alpha_0,\alpha_0;x]$ prior $\times ~ \mathcal{B}[\alpha,\beta;x]$ combination starting from de-noised curves of $\alpha_0(N)$ led to progressive agreement of 
between optimized and starting values for $\alpha_0$ at each cycle. 
This agreement between de-noised and re-optimized $\alpha_0(N)$ values was much closer than found upon re-optimizing the $\alpha_0(N)$ for the $\mathcal{N}[\mu_p,\sigma^2_p;x]$ posterior starting from the  
exponential fit. 
To examine the generality of convergence to a joint optimum, re-optimization of earlier zone-optimized $\alpha_0$ in each of 38 $(\psi_k,\psi_{k+1})$ zones at each $N$ for the $\mathcal{N}[\mu_p,\sigma^2_p;x]$ posterior was repeated 
starting from the $(0,1)$ full-range optimized single zone $\alpha_0(N)$ value.
Again the new zone-optimized values of $\alpha_0(N)$ were generally closer to the starting single zone value.  
Tightening of re-optimized $\alpha_0(N)$ to the smoothed starting curve is expected though for either reduced or increased entrapment of the optimization at local minima in $(\xi-C)^2$.
However, further tests for the binomial case showed that the locally-consistent values for $\alpha_0$ increased overall average $\xi$-to-$C$ matching. 

\subsubsection{Accuracy of Estimators Optimized for $\mathcal{B}(\alpha,\beta; x)$ as the Posterior }

For the single-zone optimized $\mathcal{B}[\alpha_0,\alpha_0;x]$ prior $\times ~ \mathcal{B}[\alpha,\beta;x]$ posterior combination at fixed values of $p_{i,true}$, 
 $C(N)$ oscillated regularly about $C=\xi$ 
at low values of $N$, but less regularly at high $N$ (Fig \ref{fig:adaptiveVaryN}).
Whereas the $\langle (C-\xi)^2 \rangle$ was reduced, 
there were some particular values of $p_{true}$ and ranges of $N$ for which the earlier un-optimized priors in combination with the $\mathcal{B}[\alpha,\beta;x]$ posterior had better $\xi$-to-$C$ agreement. 
Specific optimization of the $\mathcal{B}[\alpha_0,\alpha_0;x]$ prior for the $\mathcal{B}[\alpha,\beta;x]$ posterior did not improve ${\xi\leftrightarrow C}$ as much as it had for the $\mathcal{N}[\mu_p,\sigma^2_p;x]$ posterior (Sec. 2.5).
In the alternate plots, $C(p_{true})$ at fixed $N$ (Fig. \ref{fig:adaptiveVaryP}),
even after optimization, overconfidence, $\xi \gg C$, remained 
for the $\mathcal{B}[\alpha_0,\alpha_0;x]$ prior $\times ~ \mathcal{B}[\alpha,\beta;x]$ posterior combination,
particularly at low values of $p_{true}$.
The degree of $\xi$-to-$C$ matching 
at low $p_{true}$ was intermediate between that seen for the unoptimized $\mathcal{B}[\frac{1}{2},\frac{1}{2};x]$  (Jeffreys) and $\mathcal{B}[1,1;x]$ (Uniform) priors.
Specific optimization of the $\mathcal{B}[\alpha_0,\alpha_0;x]$ prior for the other posterior, 
$\mathcal{N}[\mu_p,\sigma^2_p;x]$, did not lead to such a large $\xi$-to-$C$ mismatch
at low $p_{true}$. 
Whereas such $\xi$-to-$C$ mismatches are only significant for 'low' values of $N$, these 'low' $N$ values are still rather large, 
and it seems safer to avoid problems by using only $\mathcal{B}[\alpha_0,\alpha_0;x]$ prior $\times ~ \mathcal{N}[\mu_p,\sigma^2_p;x]$ posterior to define $\xi$-to-$C$-optimized estimators ($\hat{p}$)$_{\xi\leftrightarrow C}$ and ($\hat{\delta}$)$_{\xi\leftrightarrow C}$.

What intrigued us more though about the second optimization of $\alpha_0$ was the marked improvement seen upon de-noising the curves of optimized $\alpha_0(N)$.
It was wondered if direct de-noising of raw histograms might similarly improve overall estimates of multinomial $p_i$ and $\sigma_{p_i}$ when the 
underlying probability densities were relatively smooth functions.

\section{Applying Direct De-noising to Multinomial Estimation}
\subsection{Direct De-noising of Experimental Histograms}

To test if such direct de-noising could improve ${\hat{p}_i \leftrightarrow p_i}$ for estimates of smooth underlying pdfs from experimental histograms, MC-generated data were analyzed. 
Six, arbitrarily selected underlying "model" pdfs, $g(x)$, were chosen to generate random data to sort into histograms to estimate $p_i$ and $\sigma_{p_i}$ bin-by-bin. 
The list of tested underlying $g(x)$ included a standard normal ($\mathcal{N}[0,1;x]$), a sawtooth, and 4 beta-pdfs, $\mathcal{B}[\alpha,\beta;x]$, with $(\alpha,\beta) \in \{(3,15),(9,11),(6,2),(5,3) \}$
to examine pdfs with varying degrees of skew.
Each underlying trial pdf $g(x)$ was used to generate a set of 10,000 random 100-bin histograms for each of several fixed sample sizes $N$.
All histograms were limited to a range of $\hat{\mu}_{g(x)} \pm 3.5\hat{\sigma}_{g(x)}$ so that the more highly skewed $\mathcal{B}[\alpha,\beta;x]$ examples of $g(x)$ had some histogram bins entirely outside of the range of possible outcomes.
Initial estimates $\hat{p}_{i,0}$ for each bin---at 
first based on the new coverage-optimized ($\hat{p}$)$_{\xi\leftrightarrow C}$ with optimized value $\alpha_0(N)$ in the prior (Eq.7, $b$$=$$2$) but later extended to other forms for estimators $\hat{p}_{i,0}$ --- were de-noised, base-line adjusted, and scaled.
De-noised bin-by-bin estimates $\hat{p}_i$ from each random histogram could be compared to $p_{i,true}$ calculated as $\Intop^{x_{i+1}}_{x_i}\hspace{-2pt}  g(x)\hspace{2pt}  dx$ over the limits of each bin.
Procedures to de-noise, scale, and base-line correct each $N$-observation random histogram were based solely on information from the random test data, independent of $g(x)$, 
but they were adjusted for simultaneous consistency over the 6 arbitrarily chosen trial pdfs.\footnote{
Initially, estimators were used for which values of the absolute maximum and minimum estimates before smoothing $\hat{p}_{max}=\hat{p}_i(N)$ and $\hat{p}_{min}=\hat{p}_i(0)$ were clearly available as parameters for the scaling and baseline adjustment after smoothing.
Later, different initial estimates were tried for which these limiting values were less clearly defined. 
An independent empirical procedure was invoked to estimate the base and peak values from unscaled curves obtained directly after smoothing.}

\subsection{Agreement between Estimates $\hat{p}_i$ and the Known Underlying pdf, $g(x)$ }

Qualitatively, estimates $\hat{p}_i$ for individual histogram bins after de-noising, scaling, and base-line adjustments appeared to be more accurate than any of the original unsmoothed estimates (Fig \ref{fig:G40_60_big}-\ref{fig:estHistE}).
The average of estimates of smoothed $\hat{p}_i$ over the 10,000 MC trial histograms at each value of $N$ were within a fraction of a percent of $p_{i,true}$ for each pdf.
The original estimators from Eqs 1 and 7 ($b$=2) led to larger values for $ |\hat{p}_i - p_{i,true}| $.

To quantify $\hat{p}_i$-to-$p_i$ matching for a single histogram, a signal-to-error (S/N) ratio: 
$$ (S/N)_h \equiv \frac{\sqrt{\langle p^2_{i,true}\rangle_{bins} - \langle p_{i,true}\rangle^2_{bins}}} {\sqrt{\langle (p_{i,true}-\hat{p}_i)^2\rangle_{bins}}}$$ 
can be used, with averages over the bins $i$ in each individual random histogram (1 MC run) indicated by brackets $\langle \rangle_{bins}$.
It is useful to compare averages of $(S/N)_h$ over all random histograms in each 10,000-run MC trial at fixed $N$ and $g(x)$ using: $(S/N)_{MC} \equiv \langle (S/N)_h\rangle_{MC}$ and $\sigma^2_{(S/N)} \equiv \langle (S/N)^2_h\rangle_{MC} - \langle (S/N)_h \rangle^2_{MC}$.
The value in the numerator of $(S/N)_h$ is always the same for a given $g(x)$ as model pdf and so $(S/N)$ is really only useful for comparing relative $\hat{p}_i$-to-$p_{i,true}$ agreement for different estimators for the same $g(x)$.
As for similar quantities, (S/N)$_{MC} \propto \sqrt{N}$ roughly:  (S/N)$_{MC} \approx M \sqrt{N} + B $, where $N$ is the number of observations, but the constants $M$ and $B$ vary for different estimators $\hat{p}$ and for different underlying densities $g(x)$.
For each underlying trial pdf, $g(x)$, 
one can estimate by interpolation the number of observations, $N_{eq}$, required for $p_i$ estimated from one form for $\hat{p}_i$ to match the (S/N)$_{MC}$ achieved using another form for $\hat{p}_i$. 
For the examples in Fig \ref{fig:G40_60_big}, the multinomial estimators based on binomial forms of ($\hat{p}$)$_{\xi\leftrightarrow C}$ would 
require $N$ to be more than $20 \times$ as large ($N$$\approx$$950$ or $N$$\approx$$1400$ instead of $N$=$40$ or $N$=$60$) to match the S/N from $\widehat{p_{i,smooth}}$.

\subsection{Different Unsmoothed Estimators $\hat{p}_i$ as Starting Points for Smoothing}

The accuracy of several other \emph{unsmoothed} multinomial estimators $\hat{p}_{i,0}$ were also compared by MC, as possible improved starting points for smoothing.
One referee suggested an accurate \emph{joint estimator}, ($\hat{p}_i$)$_{\bowtie}$, available in the package Rgbp from the "R-project" \citep{Rgbp}.  
This estimator ($\hat{p}_i$)$_{\bowtie}$, improves overall estimates over a histogram by compensating for James-Stein shrinkage of joint estimates to a mean vector (\cite{morrisLysy,efronMorris,stein}).
As with histogram de-noising, ($\hat{p}_i$)$_{\bowtie}$ estimates are improved by additional information from co-observations.
We also derived \emph{discrete multinomial estimators} $\hat{p}_i(n,N)$ (Appendix B) using prior and posterior pdfs that account for discontinuous integer outcomes.
This derivation also suggested that earlier observed, large systematic $\xi$-to-$C$ mismatches (Fig \ref{fig:varyNlowPb100} \& \ref{fig:varyNhighPb100}) for the multinomial rule-of-succession (MROS) estimator 
(Eq.1\&2, $b\ge 2$, $\mathcal{N}$ posterior) 
might have arisen from a low statistical likelihood of the high trial value for $p_i$ used in the calculations.
This suggested that we re-examine unsmoothed \emph{multinomial} estimators based on the Uniform Bayes prior. 

On testing these other \emph{unsmoothed} estimators with reduced $\xi\leftrightarrow C$ matching, higher $S/N$ values indicated that $\hat{p}_i\leftrightarrow p_i$ matching was slightly \emph{improved} relative to \emph{unsmoothed} ($\hat{p}$)$_{\xi\leftrightarrow C}$. 
\emph{Unsmoothed} multinomial $\hat{p}_i$ 
based on the initial multinomial Rule of Succession priors (eq. 1 and 3, $b\ge 2$), on discrete $\hat{p}_i(n,N)$, or on ($\hat{p}_i$)$_{\bowtie}$ 
all provide comparable (S/N)$_{MC}$ 
when estimating the same underlying pdf from random histograms (Table S1).
Thus these alternate estimators were tested further as starting points for histogram smoothing.

Applying histogram smoothing to all types of unsmoothed starting estimates, though, led to smoothed estimates $\widehat{p_{i,smooth}}$ with very similar values for (S/N)$_{MC}$ (Table S1). 
Initially, the (S/N)$_{MC}$ values for smoothed histograms from ($\hat{p}_i$)$_{\bowtie}$ varied much more than those from earlier estimators, often much higher, 
but often much lower and analysis suggested that a new procedure was needed to adjust the histogram baseline and scale after smoothing.
On modifying scaling and baseline procedures for raw $\widehat{p_{i,smooth}}$ from ($\hat{p}_i$)$_{\bowtie}$, 
($\hat{p}_i$)$_{\bowtie}$ seemed to be a starting point for smoothing that led more often to final $\widehat{p_{i,smooth}}$ with slightly higher (S/N)$_{MC}$.
However, after applying the same adjusted scaling and baseline methods to raw smoothed estimates from the other starting points, 
final smoothed histograms from these other starting points again became comparable to those from ($\hat{p}_i$)$_{\bowtie}$.
In the end, when using the \emph{unsmoothed} estimators with improved $\hat{p}_i\leftrightarrow p_i$ matching as alternate starting points for smoothed histograms, the smoothed histograms constructed from $N$=$40$-$80$ observations
had S/N values that matched those of unsmoothed histograms with $\approx 20 \times$ as many observations. 
At each $N$, the final S/N for $\widehat{p_{i,smooth}}$ from all starting estimates is nearly identical, implying that the \emph{relative} improvement is greater for unsmoothed estimators having the smallest S/N.
Generally (S/N)$_{MC}$ for smoothed estimates derived from each different unsmoothed starting estimator hovered within 10-20\% or so of each other for most examined combinations of $N$ and $g(x)$, 
with a few excursions from proximity for some particular starting estimates for some particular values of $N$ or $g(x)$.
The new heuristic method for baseline adjustment and scaling was very different than the earlier one and (S/N)$_{MC}$ values for smoothed histograms depended heavily on the details of this scaling.
Regardless of which procedure for rescaling was chosen, smoothed estimates provided higher (S/N)$_{MC}$ than unsmoothed estimates, with larger relative improvements at small $N$ and a typical sampling improvement about 10-fold (Table S1).


\subsection{Estimation of ${\sigma}_{p_i}$, a Measure of Bin-by-Bin Uncertainty in $\hat{p}_i$ }

Although the large increase in (S/N) was expected to help with the overall goal of improving Bayesian classification, it was a shift from the original focus, 
which was to obtain better estimates of the \emph{uncertainty} in $\hat{p}_i$ to allow better assessment of the uncertainty in the composite $P_o$ for classification calculated from these $\hat{p}_i$.
Qualitatively, $\sigma_{p_i}$, as assessed by agreement between $\hat{p}_i$ and $p_{i,true}$ in the MC simulations, was significantly reduced by de-noising and scaling.
The only remaining problem was that quantitative estimates of $\sigma_{p_i}$ for the de-noised $\hat{p}_i$ based on earlier estimators (Eq. 2, 4 or 8) were no longer valid.
On trying to resolve this issue of estimating $\hat{\sigma}_{p_{i,smooth}}$, some issues became more apparent about the original errors when estimating uncertainty in composite $P_o$ used for Bayes classification.

\subsection{Accurate $\hat{ \sigma}_{p_i} $ for Directly De-noised Experimental Histograms}

To estimate values for $\sigma_{p_i}$, a largely empirical approach was taken, 
since earlier results, such as the curve of $\xi$-to-$C$ optimized $\alpha_0(N)$ values, were difficult to explain.  
In contrast, many earlier estimators $\hat{\sigma}_{p_i}$ or $\hat{\delta}_i$, such as those based on probability matching priors \citep{dattaMukerjee,rousseau2000,rousseau2002}, 
were derived from first principles, often with many inherent presumptions.
MC had been used as the ultimate standard to test theoretically derived  $\hat{\sigma}_{p_i}$ or $\hat{\delta}_i$ from this earlier bottom-up approach.
Here a top-down approach was used instead.
Since MC was to be the ultimate standard,
accuracy might be improved by starting from the MC calculation and using a parameterized fit to the outcome to get $\hat{\sigma}_{p_i}$.

Use was made of a common functional form shared by earlier derived estimators $\hat{\sigma}^2_{p_i}$,
 a form that contained the value $\hat{p}_i$ (here $\widehat{p_{i,smooth}}$), the estimate for the proportion itself: 
\begin{gather}
\sigma_{i,est} = \hat{\sigma}_{p_i} = \sqrt{ \frac{ \hat{p}_i  (1-\hat{p}_i)}{A_0 N + B_0 }}\text{~.} 
\end{gather}
Given this parameterized equation to estimate $\sigma_{p_i}$ from $\widehat{p_{i,smooth}}$ based on a \emph{single, N-observation MC experiment} ($N$ observations distributed into bins of \emph{single histogram}),
optimal values for $A_0$ and $B_0$ could be determined by non-linear regression to the observed long-run MC results from many $N$-observation histograms for each of the 6 arbitrary underlying trial pdfs $g(x)$. 
Starting from arbitrary values, $A_0$ and $B_0$ were optimized by 
the Newton-Raphson method, zeroing a composite median function (more detail below) derived from 
$F(A_0, B_0, N) = \frac{\sigma_{i,est}(A_0,B_0,N)} {\sigma_{p_i,MC}}-1$ 
for each bin, at each value of $N$, and for each underlying trial pdf. 
Here, $\sigma_{p_i,MC}$ is a \emph{long-run} estimate $\langle (\hat{p}_i- p_i)^2 \rangle_{MC}$ for $\sigma_{p_i}$ for bin $i$ over \emph{many MC histograms}. 
This long-run MC estimate derives from the histogram-to-histogram variation of $\hat{p}_i$ for bin-$i$ of each underlying pdf at fixed $N$.

The sign of $F()$ indicates if the parametric estimate $\sigma_{i,est}$ for each bin of each random histogram is more prone toward overconfidence or underconfidence ($C < \xi$ or $C > \xi$).
Overconfidence, 
the underestimation of $\sigma_{p_i,MC}$ by $\sigma_{i,est}$ for particular histogram bins as indicated by a negative value of $F()$, 
had to be avoided for accurate Bayes classification. 
To determine values of $A_0$ and $B_0$ from MC results, one \emph{median} value of $\frac{\langle \sigma_{i,est}(A_0,B_0,N) \rangle_{MC} }{\sigma_{p_i,MC}} $ over histogram bins was used in $F(A_0,B_0,N)$
for each "model" pdf $g(x)$ at each $N$. 
Here $\langle \sigma_{i,est}(A_0,B_0,N) \rangle_{MC}$ is the 10,000 run MC-average of single-histogram parametric estimates (Eq. 9) for bin $i$ based on the current presumed values of $A_0$ and $B_0$ 
and $\sigma_{p_i,MC}$ is the actual long-run variance $\langle(\hat{p}_i-p_{i})^2 \rangle_{MC}$ as before. 
Use of histogram medians of MC-averages avoided instabilities in more standard least squares optimization of $A_0$ and $B_0$ due to small numbers of large outlying MC-average estimates $\langle \hat{p}_i \rangle$ in some bins for the randomly generated values from some underlying trial densities.

Non-linear optimization converged to a fairly stable fixed point,
giving common values $\bar{A_0}$ and $\bar{B_0}$ for which fairly reliable single-histogram estimates $\hat{\sigma}_{p_i}$ could be derived from Eq. 9 for de-noised estimates $\widehat{p_{i,smooth}}$ from all trial pdfs.
Table \ref{tab:A0B0Tables} lists values of $A_0$ and $B_0$ optimized for use with $\widehat{p_{i,smooth}}$ from different initial unsmoothed estimators $\hat{p}_{i,0}$.
The values $\bar{A_0}$ and $\bar{B_0}$ work for $b$=100 with the current smoothing parameters, but they change significantly with changes in $b$, in the smoothing parameters, and in the method used to correct the baseline and scale of the initial smoothed estimates.

     \begin{table}[t!]
      \begin{center}
    \scalebox{1.00} {
   \begin{tabular} { | c | c  c |  }
    \hline
  \multicolumn{1}{|c|} {Unsmoothed}           &  \multicolumn{2}{|c|} {\emph{100-bin smoothed histogram:}}    \\ 
  \multicolumn{1}{|c|} {Starting Estimator}        & {Optimized}             & {Optimized }   \\
  \multicolumn{1}{|c|} {or Estimator Set:}           &    $\bar{A_0}$:               & $\bar{B_0}$          \\ 
        \hline
  \multirow{1}{*}{Multinomial, $b_{eff}$=$100$}          &    10.00               & 561.3        \\   
        \hline
  \multirow{1}{*}{rgbp (joint), $b_{eff}$=$100$}         &     9.47               & 574.0        \\   
    \hline 
  \multirow{1}{*}{Discrete, $b_{eff}$=$100$}             &    10.57               & 655.4        \\   
        \hline
  \multirow{1}{*}{$C$-to-$\xi$ Optimized, $b_{eff}$=$2$} &    10.50               & 623.2        \\   
        \hline
  \multirow{1}{*}{Discrete \& Optim'd Combined}     &     10.51               & 633.5        \\   
        \hline
  \multirow{1}{*}{Best$^*$ 3 of 4 ($\approx$ S/N)}     &     9.93               & 578.2        \\   
        \hline
  \multirow{1}{*}{Full Set}                        &    10.14               & 590.5        \\   
        \hline
  \end{tabular}
                   }
      \end{center}
\caption{  Values $\bar{A_0}$ and $\bar{B_0}$ (Eq. 9) for single-run parametric estimates of $\sigma_{p_i}$ from a single smoothed histogram based on $\widehat{p_{i,smooth}}$ and $N$
           (co-optimized for 4 different estimators).
          Each cycle of ($\bar{A_0}$,$\bar{B_0}$) optimization used a total of 960,000 histograms from random data:  
          10,000 histograms were generated by each of 6 underlying trial PDFs for each of 16 different values for $N$.
          After 6 optimization cycles, refined values $\bar{A_0}$ and $\bar{B_0}$ exhibited consistency to $\approx 0.1-2.0\%$ of the value.
          $^*$Initially, the $C$-to-$\xi$ optimized estimator 
          exhibited anomalous unsmoothed $\langle S/N \rangle_{MC}$ (Table S1), so joint optimization of the other 3 was examined.  
 \label{tab:A0B0Tables}  }
  \end{table}

\begin{figure}
\centering
\scalebox{0.55} {
\centering
\includegraphics[width=15 cm]{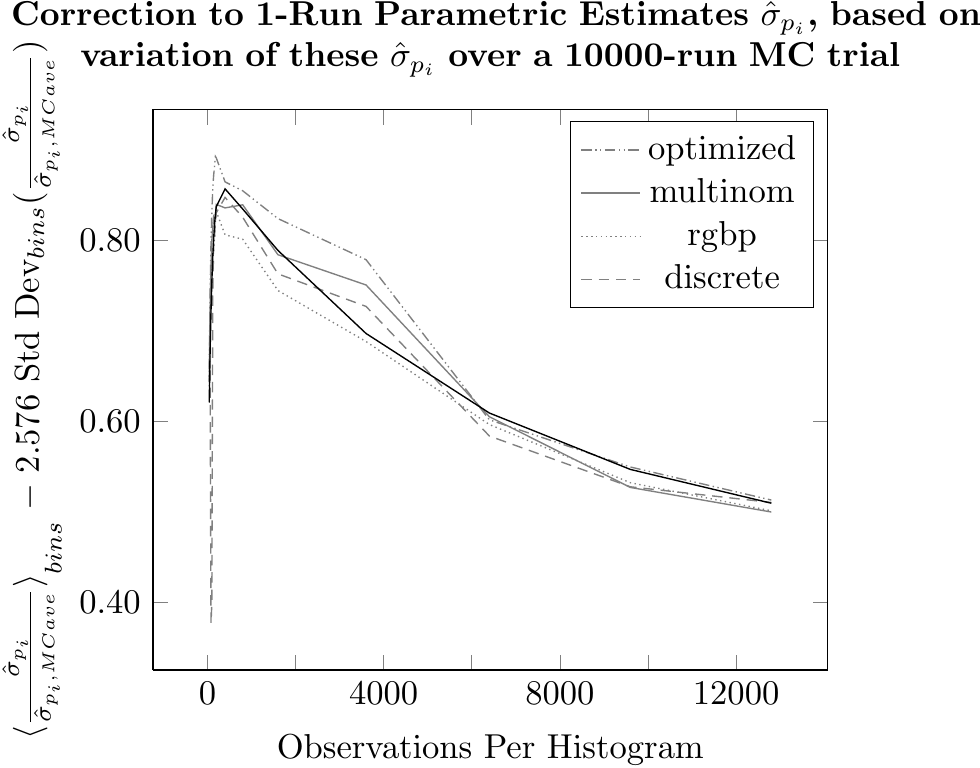}
}
        \caption{ Estimates $\hat{\sigma}_{p_i}$ from a single experimental run (1 histogram based on $N$ observations) are occasionally too small, leading to 'unwarranted overconfidence' in the certainty of the estimated value $\hat{p}_i$.  
                  Based on the run-to-run variance over 10000 random $N$-observation histograms, the underestimation factor $\langle\frac{\hat{\sigma}_{p_i}}{\hat{\sigma}_{p_i,MC}}\rangle - 2.576 $ Std Dev$( \frac{\hat{\sigma}_{p_i}}{\hat{\sigma}_{p_i,MC}} )$
                  can be used to approximate ${\hat{\sigma}^{(0.995)}_{p_i}}$, the lower limit above which the true value $\sigma_{p_i}$ will be "at 0.995 confidence."
                  A 1-run estimate ${\hat{\sigma}_{p_i}}$ is corrected by dividing by this "worst-case" underestimation factor.
                  Shown is the variation in this correction factor with number of observations per 1-run experiment, $N \in [40,12800]$.
                  Plots use averages for bins from 6 types of underlying trial pdfs (60000 total MC runs).
                  For each test histogram, 1-run bin-by-bin estimates of underlying pdf ($\hat{p}_i$) and of $\hat{\sigma}_{p_i}$ were obtained by smoothing initial $\hat{p}_i$ from 4 choices of estimator (top right).
                  The black curve is a joint fit from which approximate estimates of the correction factor can be drawn: $\Psi =  (0.45371) + (0.43287) e^{-(1.6000D-04)N} (1-e^{-(1.2296D-02)N})$.
                        \label{fig:sigRatioLimit}}
\end{figure}

When using optimal values $\bar{A_0}$ and $\bar{B_0}$, the median over bins of the ratio $\rho_i = \frac{\sigma_{i,est} }{\sigma_{p_i,MC}} $ remained close to 1, 
exhibiting slight systematic variation with sample size $N$. 
Expectedly, $\langle\rho_{i}\rangle_{MC} = \frac{\langle\sigma_{i,est}\rangle_{MC}}{\sigma_{p_i,MC}}$ for individual bins also exhibited random bin-to-bin fluctuations about the value 1.0 within histograms from each trial pdf and $N$ value in the MC run.  
These fluctuations about the value 1.0 for each single-bin ratio, $\langle \rho_i \rangle_{MC} = \frac{\langle\sigma_{i,est}\rangle_{MC} }{\sigma_{p_i,MC}} $, as quantified by the standard deviation of this ratio, $\sigma_{\langle\rho_i\rangle_{MC}}$, 
over histogram bins and underlying pdf choices, varied with varying $N$-value and with the choice of initial estimator $\hat{p}_i$ before de-noising. 
However, for bins from underlying pdfs that were sampled at the same fixed value $N$, 
even as the underlying pdf and the bin-to-bin values of $p_i$ varied, the variance values for the ratio, $\sigma^2_{\langle\rho_i\rangle_{MC}}$,
varied only slightly.  \emph{That is, $\sigma^2_{\langle\rho_i\rangle_{MC}}$ appears to be fairly insensitive to the underlying pdf being estimated by the experimental histogram}.
Since the variance in the ratio $\rho_i$, $\sigma^2_{\rho_i}$, is fairly systematic and dictated largely by $N$ (the total number of observations per histogram),
lower limiting values of $\rho_i$ based on estimates of its variance 
from long-run MC results ($\sigma^2_{\langle\rho_i\rangle_{MC}}$) can be used to adjust initial single-run estimates $\hat{\sigma}_{p_i}$ from Eq. 9 
to correct for uncertainty present in such parametric estimates of $\sigma_{p_i}$ derived from a single histogram.
Figure \ref{fig:sigRatioLimit} is a plot of the variation of estimated lower limiting values for $\rho_i$ with $N$ at the 0.01 tolerance level 
($\rho_{0.99} \approx \mu_{\rho_i} - 2.576 \sigma_{\rho_i}$) for smoothed $p_i$ histograms calculated for 4 different forms for the initial unsmoothed estimators $\hat{p}_i$.
Based on the MC observations, when one estimates lower limiting values of $\rho_i$ from the value $\sigma^2_{\langle\rho_i\rangle_{MC}}$, 
by presuming an idealized Gaussian form for $\rho_i$ ($\rho_i \sim \mathcal{N}[\mu_{\rho_i}(\approx 1),\sigma^2_{\rho_i}; x]$), 
then 1\% of the time $\rho_i$, and therefore $\hat{\sigma}_{p_i}$ based on single-run single-histogram estimates, is expected to be 1.3-1.8 times too small (reciprocal of ordinate value in Fig \ref{fig:sigRatioLimit}) for many values of $N$.
Basically, the run-to-run variation in the estimation of $\sigma_{p_i}$ leads to an expected, and predictable frequency of underestimation of $\sigma_{p_i}$ by single-histogram parametric estimators. 

The presence of expected 1.3-1.8 fold underestimates in single-run estimates $\hat{\sigma}_{p_i}$ that occur at the 0.01 tolerance level 
accentuates the fact that the typical concept of confidence levels for $\hat{p}_i$ and achieving matching between $C$ and $\xi$ fails to account for \emph{uncertainty and spread} in estimated values $\hat{\sigma}_{p_i}$ 
(or more generally the spread in estimates $\hat{\delta}^+$ and $\hat{\delta}^-$). 
However, if values for $\mu_{\rho_i}\approx$ constant and if $\sigma_{\rho_i}$ can be estimated reliably by MC results, 
then the limiting low value of $\rho_i$ at the 0.01 tolerance limit 
($\rho_{0.99} \approx \mu_{\rho_i} - 2.576 \sigma_{\rho_i}$)
provides a scale factor to correct the single-histogram estimates $\hat{\sigma}_{i,est}$ appearing in the numerator of $\rho_i$. 
Since it is not known which of the bins is among the 1\% with underestimated $\sigma_{p_i}$ values though, this means that all parametric $\sigma_{i,est}$ must be corrected.
Since most estimates of $\sigma_{p_i}$ rescaled by this procedure are too large, rescaling increases the frequency of indeterminable cases ($P_o < Z \sigma_{P_o}$) for Bayes classification, but reduces misclassification. 
Fortunately, when $p_{i}$ and $\sigma_{i}$ are estimated from de-noised histograms, 
the increase in (S/N) from de-noising greatly reduces the initial parametric estimates $\hat{\sigma}_{est}$ from Eq. 9 relative to unsmoothed estimates $\hat{\sigma}_{p_i}$ 
so that smoothed $\hat{\sigma}_{p_i}$ remains smaller even after this compensatory up-scaling.

Note that this problem with uncertainty in estimates $\hat{\sigma}_{p_i}$ is not inherent to the smoothing, but is always present for estimates $\hat{\sigma}_{p_i}$ from single-run (single-experiment) estimates, even for classical parametric estimators for $\sigma_{p_i}$.
Failure to correct for uncertainty in single-run $\hat{\sigma}_{p_i}$ presumes a tolerance of 0.5 for low estimates $\hat{\sigma}_{p_i} < \sigma_{i,true}$ that lead to $ C < \xi$ and sometimes to $C \ll \xi$;
failure to acknowledge uncertainty in $\hat{\sigma}_{p_i}$ leaves the frequency of occurrence and extent of underestimates $\hat{\sigma}_{p_i}$ unknown.  
For classifications involving several 100 bin ($b=100$) histograms, the 0.01 tolerance limit on $\sigma_{\rho_i}$ is exceeded fairly often and so some correction appears to be required to avoid misclassification.
Thus using:
\begin{gather}
\hat{ \sigma}_{p_i} = \sigma_{i,est}(\bar{A_0},\bar{B_0},\widehat{p_{i,smooth}},N) = \frac{1}{\rho_{0.99}(N)}\sqrt{\frac{\widehat{p_{i,smooth}}(1-\widehat{p_{i,smooth}})}{\bar{A_0} N + \bar{B_0}}} 
\end{gather}
reproduces the "observed" overall MC variance in $\hat{p}_i$ from run to run for a particular value of $N$ when considering all bins in the combined collection of 6 known model pdfs, 
and includes $\rho_{0.99}(N)$ as a correction to adjust for underestimates due to run-to-run uncertainty in $\hat{\sigma}_{i,est}$.  
Approximate functional forms for $ \rho_{0.99}(N) = \mu_{\rho_i} - 2.576 \sigma_{\rho_i}$ (Fig.\ref{fig:sigRatioLimit}) from a single joint fit to $\rho$ from the underlying pdfs of the test set allows one to apply the analogous corrections to new histograms without re-running MC for each case.
Scaling by $\rho_{0.99}(N)$ values from such fits 
avoided anticipated large errors in Bayesian classification from the occasional 20-50\% underestimates of confidence interval widths due to statistical variation in $\hat{\sigma}_{p_i}$.

Although different in detailed value, the overall shape of $\rho_{0.99}(N)$ was fairly similar for smoothed histograms derived from different starting estimators $\hat{p}_{i,0}$ (Fig.\ref{fig:sigRatioLimit}).
Larger changes in $\rho_{0.99}(N)$ values were observed upon improving the heuristic procedure for scaling and baseline-adjustment after histogram smoothing,
but again the overall shape of the $\rho_{0.99}(N)$ curves remained similar. 
All $\rho_{0.99}(N)$ curves exhibited maxima near about $N$=$180$ observations, a common feature that likely results from
the number of bins in the histogram and from the detailed histogram smoothing parameters, which were all held constant here.

The closer the curve in Fig.\ref{fig:sigRatioLimit} to 1 for a particular starting estimator $\hat{p}_{i,0}$, the smaller the required correction to the initial $\hat{\sigma}_{p_i}$ from Eq. 9.  
Invariance of \emph{relative} positions of $\rho_{0.99}(N)$ plots for each pre-smoothing estimator $\hat{p}_{i,0}$ throughout the process of optimizing the values of $A_0$ and $B_0$ for Eq. 9 suggests 
slight systematic differences in the run-to-run variation in estimates $\hat{\sigma}_{p_{i,smooth}}$ from different $\hat{p}_{i,0}$, but the effect is small and decreases as $A_0$ and $B_0$ become better determined.
The $(\hat{p_i})_{\xi\leftrightarrow C}$ optimized estimators require the smallest correction to initial $\hat{\sigma}_{p_i}$ (run-to-run variance in single-histogram based estimates $\hat{\sigma}_{p_i}$ from Eq. 9 is smallest);
the $\hat{\sigma}_{p_i}$ for joint estimators ($\hat{p}_i$)$_{\bowtie}$ require the largest correction (highest run-to-run variance in $\sigma_{i,est}$ from Eq. 9); and the discrete and multinomial estimators require intermediate sized corrections. 
The corrections to $\hat{\sigma}_{p_i}$ for $\widehat{p_{i,smooth}}$ from all starting estimators $\hat{p}_{i,0}$ became more equivalent at higher $N$. 
Differences in the relative run-to-run consistency of single-histogram estimates $\hat{\sigma}_{p_i}$ from different unsmoothed $\hat{p}_{i,0}$, as indicated by the relative size of up-scaling corrections to $\hat{\sigma}_{p_i}$ [$\rho_{0.99}(N)$], 
appear unrelated to relative ultimate $\widehat{p_{i,smooth}}$-to-$p_i$ consistency [S/N],
which is fairly high for all starting unsmoothed estimators.  
Given that results from different $\hat{p}_{i,0}$ were similar, the discrete estimator seemed less practical since it required a one-time co-optimization of admissible values $\hat{\theta}_i$ for all possible outcomes at a given $N$ and this became slow for $N > 1000$.
The starting estimator ($\hat{p}_i$)$_{\bowtie}$ was also more computationally intensive without obvious advantage to this point.

Next, we examine plots to assess the accuracy of the final estimators $\widehat{\sigma_{p_i,smooth}}$ from Eq. 10 and compare $\xi$-to-$C$ and $\hat{p}_i$-to-$p_{true}$ accuracy for different underlying trial pdfs.


\section{Plots of Multinomial Estimates $\widehat{p_{i,smooth}}$ and of Coverage from $\widehat{\sigma_{i,smooth}}$}
\subsection {Test Conditions}

For computational tests of the final estimators, the trial underlying pdfs $g(x)$ to be examined were chosen to be
the 6 pdfs from the original basis set 
along with an additional 8 beta-pdfs, $\mathcal{B}[\alpha,\beta;x]$, each with differing values of $\alpha$ and $\beta$ and hence differing amounts of skew.
Relative estimator accuracy was compared for de-noised estimator pairs $\hat{p}_i$ and $\hat{\sigma}_{p_i}$ from 4 different \emph{unsmoothed} estimators:  
($\hat{p}_i$)$_{\xi\leftrightarrow C}$ from $\mathcal{B}[\alpha_0,\alpha_0;x]$ prior $\times~\mathcal{N}$-posterior with values for $\alpha_0$ optimized at a nominal binomial confidence level of $\xi=0.95$;
unoptimized multinomial estimators (Generalized Rule of Succession, $\mathcal{B}[1,(b-1); x]$ prior $\times~\mathcal{N}$-posterior);
the Rgbp ($\hat{p}_i$)$_{\bowtie}$ joint estimator \citep{Rgbp}; and the discrete domain estimators $\hat{p}(n_i,N)$ based on the purely combinatorial prior.
Tests using estimators based on the Jeffreys-Bayes prior had been examined at earlier stages of development,
but the resulting estimates $\hat{p}_i$ and $\hat{\sigma}_{p_i}$ were no better and often less accurate than estimates using the uniform Bayes priors for the ranges of $p_i$ values relevant to this histogram analysis.\footnote{
After the recent improvements in the scaling and baseline adjustment methods, it is possible that smoothing of starting estimators derived from Jeffreys-Bayes priors and beta-function posteriors may lead to disproportionate improvements,
but the similarity of the final accuracy in the following results from the variety of other starting $\hat{p}$ used for smoothing suggested that this was not likely.}

\subsection {Qualitative Comparison}
In general, no difference in the results could be detected between trial pdfs that had or had not been pdfs in the original basis set used to develop the scaling method and to establish the parameters for error estimation.
Any general trends in accuracy and run-to-run reliability appeared to reflect similarities in the general shapes of the underlying pdf more so than provenance from the basis set, and so only a few unique cases are presented.

\begin{figure}
\centering
\scalebox{0.73} {
\centering
\includegraphics[width=15 cm]{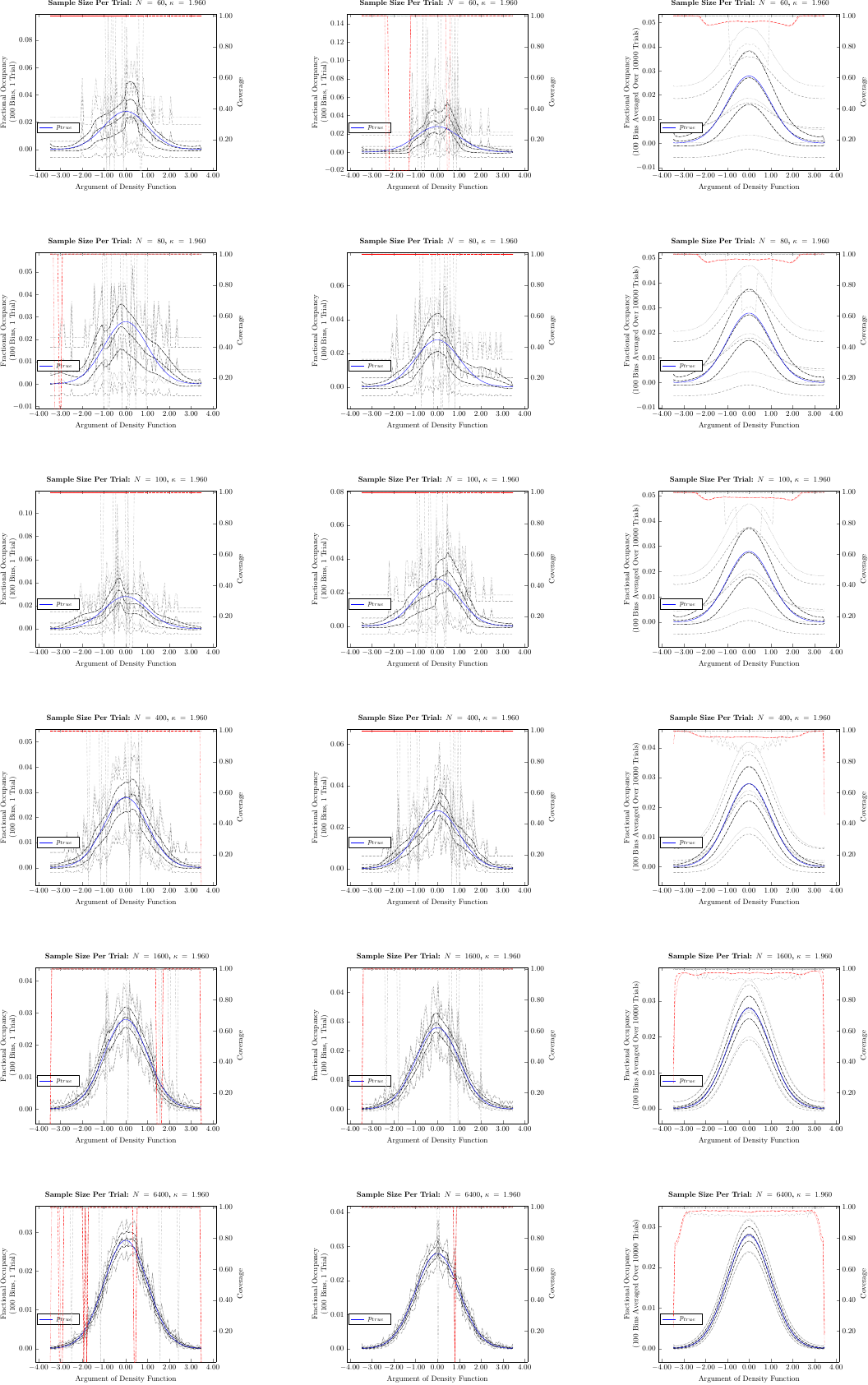}
}
                \caption{ Estimates of a randomly generated underlying Standard Normal PDF using a 100-bin histogram. 
                         Each line-type shows separate plots of $\hat{\delta}^{+}$, $\hat{p}_i$, and $\hat{\delta}^{-}$ at $\xi$=0.95 (normal posterior PDF) for a different estimator.
                         Shown are smoothed (black lines) and unsmoothed (gray lines) estimates by the \emph{Multinomial Rule-of-Succession} (dashed lines) or the \emph{RGBP Joint} estimator (dotted lines).
                         In each row (single sample size $N$),
                         two graphs at the left are separate single MC runs, and  
                         the graph at the right is the average of 10,000 MC runs.
                         Coverage here is the fraction of total (1 or 10,000) MC trials for which the underlying PDF ($p_i$, blue) lies in ($\hat{\delta}^{-}$,$\hat{\delta}^{+}$).
                         It is shown in red for smoothed estimates and in gray for unsmoothed ones, with line types corresponding to those of the density estimates.
                        \label{fig:estHistA}}
\end{figure}

\begin{figure}
\centering
\scalebox{0.73} {
\centering
\includegraphics[width=15 cm]{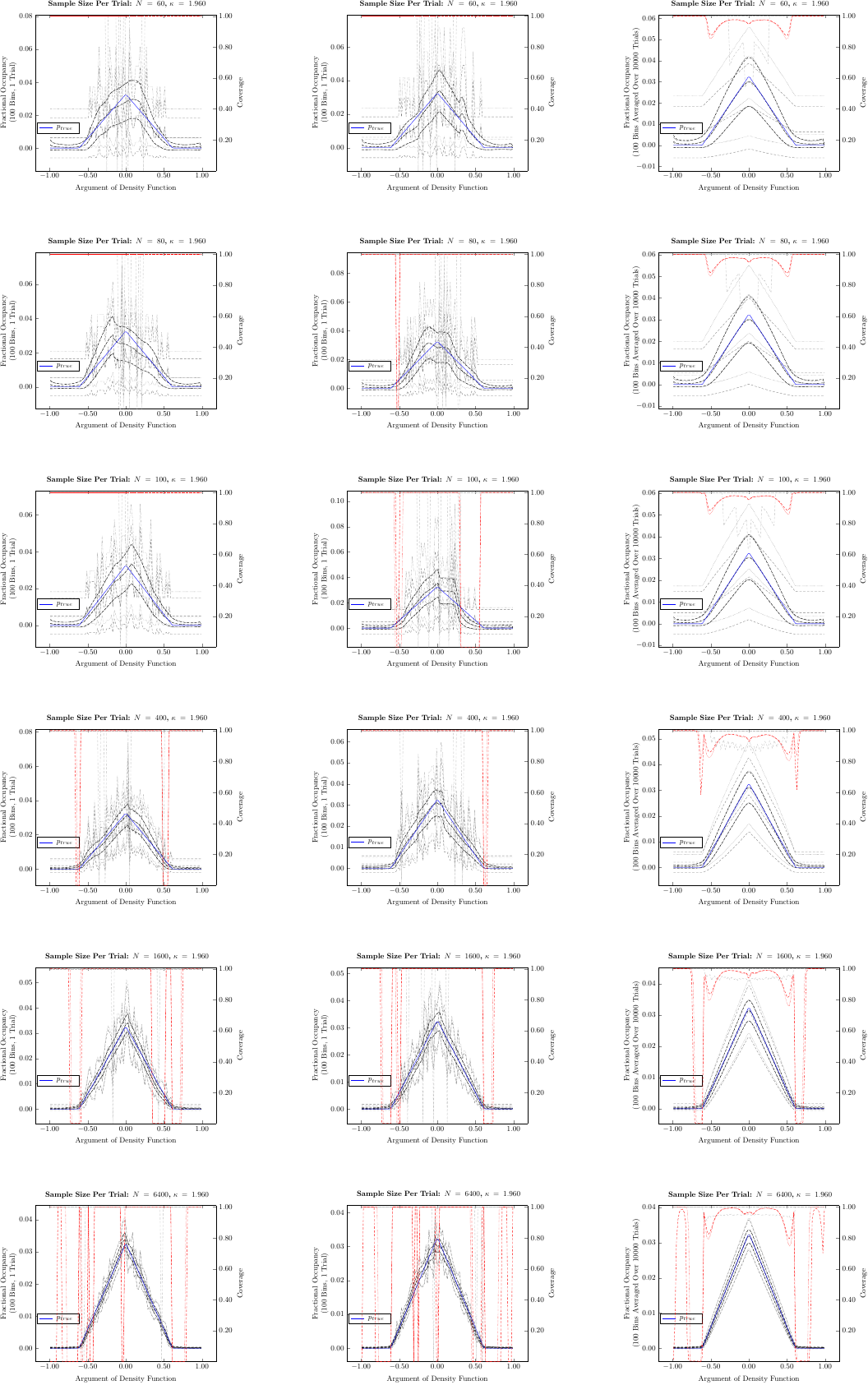}
}
                \caption{ Estimates of a randomly generated underlying Sawtooth PDF using a 100-bin histogram. 
                         Each line-type shows separate plots of $\hat{\delta}^{+}$, $\hat{p}_i$, and $\hat{\delta}^{-}$ at $\xi$=0.95 (normal posterior PDF) for a different estimator.
                         Shown are smoothed (black lines) and unsmoothed (gray lines) estimates by the \emph{Multinomial Rule-of-Succession} (dashed lines) or the \emph{RGBP Joint} estimator (dotted lines).
                         In each row (single sample size $N$),
                         two graphs at the left are separate single MC runs, and  
                         the graph at the right is the average of 10,000 MC runs.
                         Coverage here is the fraction of total (1 or 10,000) MC trials for which the underlying PDF ($p_i$, blue) lies in ($\hat{\delta}^{-}$,$\hat{\delta}^{+}$).
                         It is shown in red for smoothed estimates and in gray for unsmoothed ones, with line types corresponding to those of the density estimates.
                        \label{fig:estHistB}}
\end{figure}

\begin{figure}
\centering
\scalebox{0.73} {
\centering
\includegraphics[width=15 cm]{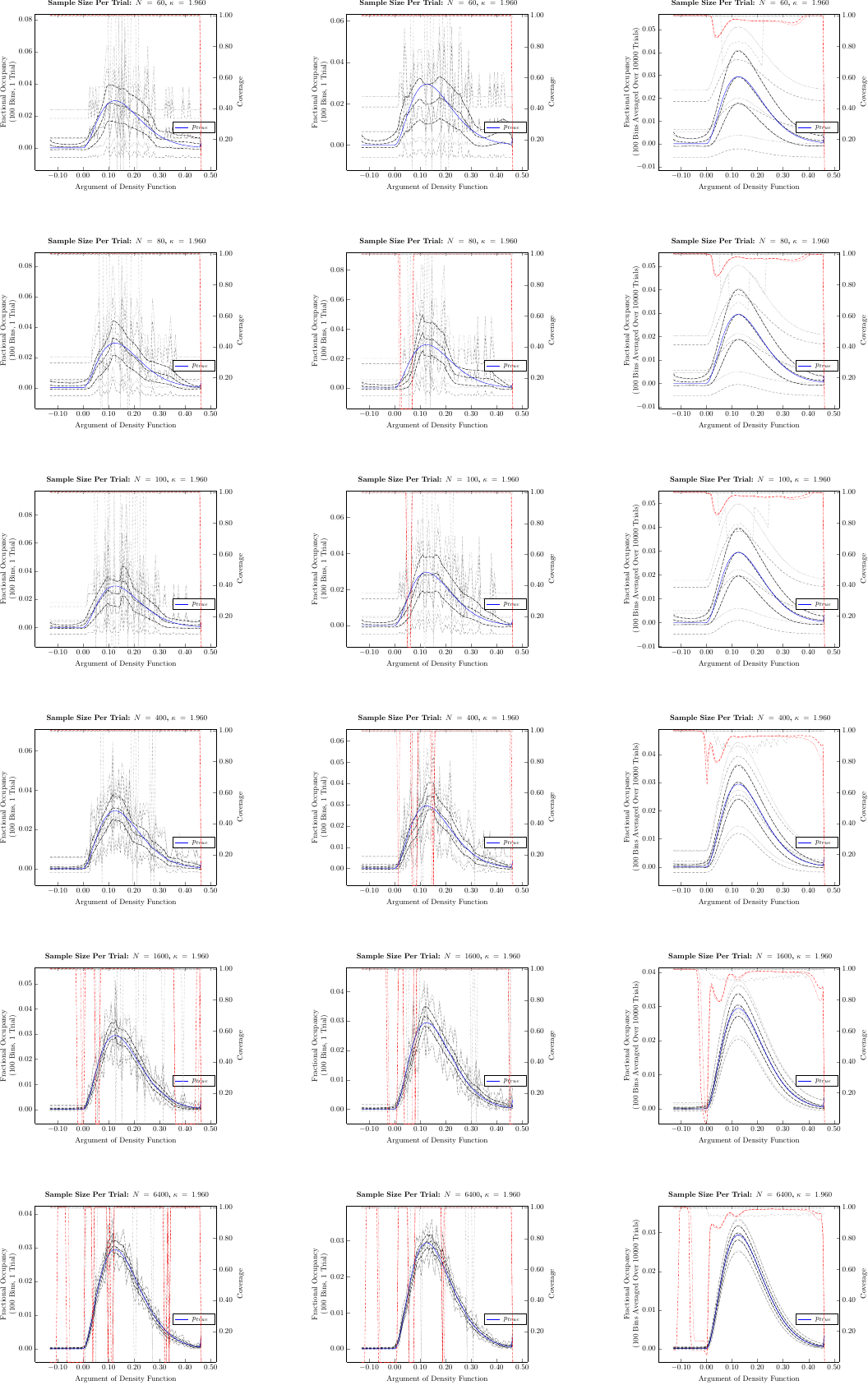}
}
                \caption{ Estimates of a randomly generated underlying $\mathcal{B}[3.0,15.0;x]$ PDF using a 100-bin histogram. 
                         Each line-type shows separate plots of $\hat{\delta}^{+}$, $\hat{p}_i$, and $\hat{\delta}^{-}$ at $\xi$=0.95 (normal posterior PDF) for a different estimator.
                         Shown are smoothed (black lines) and unsmoothed (gray lines) estimates by the \emph{Multinomial Rule-of-Succession} (dashed lines) or the \emph{RGBP Joint} estimator (dotted lines).
                         In each row (single sample size $N$),
                         two graphs at the left are separate single MC runs, and  
                         the graph at the right is the average of 10,000 MC runs.
                         Coverage here is the fraction of total (1 or 10,000) MC trials for which the underlying PDF ($p_i$, blue) lies in ($\hat{\delta}^{-}$,$\hat{\delta}^{+}$).
                         It is shown in red for smoothed estimates and in gray for unsmoothed ones, with line types corresponding to those of the density estimates.
                        \label{fig:estHistC}}
\end{figure}

\begin{figure}
\centering
\scalebox{0.73} {
\centering
\includegraphics[width=15 cm]{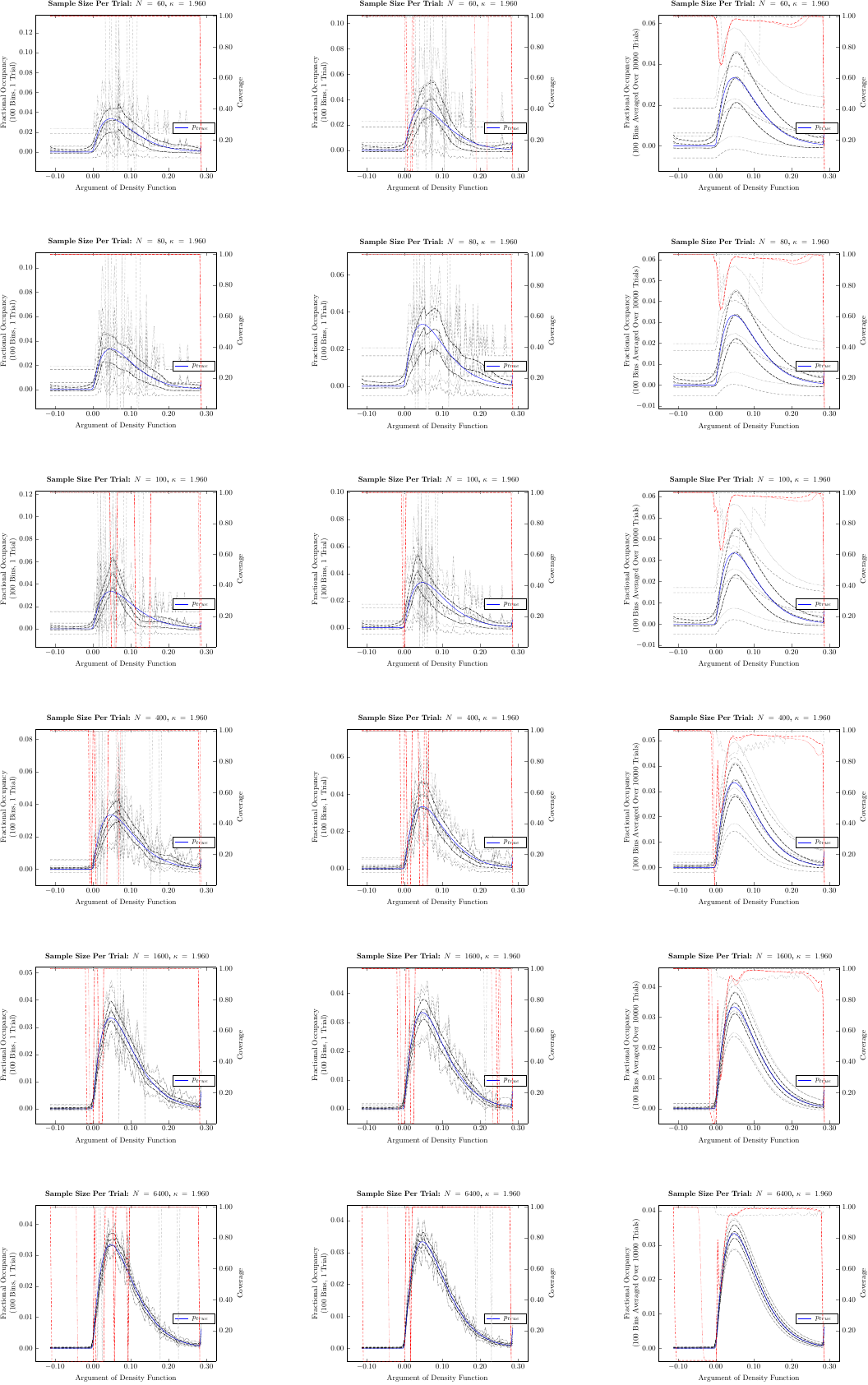}
}
                \caption{ Estimates of a randomly generated underlying $\mathcal{B}[2.0,21.0;x]$ PDF using a 100-bin histogram. 
                         Each line-type shows separate plots of $\hat{\delta}^{+}$, $\hat{p}_i$, and $\hat{\delta}^{-}$ at $\xi$=0.95 (normal posterior PDF) for a different estimator.
                         Shown are smoothed (black lines) and unsmoothed (gray lines) estimates by the \emph{Multinomial Rule-of-Succession} (dashed lines) or the \emph{RGBP Joint} estimator (dotted lines).
                         In each row (single sample size $N$),
                         two graphs at the left are separate single MC runs, and  
                         the graph at the right is the average of 10,000 MC runs.
                         Coverage here is the fraction of total (1 or 10,000) MC trials for which the underlying PDF ($p_i$, blue) lies in ($\hat{\delta}^{-}$,$\hat{\delta}^{+}$).
                         It is shown in red for smoothed estimates and in gray for unsmoothed ones, with line types corresponding to those of the density estimates.
                        \label{fig:estHistD}}
\end{figure}

\begin{figure}
\centering
\scalebox{0.73} {
\centering
\includegraphics[width=15 cm]{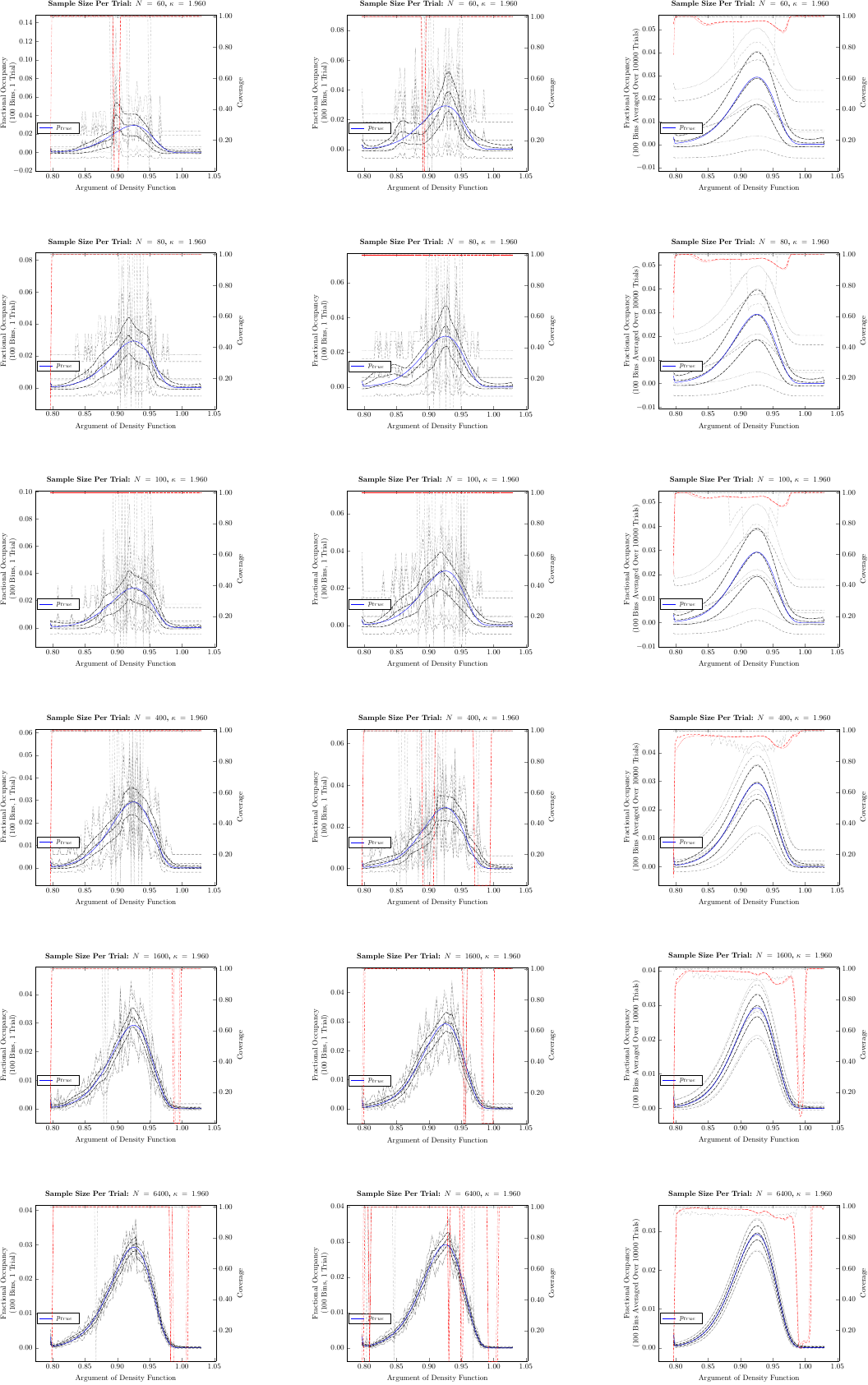}
}
                \caption{ Estimates of a randomly generated underlying $\mathcal{B}[63.0,6.0;x]$ PDF using a 100-bin histogram. 
                         Each line-type shows separate plots of $\hat{\delta}^{+}$, $\hat{p}_i$, and $\hat{\delta}^{-}$ at $\xi$=0.95 (normal posterior PDF) for a different estimator.
                         Shown are smoothed (black lines) and unsmoothed (gray lines) estimates by the \emph{Multinomial Rule-of-Succession} (dashed lines) or the \emph{RGBP Joint} estimator (dotted lines).
                         In each row (single sample size $N$),
                         two graphs at the left are separate single MC runs, and  
                         the graph at the right is the average of 10,000 MC runs.
                         Coverage here is the fraction of total (1 or 10,000) MC trials for which the underlying PDF ($p_i$, blue) lies in ($\hat{\delta}^{-}$,$\hat{\delta}^{+}$).
                         It is shown in red for smoothed estimates and in gray for unsmoothed ones, with line types corresponding to those of the density estimates.
                        \label{fig:estHistE}}
\end{figure}

To appreciate both the gain in accuracy and the inherent differences between the different procedures for estimating values for $p_i$ in a 100-bin histogram,
it is useful to examine both individual estimates from separate single $N$-observation random trials and the average of estimates from a larger number of $N$-observation trials.
These are shown in Figures \ref{fig:estHistA}-\ref{fig:estHistE} using $\bar{A}_0 = 10.51$ and $\bar{B}_0 = 633.0$ for all $\hat{p}_{i,0}$ and a selected sampling of examined pdfs.

\begin{figure}
\centering
                \begin{subfigure}[h]{0.68\textwidth}
                 \includegraphics[width=\textwidth, clip=true, trim= 0cm 0cm 0cm 0cm ]{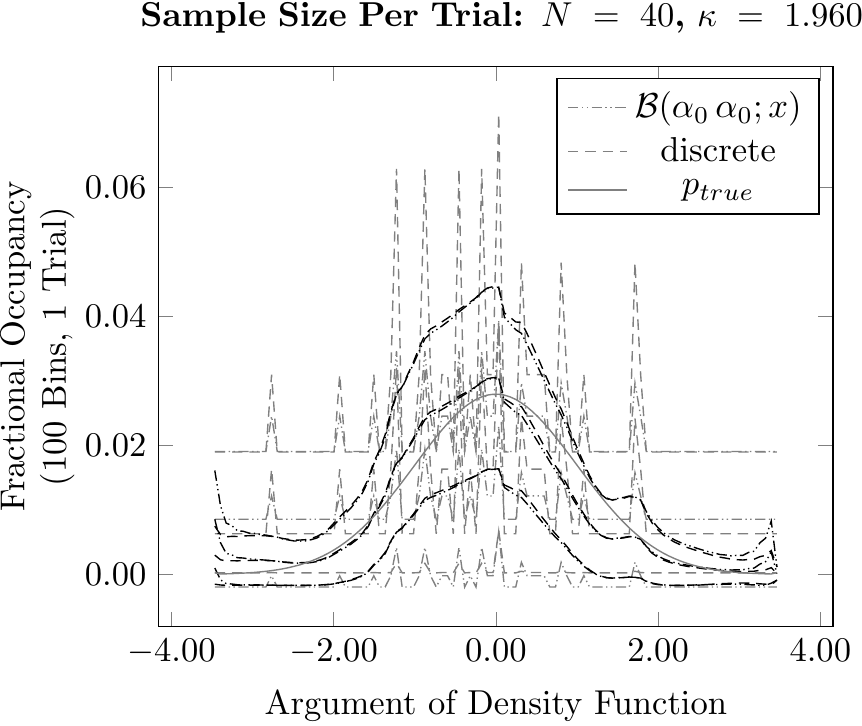}
                \end{subfigure}

                \vspace{16pt}

                \begin{subfigure}[h]{0.68\textwidth}
                 \includegraphics[width=\textwidth, clip=true, trim= 0cm 0cm 0cm 0cm ]{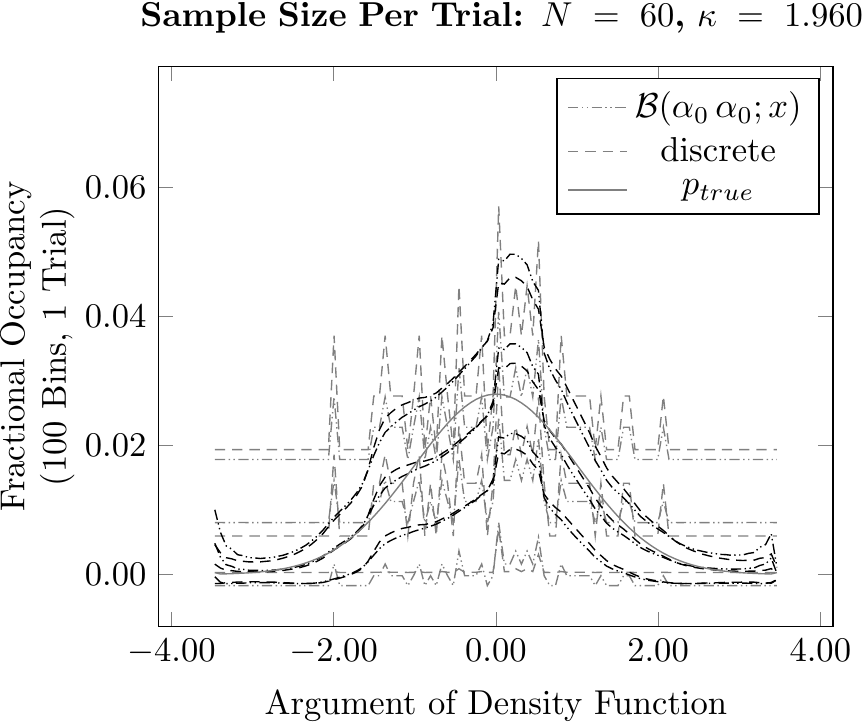}
                \end{subfigure}
                \caption{ Estimates of a randomly generated Standard Normal PDF using a 100-bin histogram at small sample sizes. 
                         Different line-types, corresponding to estimates from the \emph{Optimized} $\mathcal{B}(\alpha_0,\alpha_0;x)$ or the \emph{Discrete} estimator (original, gray; de-noised, black), 
                         show the estimated upper limit $\delta^{+}$, expected value $\hat{p}_i$, and lower limit $\delta^{-}$ of the 95\% confidence range.  
                         The underlying PDF is in solid gray.
                        \label{fig:G40_60_big}}
\end{figure}

Application of the de-noising and scaling procedure to individual random trials appears to work.
Improvement is especially pronounced at fairly small $N$ for which the combination of large $\sigma_{p_i}$ and possible bias to $\frac{1}{b_{eff}}$ 
led unsmoothed estimates to be so noisy as to render the true underlying pdf, $g(x)$, unrecognizable.
At small $N$ ($N\approx40$),  there is an increase in $S/N$ from $S/N \approx$1 for unsmoothed estimates to $S/N \approx$3 for smoothed estimates.
The required number of observations for the unsmoothed estimators to match this increased $\langle$(S/N)$\rangle_{MC}$ is about $20\times$-$25\times$ greater than the actual $N$.
Improvement is also significant for fairly large $N$ (Table S1). 

De-noising reduces $|\hat{\delta}^+ - \hat{\delta}^-|$ while usually increasing or maintaining $C$ for the narrower confidence interval.  
The joint estimator $(\hat{p_i})_{\bowtie}$ gives the most consistently high $C$ among \emph{unsmoothed} estimators.
Plots show that the $\hat{\delta}_i$ from \emph{unsmoothed} $(\hat{p_i})_{\bowtie}$ provide high coverage, but at the cost of variable and often large $|\hat{\delta}^+ - \hat{\delta}^-|$.
Smoothing decreases the interval lengths for all tested starting estimators, and this leads to coverage errors for small $p$ at the periphery of the histograms, but it improves $\hat{p}$-to-$p$ matching significantly.
For histograms with $N < 200$, \emph{unsmoothed} $(\hat{p_i})_{\bowtie}$ gives $ S/N < 2$ but good $\xi$-to-$C$ matching, but all smoothed histograms have $S/N \approx 3$ or $S/N > 3$ for $N$ as low as 40 (Table \ref{tab:P0Tables}).

During our prior modeling of drug activity, it had been hoped that any noise in estimated pdfs would cancel upon averaging estimation parameters over many observations.
However in hindsight, from the magnitude of noise observed in these MC trials (Figs. \ref{fig:G40_60_big}-\ref{fig:estHistE}), and from the observed bin-to-bin variance of the estimates found in these trials,
it seems more likely that the noise would not always have canceled for the sample sizes that were actually used.
Excess residual noise and bin-to-bin variance and the resulting concealment of systematic error 
could have contributed substantially to the earlier observed inconsistency for different classification tests when estimating $\sigma_{P_o}$ for the composite Bayes probability of classification.

\subsection{Systematic Error and Bias in the Unsmoothed Estimates}
Looking more closely at average effects observable over large numbers of random trials at a fixed sample size $N$ (the right column of graphs in each of Figs \ref{fig:estHistA}-\ref{fig:estHistE}), 
there was initially a significant systematic displacement of estimates $\hat{p}_i$ to values larger than $p_{i,true}$ before smoothing and baseline correction.  
For unsmoothed $b_{eff}$=2 estimators, the systematic displacement was originally so large that the average difference between $\langle \hat{p}_i \rangle_{MC}$ and $p_{i,true}$ was initially greater than the value of $p_{i,true}$ itself throughout most bins of a histogram compiled for small $N$.
Systematic displacement, which decreases $\langle$(S/N)$\rangle_{MC}$ and indicates lack $\hat{p}$-to-$p$ matching, had not been as large with unsmoothed multinomial Rule-of-Succession (MROS) estimators ($b>2$, Eqs. 1, 3, 7 ) or with the Rgbp joint ($\hat{p}_i$)$_{\bowtie}$ or discrete $\hat{p}(n_i,N)$ estimators.
However, both systematic displacement and $C$ decrease if initial $b_{eff}$=$2$ estimates are rescaled to $\Sum p_i = 1$ \emph{before} smoothing and this rescaling leads the resulting S/N values to be identical to those for the $b_{eff}>2$ MROS estimators.
For smoothed histograms, this systematic displacement is further reduced by the procedures for scaling and baseline adjustment. 
Comparison of the plots at the right side of the figures to individual random histograms at the left was done 
to assure that the observed large reduction in the mean average absolute residual between $p_{i,true}$ and $\langle \widehat{p_{i,smooth}} \rangle_{MC}$ caused by the de-noising procedure 
was not a computational artifact.  

On the one hand, the systematic offset of $\hat{p}_i$ to high values before de-noising at small values for $p_{i,true}$ is understandable in terms of bias toward the limiting value $\lim_{N \to 0} \hat{p} = \frac{1}{2}$ for $\hat{p}$ from different choices of binomial prior at small $N$.
Similar bias toward $\frac{1}{b}$ occurs for Dirichlet priors at larger $b$.
On the other hand, the form of $\hat{p}$ derives directly from a well defined moment integral for an expectation value.  
Estimates derived directly from moment integrals over an underlying pdf are supposed to be unbiased estimators:  
the value of the estimate for the parameter ($\hat{p}_i$) is precisely equal to the expected value for the parameter ($\langle p_i \rangle$) over the range of possible values.
Apparently this statement about bias is only true asymptotically as $N\rightarrow\infty$.  
As $N\rightarrow 0$,
$\hat{p} \rightarrow \frac{1}{2}$ for binomial estimators, $\hat{p} \rightarrow \frac{1}{b}$ for multinomial estimators, and $\hat{p}=\frac{n}{N} \rightarrow \frac{0}{0}$ for the maximum likelihood estimator. 
These limiting values more closely reflect observed bias toward $\frac{1}{2}$, $\frac{1}{b}$, or toward the "undefined" first few random observations (for $\frac{0}{0}$).

At first glance, use of $\mathcal{U}[0,1;x]$ as a Bayes prior seems to weight to any possible estimate $p \in (0,1) $ equally. 
Unfortunately the $\mathcal{U}[0,1;x]$ prior has a fixed mean value of $\frac{1}{2}$ and apparently this biases $\hat{p}$ from this prior toward $\frac{1}{2}$ for small $N$.
Similarly, although the Jeffreys binomial prior $\mathcal{B}[\frac{1}{2},\frac{1}{2};x]$ is weighted symmetrically to accentuate values for $p$ near $0$ or $1$, it too has a mean value of $\frac{1}{2}$ 
and its associated $\hat{p}$ is biased, albeit by "half-as-much", to the value $\frac{1}{2}$ for small $N$.
This bias to $\frac{1}{2}$ before renormalization is probably more readily apparent at large $b$ (for example, 100-bin histograms) since the average $p_{i,true}$ is small.
The bottom line, which should not have been a surprise, is that estimators based on Bayes priors are biased toward the mean value of the presumed priors for small $N$.

Unfortunately, the larger optimized values found for $\alpha_0$ in the $\mathcal{B}[\alpha_0,\alpha_0;x]$ adaptive prior cause even larger bias toward $\hat{p}_i=\frac{1}{2}$.
This is apparent for small values of $p_{i,true}$ in histograms averaged over many trials at fixed $N$.
This bias at small $p_{i,true}$ appears both as an increased systematic difference $(\hat{p}_i - p_{i,true})$ before renormalization 
and also as an increase in the confidence interval width $|\hat{\delta}^+ - \hat{\delta}^-|$ (Fig \ref{fig:intervalLengths}).
At low $p_{i,true}$, $\xi$-to-$C$ matching improves for $\xi\leftrightarrow C$ optimized $\mathcal{B}[\alpha_0,\alpha_0;x]$ prior $\times ~ \mathcal{N}$ posterior relative to that for $\mathcal{U}[0,1;x]$ prior $\times~\mathcal{N}$ posterior.
However, decreased $\xi$-to-$C$ matching for $\mathcal{U}[0,1;x]$ prior $\times~\mathcal{N}$ posterior is due to improved $C$; $C \gg \xi$.  
$C$ is closer to $\xi$ for $\hat{\delta}_i$ from the $\mathcal{B}[\alpha_0,\alpha_0;x]$ prior $\times~\mathcal{N}$ posterior at low $p_{i,true}$ because $C$ is reduced.
Despite the large systematic $\hat{p}_i$-to-$p_i$ offset in the raw estimates from $\xi\leftrightarrow C$ optimized $\mathcal{B}[\alpha_0,\alpha_0;x]$ prior $\times ~ \mathcal{N}$ posterior, estimates $\hat{p}_i$ from priors using $\xi\leftrightarrow C$ optimized values of $\alpha_0$ serve as useful starting points for the de-noising, 
base-line adjustment, scaling procedure to get $\widehat{{p}_{i,smooth}}$.  
Seeing that bias in the initial estimator coincides with a systematic $\hat{p}_i$ to $p_{i,true}$ displacement, 
the key to reducing bias in $\widehat{{p}_{i,smooth}}$ appears to be the subsequent scaling and base-line corrections. 

It has not yet been tested whether the latest direct de-noising and scaling procedures provide similar improvement to initial $\hat{p}_i$ based on $b_{eff}=2$ Jeffreys priors.
The optimal empirical values for $\bar{A_0}$ and $\bar{B_0}$ and the functional form for $\rho(N)$ for these are expected to differ only slightly. 
Iterative optimization, using 10,000 trials for each of several different $N$ and for each of several underlying test density functions, takes about a week per cycle and all of our available computational capacity.
Given the similarity of the accuracy obtained from other diverse starting estimates, significant improvement, while possible, is not anticipated.

\subsection{Remaining Systematic Local $\xi$-to-$C$ Discrepancy in De-noised $\widehat{p_{i,smooth}}$}
A key problem that appears to have been introduced by the de-noising procedure is the occurrence of localized groups of histogram bins toward the edges of the histograms that exhibit high values for the $\xi$-to-$C$ discrepancy.
The coverage value falls toward $0$ for some histogram bins lying toward the edges of the histogram, even for rather large sample sizes.  
This large systematic decrease in $C$ for a few bins near the edges of the histogram might largely be due to a 'data edge effect' inherent to the de-noising procedure.

At least two factors inherent to the de-noising process might contribute to such 'edge effects'.
For one, the de-noising procedure becomes less accurate toward the two edges of each histogram.
At these boundaries, there are fewer segments of measured data available for the multiple least squares line-fits that are used to estimate a value at each point in the de-noised histogram curve. 
Secondly, there is a histogram construction artifact: observations outside the range of the histogram were included in the terminal bins but this was not taken into account during the smoothing.
This small contribution does not rationalize low $C$ seen centrally in histograms in regions near $x$=$0$ at the edges of the functional ranges when $g(x)$ is a highly skewed $\mathcal{B}(\alpha,\beta;x)$ pdf.
As a generalization, it seems that $\xi$-to-$C$ difficulties arise when the smoothing process encounters regions with sparse occurrences.
Regions of residual, systematically low $C$ had been observed elsewhere in similar plots at earlier stages of developing the procedures here,
but these were largely eliminated by improved methods for baseline correction and scaling of initial $\widehat{p_{smooth,0}}$.
Care should be taken to watch for and correct such errors if the procedures here are extended to other histogram sizes or to different smoothing parameters.

Problems with highly skewed pdfs that give rise to adjacent sparse regions might be avoidable.  
If high skew is detected, the measurement scale may transformed to one for which observations are more equitably distributed throughout the histogram.
Aside from this, the largest remaining $\xi$-to-$C$ discrepancies, which tend to occur in or near sparsely filled histogram regions,   
might be of little consequence in many applications that use estimates of pdfs based on histograms, since few observations are affected.
However, it is important to avoid using information from bins for which large $\xi$-to-$C$ discrepancy is expected.


\section{Discussion}

Methods to estimate proportions and their uncertainty from experimental occurrence frequencies have been compared.  
The goal was to understand apparent errors in estimates of uncertainty, $\sigma_{P_o}$ for a statistical classification method to assess drug effectiveness.
Poor estimates of uncertainty for Bayesian composite probabilities $P_o$ suggested inconsistent estimates of error intervals $\hat{\delta}_i$ for component probabilities $\hat{p}_i$.
To improve $\hat{\delta}_i$, methods were first examined to improve agreement between calculated coverage $C$ and target, nominal confidence levels $\xi$.  
This was done by optimizing a single parameter $\alpha_0$ in the Bayes prior pdf used to derive the estimators.
New methods for $\xi$-to-$C$ optimization led to a de-noising method 
that could also be applied directly to experimental histograms to improve estimates $\hat{p}_i$ directly.
Such de-noised histograms $\widehat{p_{i,smooth}}$ led to more accurate estimates of known underlying pdfs
than any of the earlier multinomial estimates $\hat{p}_i$.
On re-examining $\hat{p}_i$-to-$p$ matching in addition to $\xi$-to-$C$ discrepancy as a measure of reliability in this multinomial case, 
it became understood that both of these criteria for reliability are important and that the initially tested un-smoothed multinomial Rule-of-Succession (ROS) estimators, 
while apparently poor in some tests by $\xi$-to-$C$ matching, were equivalent to rescaled unsmoothed binomial ROS and similar to unsmoothed optimized estimators ($\hat{p}_i$)$_{\xi\leftrightarrow C}$ by the $\hat{p}_i$-to-$p$ criterion.
Aside from $\hat{p}_i$-to-$p$ and $\xi$-to-$C$ matching, run-to-run consistency of estimates $\hat{\sigma}_{p_i}$, and thus of $\hat{\delta}_i$, affected the accuracy of earlier Bayes classification trials.
A factor, $\rho_{0.99}$, was introduced to correct for expected run-to-run variation in estimates $\hat{\sigma}_{p_i}$ that would be expected to cause error in estimates from a single histogram.
Smoothed (de-noised) histogram estimators were generally much better by all criteria regardless of the choice of initial estimators $\hat{p}_{i,0}$ used to construct the unsmoothed histogram.
However, when tested by MC simulation, a large increase in the $\xi$-to-$C$ discrepancy was often found for a few of the bins toward the edges of de-noised histograms.

The choice of optimal starting estimate $\hat{p}_{i,0}$ for de-noising is still somewhat unsettled. 
Using joint estimator ($\hat{p}_i$)$_{\bowtie}$ as an \emph{unsmoothed} estimator led to maximal coverage for $\hat{p}_{i,0}$, 
but using ($\hat{p}_i$)$_{\bowtie}$ as a starting point for smoothing led to single-histogram estimates $\hat{\sigma}_{p_i}$ for $\widehat{p_{i,smooth}}$ with slightly lower run-to-run consistency, 
requiring slightly larger than average up-scaling corrections by $\rho_{0.99}$ to account for this variation.
Comparatively, the size of $\rho_{0.99}$ corrections to $\hat{\sigma}_{p_i}$ for $\widehat{p_{i,smooth}}$ from the multinomial generalized ROS estimator or from the discrete estimator are slightly smaller, 
but those from the discrete estimator using the combinatorial prior of Appendix B appear to be less consistent from $N$ to $N$.
Corrections to $\sigma_{p_i}$ for ($\hat{p}_i$)$_{(\xi\leftrightarrow C)}$ appear to be the smallest. 
However, all differences between corrections to $\hat{\sigma}_{p_i}$ for $\widehat{p_{i,smooth}}$ from different starting estimators are fairly small. 
One remaining, unaddressed point of concern is that 
the similar final accuracy of de-noised estimates, despite varying accuracy in the initial unsmoothed estimates,
might result from unintentional $\hat{p}_{i,0}$-dependent bias of the post-smoothing baseline and scaling procedures.  

Despite $\xi$-to-$C$ inconsistencies toward the edges, the de-noised histograms for multinomial estimates ($\widehat{p_{i,smooth}}$) turned out to be practically useful.
Extensive tests by application to Bayes Classification, have so far only been done for $\widehat{p_{i,smooth}}$ and $\widehat{\delta_{i,smooth}}$ derived from ($\hat{p}_i$)$_{(\xi\leftrightarrow C)}$ 
(numerically optimized $\xi$-to-$C$ matching, $\mathcal{N}$ posterior) as the starting Bayes estimator and with the original base-line correction and scaling procedure.
Further improvement is suggested by the comparative results for updated procedures examined here.


When Bayes Classification of drug sensitivity was examined using the earlier form for $\widehat{p_{i,smooth}}$ from optimized estimators ($\hat{p}_i$)$_{\xi\leftrightarrow C}$ together with the corresponding up-scaled $\hat{\sigma}_{p_i}$, 
the reliability of the uncertainty estimates improved relative to unsmoothed estimators \citep{friedmanBayesICA}.
This was particularly apparent in cases for which limited experimental information was available for assessing drug activity.
For some difficult classifications, before improvements by using de-noised elementary $\hat{p}_i$, 
several \emph{false} predictions were still found to have composite $P_o$ lying $4~\hat{\sigma}_{P_o}$ to $7~\hat{\sigma}_{P_o}$ from the decision cut-off point, suggesting a falsely high level of certainty.
There was no clear correlation between the number of standard deviations from the decision cut-off point and the accuracy of the prediction.
With the improved estimates of elementary probabilities and uncertainties, more predictions for the difficult known test cases fell close enough to the decision cut-off (within $2~\sigma_{P_o}$) to signal that the outcome was less certain.  
The few remaining erroneous predictions were only slightly beyond $2~\sigma_{P_o}$ from the decision cut-off point and most predictions beyond $2~\sigma_{P_o}$ were correct \citep{friedmanBayesICA}.\footnote{
Values of $\sigma_{P_o}$ were increased (Z decreased) to the equivalent "infinite sample limit" to account for varying numbers of samples in each set.
Each sample being classified had independent cell-by-cell measurements of the same feature for each cell in the sample; 
this allowed construction of a distribution \emph{over different cells from within each sample} and a sample-to-sample comparison of these single-sample distributions.
Classification is more difficult using distributions \emph{over different samples} of single observations from each sample,
with no information about the spread of measured values within each individual sample. 
}

More critically, however, accurately defined histogram shapes allowed one to tailor separate basis sets for each test set subsample.  
A combination of basis (learning) set subsamples can be found that causes 
the joint distributions for this combination 
to approach the observed underlying distributions in each individual test set subsample.  
Each combination of basis set subsamples is found without regard to basis-set subsample classification, only using pdf shape.
The tailored basis sets from this procedure led to the above described accuracy for the statistical classification of drug sensitivity for each test set sample.
Based on the noise and systematic error in estimated histograms (Figs \ref{fig:estHistA}-\ref{fig:estHistE}), without de-noising, 
such matching of the distribution shape of each test set sample to that of a combination of basis set subsamples is unlikely to have provided meaningful results.

The insensitivity of Bayes classification for real experimental data to the large $\xi$-to-$C$ discrepancies found for a few histogram bins in the MC tests 
is likely because there were few or no observations sorted into these bins from either the test set or basis set in the real data for which the underlying distributions were unknown.
%
The earlier, apparently unreliable estimates $\hat{\sigma}_{P_o}$ for predictions of drug effectiveness stemmed from unreliable estimates for both $p_i$ and $\sigma_{p_i}$ due to discrete sampling noise, from the overly wide ranges for confidence intervals for unsmoothed $p_i$, 
and from systematic offsets due to the inherent bias of $\langle p_i \rangle$ to $\frac{1}{b}$ when using the generalized multinomial ROS or to $\frac{1}{2}$ when using either the Jeffreys or Uniform prior pdf with $b_{eff}=2$.
A further large contribution stemmed from an inappropriate presumption of the identity of the underlying distribution of measured values between the learning set and the test set,
based solely on the apparent similarity of experimental measurement conditions.
Bayes classification presumes IID observed values.
Whereas experimental values had been adjusted to be independent, prior to basis set tailoring, they had only been presumed, but not specifically adjusted, to be identically distributed.

Whereas improvements from de-noising and scaling are significant in estimates $\hat{p}_i$ from large multi-bin histograms, 
it is difficult to imagine a way to apply these methods to the binomial case without requiring a large number of additional measurements.
On the other hand, one may imagine using MC calculations to derive corrections to single run parametric estimates $\hat{\sigma}_p$ based on Eq. 2 or 4 to account for expected run-to-run variation in these estimates.
However, in this simpler binomial case, it may be possible ultimately to derive exact expressions to account for this run-to-run variation in $\hat{\sigma}_p$ for each estimator type.

Barring such further developments, for this binomial case, 
if use of $\mathcal{N}[\mu_p, \sigma^2_p; x]$ as posterior pdf is desired, and if $p$ values of interest are far from 0 or 1, then it makes sense to use this in combination with a $\mathcal{U}[0,1;x] = \mathcal{B}[1,1; x]$, Uniform Bayes prior pdf (Eq. 1 \& 2, $b=2$).
Alternatively, if use of the $\mathcal{B}[n+\alpha_0,N-n+\alpha_0;x]$ Bayes \emph{posterior} pdf is desired, then the choice is less clear.  
If initial estimates indicate that the $p_i$ value interest is close to 0 or 1, then the  $\mathcal{B}[\frac{1}{2},\frac{1}{2}; x]$ Jeffreys Bayes prior ($\alpha_0=\frac{1}{2}$) yields more reliable error estimates.
Otherwise, the Uniform prior ($\alpha_0=1$) yields more reliable error estimates.
Either the combination of a Jeffreys Bayes \emph{prior} pdf with a $\mathcal{B}[n+\alpha_0,N-n+\alpha_0;x]$ Bayes \emph{posterior} pdf
or the combination of a Uniform Bayes \emph{prior} pdf with a  $\mathcal{N}[\mu_p, \sigma^2_p; x]$ \emph{posterior} pdf both yield confidence intervals with short nearly equal interval lengths.
The combination of the Uniform Bayes \emph{prior} with the $\mathcal{B}[n+\alpha_0,N-n+\alpha_0;x]$ Bayes \emph{posterior} has slightly broader confidence intervals for smaller values of $p_{i,true}$ (Fig \ref{fig:intervalLengths}).
If more certainty about the coverage of the error estimate is required, 
the adaptive \emph{priors} $\mathcal{B}[\alpha_0,\alpha_0; x]$ \emph{optimized} for $\xi$-to-$C$ matching offer an alternative at the cost of much further increased interval lengths for small values for $p_{i,true}$ for some $N$.
The combination of the discrete prior and posterior gives results similar to the earlier estimators for which priors and posteriors presumed continuity.  
While avoiding presumptions about continuity for the Bayes posterior and avoiding probability density for the posterior lying outside the admissible range $(0,1)$,
the derivation of this discrete prior 
still presumes that the absence of \emph{observable} knowledge about the existence of continuity for an underlying pdf requires the most accurate description for this pdf to lack continuity.  
(That is, it presumes that estimator outcomes need to be "quantized" by the discrete number of observations used to estimate an underlying pdf.)  

All elementary $\mathcal{B}$ pdf priors leading to necessarily non-zero valued best estimates for \emph{binomial} $p$ are biased toward $p=0.5$, but this bias increases as the effective value for $\alpha_0$ increases in the prior.
In the \emph{multinomial} case, similar, stronger bias toward $p_i=\frac{1}{b}$ and inconsistency between $C$ and $\xi$ had been detected for non-informative Dirichlet priors, leading us initially to abandon their use.
Considering how initial bias manifests itself for the Jeffereys and Uniform, priors, the earlier recognized asymptotic estimators, such as those of Wilson and Wald can also be considered biased.
For these estimators, the value $\alpha_0$ is effectively $0$, and the limiting estimate $ \displaystyle\lim_{N \to  0} \hat{p}$ is the undefined quantity $\frac{0}{0}$.
Such estimators are 'randomly' biased toward the "undefined results" of the first few random observations.  
If the first random occurrences happen to be in the "wrong" direction, $\langle p \rangle < p_{i,true} -Z~\sigma$ or $\langle p \rangle > p_{i,true} +Z~\sigma$, 
in the direction away from $p_{i,true}$ (\emph{e.g.} 3 or 4 heads in a row when flipping a fair coin),
it takes more trials (larger $N$) for this earlier class of estimator to recover to provide estimates close enough to $p_{i,true}$ to get reliable $\xi$-to-$C$ matching.

In spite of these general difficulties for the elementary Bayes priors for either the binomial or multinomial cases, 
the procedures presented here for de-noising and scaling \emph{multinomial} estimates appear to offer a way to reduce the inherent bias of unsmoothed elementary estimators for the multinomial case.  
Reduced bias from de-noising stems from using joint information about $\hat{p}_i$ available from the measured occupancies of histogram bins in the vicinity of bin $i$.
Reduced bias is possible here largely due to an invariant and relatively smooth form for the underlying pdf being estimated by the histogram.
 This presumption of smoothness of the \emph{underlying} pdf seems to contradict earlier arguments about the lack of smoothness of the \emph{observable} pdf that warranted the introduction of discrete estimators.
 However, such smoothness of the \emph{underlying} pdf more accurately reflects the actual underlying process by which the analyzed random trial distributions were generated here.
The \emph{observable} pdf is different than and can at best only approach the unknown \emph{underlying} pdf.
Run-to-run variation in the \emph{observable} pdf at fixed $N$ forbids discernment between closely related forms for a smooth unknown \emph{underlying} pdf.

Improvement in the estimates $\widehat{p_{i,smooth}}$ and $\widehat{\sigma_{i,smooth}}$ is fairly general for all examined underlying $g(x)$ and for all choices of initial estimators used to construct the experimental histogram,
offering a significant reduction in required sample size to achieve a fixed $\langle S/N \rangle$ level.  
The reduction in required sample size by nearly an order of magnitude
suggests further application of $\widehat{p_{i,smooth}}$ and $\widehat{\sigma_{i,smooth}}$ to estimating sample means and medians, downweighting of outliers, and to the idealization of raw measurements by adjustment to the nearest smooth distribution.
These extended applications are currently being tested and preliminary routines are included with the accompanying source code.

\newpage


\newpage

\appendix
\section*{Appendix A: De-Noising by Assuming Approximate Local Linearity}
\label{app:de-noising}
De-noising of the $\alpha_0(N)$ curve (Fig \ref{fig:alphaVsNhorrid}), obtained when $\alpha_0$ for the $\mathcal{B}[\alpha_0,\alpha_0;x]$ prior $\times$beta-pdf Bayes posterior combination was optimized for $C$-to-$\xi$ matching,
was expected to improve the estimates of $\alpha_0$ at each $N$.
The absence of a readily apparent functional form for this curve
led us to consider de-noising this curve by simple linear least squares approximations about contiguous zones of data points (ranges of values for $N$ about some point $n_i$ on the curve).  
Such piecemeal linear least squares fits could be carried out on different $L$-point data segments of fixed arbitrary datalength $L$ as the portion of data being considered $d: (i-[L/2],~\dots,~i+[L/2])$ moved along ordered values $n_i$.
Smoothing would occur by approximating values for $\alpha_0$ at points within each segment by the corresponding points on the least squares best fit line. 

It was realized, though, that the quality of the approximation to the optimized $\alpha_0$ function for a given value of $n_i$ would vary depending on the position of $n_i$ in the local data segment, $d$, being fit to the line.
For instance, for if $\alpha_0$ were a sawtooth function, the approximation to $\alpha_0$ for the point $n_i$ at the cusp of the sawtooth would be quite poor if $n_i$ were at the center of the local data segment used for the linear fit, 
but it would be rather good if the point $n_i$ were at the end of either of the two data segments that terminate at the cusp.  
Thus, rather than using the approximation for $n_i$ only when it was the central point in each data segment, $d$, being fit to a line,
it was considered worthwhile to allow every possible data segment of datalength $L$ passing through $n_i$ to contribute to a final de-noised, \emph{weighted-average} approximation for that point.

The weight in such an average may be based on a 'goodness-of-fit' value, $Q_{n_i\leftrightarrow n_j | d}$, for point $n_i$ relative to the goodness-of-fit values for all other points in the same local data segment $d$.
To define a relative goodness-of-fit value for a particular point $n_i$ on a particular $d$, one examines the deviation between the point's observed and approximated values relative to the analogous deviation found for all other points in $d$.
From the estimated value for the standard deviation, $\sigma_d$, over all points for this particular data segment, $d$, 
one may use $Q_{n_i\leftrightarrow n_j | d}\equiv\frac{1}{\sigma_d}e^\frac{-(\alpha_{obs,i,d}-\alpha_{approx,i,d})^2}{2{\sigma^2}_d}$ as the weight on the interpolated value for point $i$ based on the linear fit to that particular segment.  
This constitutes a point-to-point weight for the relative fit within a single choice for local data segment $d$.

The quality of the overall linear fit to \emph{all} points in a single choice for $d$ that includes the point $n_i$ may also vary relative to the quality of the overall fit for other choices for $d$ through $n_i$.
Thus, additional segment-to-segment weights, $Q_{d\leftrightarrow d'}$, based on the comparison between different choices of $d$ can also be applied.
Such segment-to-segment weights may be based on the probability of whether the observed value for $\sigma_d$ for the overall fit for a given segment is better than or worse than the average value ${\sigma}_{ave}$ of the overall $\sigma_d$ values found for all data segments considered.
If the values for $\sigma_d$ can be taken to be approximately normally distributed about a mean value for a given collection of considered line segments, such a weight can be taken to be of the form:
$Q_{d\leftrightarrow d'}\equiv\frac{1}{2}\left( 1- erf(\frac{\sigma_{obs,d}-\sigma_{ave}}{\sqrt{2} \sigma_{rms} }) \right)$ where $\sigma_{rms}$ is the observed standard deviation of $\sigma_d$ over all considered line segments $d$.
Thus this second weight for the weighted average of each approximation is a measure of how good the fit of each local data segment is relative to the 'average' fit of a data segment for a given set of data.  
Since the final weighted average to determine a de-noised estimate at each point is taken individually at that point, the weights on the results for each data segment at each point are only relative to weights from the other data segments that include that particular point.  
However, the values of the overall mean ($\sigma_{ave}$) and standard deviation ($\sigma_{rms}$) of $\sigma_d$ used as benchmarks for comparison may be chosen to depend either on the whole data set or on the local set of data segments.

Since the values used for each $\sigma_d$ are normalized on a per-point basis in each considered data segment, 
it is also possible to use similar weighting to extend the least squares analyses to include data segments of different datalengths $L$.
The use of such weighting would at first appear to obviate the need for careful selection of which $L$ values to consider, however for larger values of $L$,
there are more line segments passing through $n_i$ that contribute to the approximation for that particular point, skewing the final approximation for $\alpha_0$ to estimates based on larger $L$.  
Whereas the fit is expected to be poorer for the longer segments, leading to down-weighting for approximations from these data segments, the increased number of such segments of longer $L$ affects the values of overall $\sigma_{ave}$ and $\sigma_{rms}$. 
These values of $\sigma$ are the benchmarks for relative goodness-of-fit, 
and so the net effect of the increased 'sampling' at each $n_i$ due to larger $L$ values is to increase $\sigma_{rms}$ and thus to increase the contribution to the final overall weight from longer local data segments.
This bias toward the longer data segments is further accentuated if the standard normal probability density and $erf$ functions used to calculate the weights are replaced by Student-equivalents to account for the discrete number of sampled values for each line segment. 
However, the Student-equivalents of these weighting functions are advantageous since they render the final approximated $\alpha_0(N)$ functions to be less sensitive to the choice of the smallest $L$ value to use for local data segments.  
Adjusting final weights by $1/L$ to account for differences in segment lengths leads to a much noisier final estimate (weighted average),
implying that the resulting weighted average is skewed too much toward the estimates from segments of shorter datalength.
Empirically, for the cases examined, $1/\sqrt{L}$ is found to be a more appropriate adjustment.
Possible ways are being considered for making the smoothed estimate less dependent on the choice of upper limit for $L$, but so far improvements are not as clear-cut, so these have not yet been implemented.

Aside from the weights based on (i) an individual point's goodness-of-fit relative to that of other points in the same data-segment ($Q_{n_i\leftrightarrow n_j | d}$), and (ii) an individual data-segment's goodness of fit relative to that of other data segments passing through $n_i$ ($Q_{d\leftrightarrow d'}$), 
a third weight was also found to be useful.
This weight, $Q_{x(1-x)}$, is based on a point's position within the data segment.  
As noted for the sawtooth function, it is often useful for the point being approximated to be at the terminus of the data segment used for linear approximation.
Each data segment has only two such positions. 
The sampling bias due to the relative excess of interior points over end points can lead to the truncation of sharp features, as one might expect from such overweighting of interior points.  
To compensate for such overweighting of the interior points, one may add an extra weight that increases for points toward the termini of the data segment.  

The shape of the binomial Jeffreys Bayes prior pdf $\mathcal{B}[\frac{1}{2},\frac{1}{2};x]$, which is actually a function of the form $\frac{C}{x^\frac{1}{2}(1-x)^\frac{1}{2}}$ can serve as an inspiration for such a weight,
 if one considers the value $x$ in this function to be a measure of the fractional position of the point in question along the data segment under consideration.  
Difficulties that arise when $x$ has a value of $0$ or $1$ in such a function suggest that those values should be avoided.
For a data segment consisting of $L$ points, one therefore sets the fractional position used for $x$ in the weight to $\frac{i}{L+1}$ where $i$ is the integer order position (from $1$ to $L$) of the point along the data segment.
The resulting weight for allowing adequate representation as $n_i$ approaches the terminus of the data segment would then be $Q_{x(1-x)}\equiv\frac{1}{\left(x(1-x)\right)^{W_T}}$, where $W_T$ is a suitably chosen exponent.
In practice, values of $W_T \gg 3$ cause essentially only segments with $n_i$ at the termini to contribute to the final weighted average of least squares approximations for $\alpha_0(n_i)$.  Values of $W_T$ between $\frac{1}{2}$ and $2$ appear to be useful.  
If the chosen value of $W_T$ is too small, sharp features are truncated.  If it is too large, and if fairly long data-segment lengths are included, the sharp features can be overshot, leading to excessive spikiness at such regions.

The computation of least squares fits is done most efficiently one data segment-length $L$ at a time. 
To apply the de-noising to fits with variable segment lengths, it was found convenient to perform the de-noising in two passes, 
the first pass to obtain values of $\sigma_{ave}$ and $\sigma_{rms}$ over all segment lengths and the second pass to calculate the weighted average using these values in the overall weights.

\renewcommand{\thefigure}{A1}
\begin{figure}
\centering
\begin{subfigure}[h]{0.45\textwidth}
\centering
\includegraphics[width=\textwidth, clip=true, trim= 0cm 0cm 0cm 0cm ]{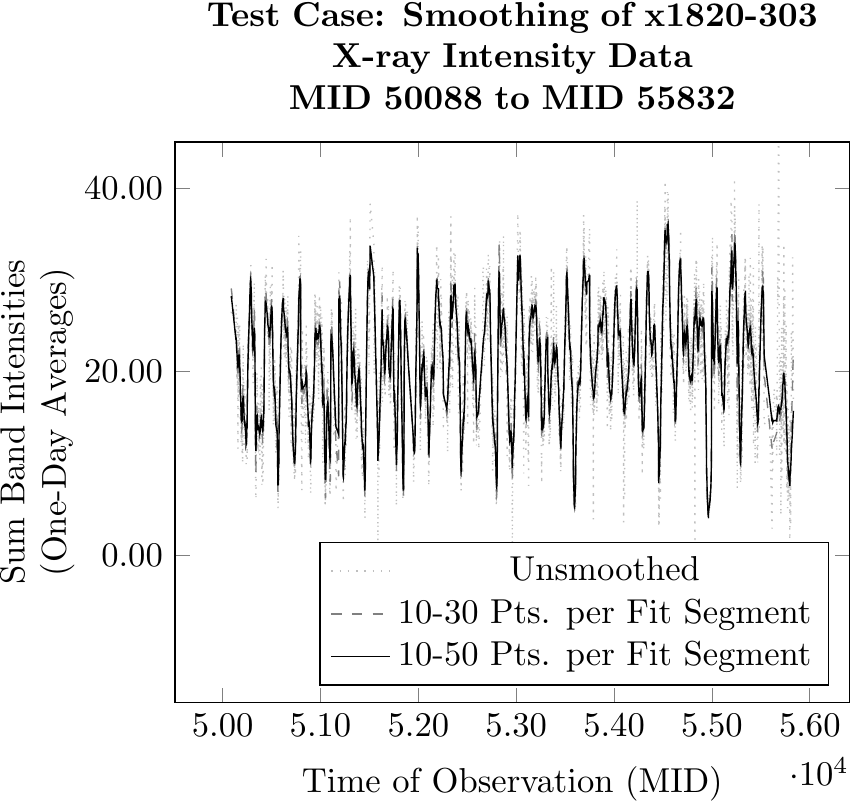}
\end{subfigure}
\begin{subfigure}[h]{0.41\textwidth}
\centering
\includegraphics[width=\textwidth, clip=true, trim= 0cm 0cm 0cm 0cm ]{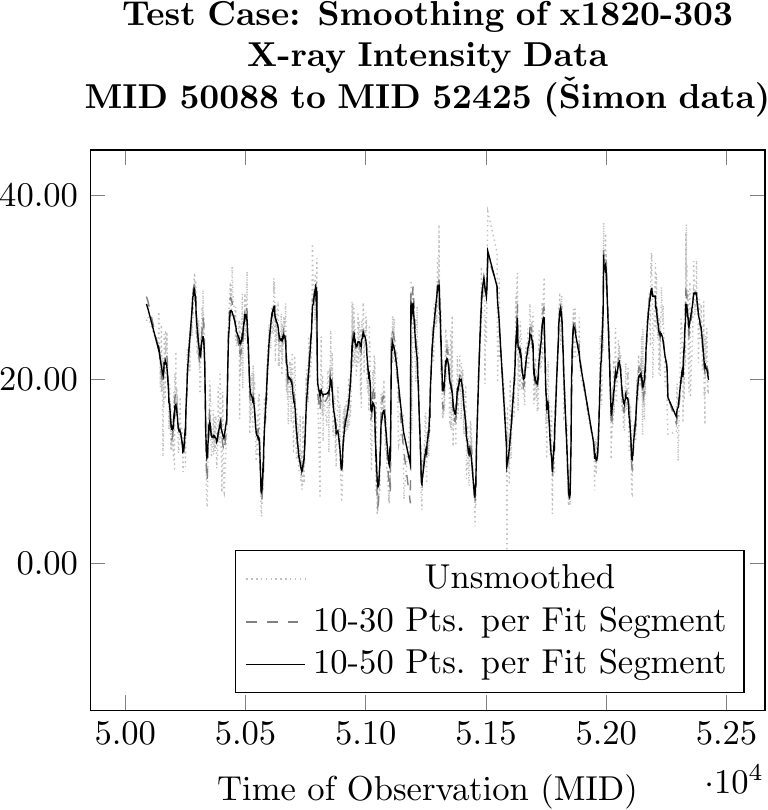}
\end{subfigure}
                \caption{ Tests of multi-range curve de-noising (smoothing) using data from astronomy. 
                          To account for differing segment lengths in the weighted averages used for each smoothed point, calculations of segment-to-segment weights 
                          used equivalents of the standard normal distribution and $erf$ functions derived using a Student-\textbf{t} probability distribution instead of Gaussian exponential functions.  
                          Similar curves could be obtained though by a judicious (albeit arbitrary) choice of cut-off values for the lower number of observation points per group.
                          Ranges of the local line fits shown here include all adjacent groups either from 10 to 30 observation points per group or from 10 to 50 observation points per group.\label{fig:astro}}
\end{figure}
\renewcommand{\thefigure}{A2}
\begin{figure}
\centering
\begin{subfigure}[h]{0.53\textwidth}
\centering
\includegraphics[width=\textwidth, clip=true, trim= 0cm 0cm 0cm 0cm ]{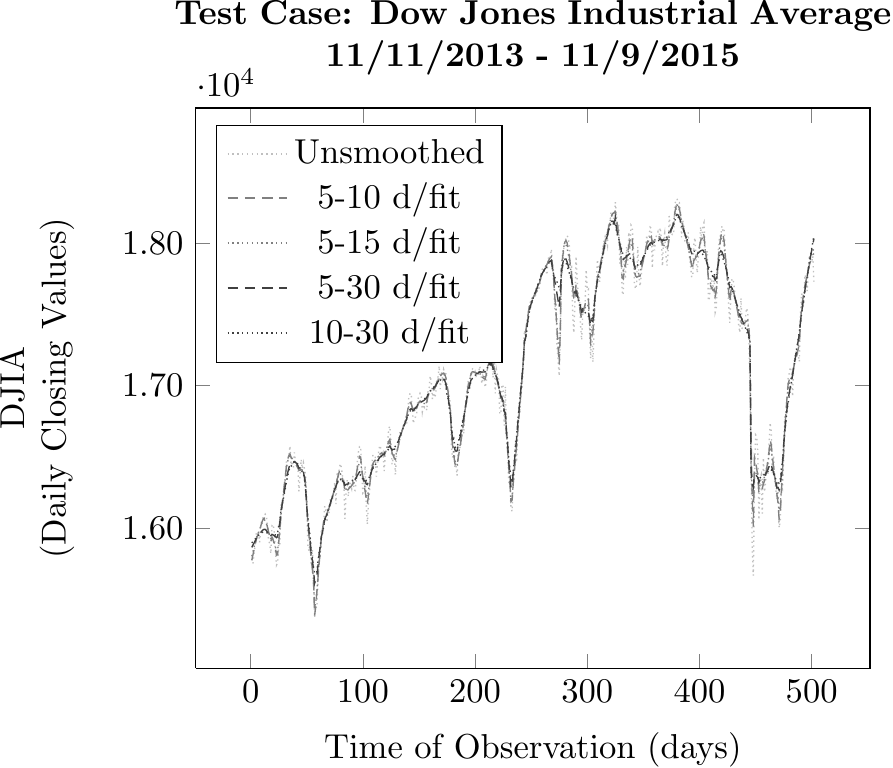}
\end{subfigure}
                \caption{ Tests of multi-range curve de-noising (smoothing) for daily closing values Dow Jones Industrial Average. 
                          Again the effect of changing the lengths of the longest and shortest data segments (number of days per fit) is compared.
                          The effect of the short end length is mitigated by using $N$ dependent Student distribution equivalents of probabilities for the weights used in weighted averages for point-by-point estimated (smoothed) values.
                          In the current version of smoothing, the degree of fit is largely determined by an arbitrarily chosen, longest segment length, 
                          but fixes are being considered.\label{fig:djia}
                          }
\end{figure}
On doing a rudimentary search of the literature to find the origin of this method, it was found that 'moving' least squares is, in fact well known in a singly moving form \citep{levin98,boseAhuja,fleishman}.
The position of the point being approximated in these earlier descriptions appears categorically to be at the center of the data segment under consideration.
In the past, the cusp problem appears to have been attacked by increasing the order of the polynomial being fit to greater than linear order.  
Thus a commonly used local regression analysis, "loess" \citep{loess} is also based on estimating values for the central abscissa point in each local zone, 
weighting each neighboring point in a \emph{single polynomial} regression analysis by the distance along the abscissa to this central point 
rather than by using comparative goodness of fit for \emph{multiple linear} regression analyses.
For this reason we suggest the name GOOFY-loess (GOodness Of Fit Yclept-loess) for the procedure introduced here.
Earlier de-noising methods do not appear to be as effective at avoiding the truncation of sharp features, most likely because such sharp features are not well modeled as low order polynomials.
The use of Gaussian weights that were somewhat related to the comparative goodness-of-fit was also suggested earlier \citep{levin98}, 
although as far as could be discerned use of the average value $\sigma_{ave}$ and the overall $\sigma_{rms}$ as the basis for the comparison was not.

As mentioned, the level of smoothing in the method described here can be adjusted by altering the lengths of the shortest and longest data segments ($L_{min}$,$L_{max}$) being considered as least squares zones for linear fitting.
However the $L_{min}$ is less critical when Student-equivalents are used and possible new weights that may decrease the influence of the choice of $L_{max}$ are being tested.
The fidelity of the de-noised approximation near sharp peaks and troughs can be adjusted by altering the weight exponent $W_T$ used for the positional weighting.
Since this de-noising procedure maintains a rather high degree of fidelity, it may be performed several times in succession without introducing obvious distortions.
In initial test cases with noisy data, when the limiting segment lengths and positional weighting parameters were well chosen, distortions that might be considered serious did not begin to appear until the fourth or fifth cycle of de-noising.
However, distortions occur at earlier cycles when smoothing is applied to cases for which S/E is already fairly high before smoothing.
For 100-bin histogram estimates of underlying general pdfs, S/E continued to increase with up to 3 cycles of smoothing for 40$<N<$400 ($N$, the total number of observations sorted into the histogram),
up to 2 cycles for 400$<N<$4000, and with up to 1 cycle 4000$<N$ up to the highest examined value $N=12800$.  
For the sawtooth pdf, the examined underlying trial pdf that is predominantly linear, S/E continued to increase for further cycles of smoothing beyond these limits.

Gnu-fortran source code that implements this curve de-noising is included in the supplemental material.
Test data from astronomy \citep{simonAstro,website:ASMlightCurves}, an area where such curve smoothing is commonly invoked, provided a general feel of the effects of changing the de-noising parameters.
These extraneous data were found to be useful for the development and application of the source code.
Figure \ref{fig:astro} shows the results of smoothing the astronomic data for x1820-303 
(one day averaged light curves of sum band intensities for x1820-303 between MID 50088 and MID 52425; 
apparently the same data that were originally analyzed in \citep{simonAstro} and called 4U 1820-30) 
and of smoothing an extended data set for x1820-303 that includes some more recently measured time points.
This Figure and Figure \ref{fig:djia} both illustrate the effects of varying $L_{min}$ and $L_{max}$.

\section*{Appendix B: Binomial and Multinomial Estimators $\hat{p}(n,N)$ and $\hat{\delta}(n,N)$ that Presume a Discrete Set of Estimation Outcomes}
\label{app:DiscreteEstimator}

\subsection {Some Discrete Posteriors and their Priors for the Binomial Case}

To estimate $p$ from an $N$-observation binomial trial with a given \emph{integer} number of successes $n \in \mathbb{Z}$, the conditional probability of assigning the estimator $\hat{p}$ a particular value $\theta$ is zero, $P(\hat{p}$$\leftarrow$$\theta|{n}) = 0$, for all but a discrete number of admissible values,
 $\theta_j \in \Theta_N = \{\theta_0, \theta_1, \dots \theta_N\}$.  
A discrete \emph{posterior} is defined by mapping $p$$\equiv$$p_{true}$ onto $\Theta_N$, which is equivalent to assigning the value of an element from $\Theta_N$ to the estimator $\hat{p}$:
$P(p$$\rightarrow$$\Theta_N|n)$=$P(\hat{p}$$\leftarrow$$\theta^*|n)$; $\theta^*$ can be chosen as $\theta^*(p)= \argmin_{\theta_j} \{|p-\theta_j|\}$.
That is, mapping $\hat{p}\leftarrow\theta^*$ can be defined as selecting admissible value $\theta_j$ closest to the unknown "true" underlying value $p$ ($\theta^*: p \rightarrow \Theta_N$).
The discrete Bayes \emph{prior} $\bar{\pi}$ 
is the set $\Theta_N$ of admissible values $\theta_j$ (domain of $\bar{\pi}$) paired element-by-element with a set of \emph{prior} probabilities $\bar{\pi}_j$ (weights, range of $\bar{\pi}$) 
assigned to each $\theta_j$.

The area bounded by the discrete posterior $P(p$$\rightarrow$$\Theta_N|n)$ is a union of varied, adjacent, non-overlapping rectangular regions each with a $p$ coordinate (centered) about a different $\theta_j$, but with the regions truncated at $p$$=$$0$ or $p$$=$$1$ for $j$$=$$0$ or $j$$=$$N$.
The \emph{starting} $p$-value for the rectangular region for which $\theta^*$$=\hspace{2pt}$$\theta_0$ is at $p=\phi_0\equiv 0$.
When $j$$\neq$$ 0$, $\theta^*$$=\hspace{2pt}$$\theta_j$, the \emph{starting} $p$-value is $p=\phi_j\equiv\frac{1}{2}(\theta_{j-1}+\theta_j)$, the half-way point between defining values $\theta_{j-1}$ and $\theta_j$. 
If starting point $\phi_{N+1}$ is assigned the value $1$, then the width of each rectangular region in the posterior is $W_j=\phi_{j+1}-\phi_{j}$.
Weighting by this width gives the \emph{a priori} probability of assigning each admissible value $\theta_j$ to $\hat{p}$: 
$\bar{\pi}_j=W_j \hspace{2pt} \bar{\pi}_{\diamond}(j, W_j, \theta_j)$, 
a component of the discrete prior $\bar{\pi}$.
For the posterior $P(\hat{p}$$\leftarrow$$\theta^*$$=$$\theta_j|X$$=$$n)$, the height of the rectangular region associated with admissible value $\theta_j$, is proportional to $\binom{N}{n}\hspace{3pt}\bar{\pi}_{\diamond}()\hspace{2pt}{\theta_j}^n (1-\theta_j)^{N-n}$
but is normalized so that the total area of all rectangles for each $\theta_j$ at outcome $n$ sums to 1. 
The binomial term $\binom{N}{n}$ is constant for all rectangles at fixed $n$ and divides out in the normalization.

This posterior for a given outcome $n$ defines equal-tailed confidence intervals $\delta_{|n} = (\delta_{|n}^-,\delta_{|n}^+)$ as the limiting values (Eq. 6) from either edge, $p=0$ or $p=1$, up to which the sum over rectangular areas, full or partial, matches $\frac{1-\xi}{2}$.
Confidence interval $\delta_{|n}$ about the particular admissible value $\theta_{j=n}$ depends on outcome value $n$ and is distinct from the "$p$-success" region for $\theta_j$, $\Phi_j = ( \phi_j, \phi_{j+1})$, for which $p \in \Phi_j$ is closest to $\theta_j$.
Values of $p$ in adjacent confidence intervals $\delta_{|n}$ and $\delta_{|n+1}$ depend on $\xi$ and generally overlap, but $\Phi_j \cap \Phi_{j+1} = \emptyset$.
Admissible values $\theta_j$ do not vary with outcome $n$. 

Conditional expectation values $\langle \theta | n \rangle$ may be defined as posterior weighted averages of admissible values $\theta_j$.
Iterative updates for admissible values of the form $\theta^{'}_{j=n}\leftarrow \langle \theta | n \rangle $ usually do \emph{not} provide self-consistent $\theta_j$ values.
Such iterative values for $\theta_j$, $\delta_{|n}$, and $W_j$ are unstable, with many $\theta_j \rightarrow 0.5$ and $W_j \rightarrow 0$ for $j \approx N/2$.
Instead, a self-consistent value for each $\theta_{j=n}$ that is stable to iteration can be obtained from the \emph{median} of the $n$-th confidence interval $\theta^{'}_{j=n} \leftarrow \tilde{\theta}_{n} = \frac{1}{2}(\delta_{|n}^++\delta_{|n}^-)$ defined at each $n$ by the provisional posterior.
On iterative updates of $\theta_{j=n}$ to $\tilde{\theta}_{n}$ at each $n$, all $\theta_j$ and interval limits $\delta_{|n}$ reach fixed values. 
The centers of the rectangular zones defined by the slightly off-center iteratively stable $\tilde{\theta}_{n}$ appear better to reflect underlying $p$ and may be used as $\theta_n$.
\emph{For $\theta_j$ defined this way, tests confirm that $C \approx \xi$, 
with $|C-\xi|$ and $\langle (\hat{p}-p)^2\rangle$ comparable to or often smaller than from presuming continuity of $\theta$.}
Self-consistent values for $\theta_{j=n}$ depend on $\delta_{|n}$, so the list of self-consistent admissible values $\theta_j$, comprising the range for the estimator $\hat{p}(n,N)$, depends on $\xi$ as well as on $\bar{\pi}_{\diamond}(j,W_j,\theta_j)$.
Supplemental material includes computer code to obtain and test self-consistent $\Theta_N$ sets.

\subsection {Modified Estimates for $p_i$ for Multinomial Histograms}

The critical argument concerning the need for discrete estimators is about whether one is justified to presume a higher level of precision in estimates $\hat{p}(n,N)$ than might ever
be discerned in a counting experiment that entails only $N$ total observations.
In experiments for which observations of instances are counted to assess a value for $p$ or a confidence range $\delta$, 
the underlying value for $p_{true}$ might ultimately be determinable to fairly high precision if $N$ is allowed to be large enough, 
and if $p_{true}$ is sufficiently invariant over the course of the measurement.
However, given a \emph{limited, fixed} integer value for the total number of observations $N$, such precision can never be achieved; 
only a fixed number of estimated values $\hat{p}$, the set of admissible values $\Theta_N \in \{\theta_0 \dots \theta_N\}$, is possible.
The values available to $p_{true}$ might be continuous, but due to experimental limitations, the values available to the estimator $\hat{p}$ cannot be continuous.
This quandary about continuity of $\hat{p}$ is similar to the one about how precisely one can determine positions for sub-atomic particles.
Unlike the precise position of say an electron, 
a more precise estimate $\hat{p}$ for $p_{true}$ might ultimately be found if $N$ is allowed to increase,
because $p_{true}$ might remain sufficiently constant over the course of the measurement, 
but a more precise value cannot be found for the same fixed value $N$.

Presuming a need for such discrete estimators, one can define the posterior probabilities by constructing probability tables for $P(\hat{p}\leftarrow \theta^*_j|n)$:  the probability, given $n$ observations of 'successes' in a binomial Bernoulli trial, 
that the admissible value $\theta^*_j$ assigned to the estimate $\hat{p}$ is the $\theta_j$ from the list closest to the true underlying $p_{true}$.
Alternatively, one may think of the value $j$ used as the subscript for admissible value $\theta^*_j$ as the value for observed $n$ that would have caused $\theta^*_j$ to be closest to $p_{true}$.
Table \ref{tab:DiscreteProbTable}  is a concrete example for $N=5$.
Use of the form $\binom{N}{n} \theta^n_j (1-\theta_j)^{N-n}$
is inherent to the presumption that only $N+1$ values of $\theta_j$ are possible at the precision available from a Bernoulli trial that assesses the possible success of only $N$ total observations.
For a valid probability table, all entries in the table must sum to 1.
The necessity of weights $\bar{\pi}_j$ between columns of the table, is inferred since without them, the sum would instead be 1 for entries in each column.

\renewcommand{\thetable}{B1}
     \begin{table}[t!]
    \scalebox{0.68} {
    \centering
   \begin{tabular} { | r | c | c | c | c | c | c | }
    \hline
               & \multicolumn{6}{|c|}{                 }\\ 
\multicolumn{1}{|c|}{Observed}     & \multicolumn{6}{|c|}{Probability Table for N=5:  Conditional Probability $P(\hat{p} \leftarrow \theta^*_k~|~n)$} \\
\multicolumn{1}{|c|}{Number of}    & \multicolumn{6}{|c|}{(probability that $\hat{p}$ would be assigned admissible value $\theta_n$, when true} \\
\multicolumn{1}{|c|}{Success}      & \multicolumn{6}{|c|}{ underlying $p$ is closest to $\theta^*_k$, given $n$ observed outcomes were "\emph{successes}")} \\
 \multicolumn{1}{|c|}{Occurrences} & \multicolumn{6}{|c|}{              } \\ \cline{2-7}
               &     $\theta_0$ (k=0) :                            &    $\theta_1$ (k=1) :                            &    $\theta_2$ (k=2) :                            &    $\theta_3$ (k=3) :                            &    $\theta_4$ (k=4) :                            &    $\theta_5$ (k=5) :                                  \\
    \hline
               &                                                  &                                                  &                                                  &                                                  &                                                  &                                                    \\
  {n=0:}       & $\bar{\pi}_0 \binom{5}{0} {\theta_0}^0 (1-\theta_0)^5$   & $\bar{\pi}_1 \binom{5}{0} {\theta_1}^0 (1-\theta_1)^5$   & $\bar{\pi}_2 \binom{5}{0} {\theta_2}^0 (1-\theta_2)^5$   & $\bar{\pi}_3 \binom{5}{0} {\theta_3}^0 (1-\theta_3)^5$   & $\bar{\pi}_4 \binom{5}{0} {\theta_4}^0 (1-\theta_4)^5$   & $\bar{\pi}_5 \binom{5}{0} {\theta_5}^0 (1-\theta_5)^5$     \\
               &                                                  &                                                  &                                                  &                                                  &                                                  &                                                    \\
    \hline
               &                                                  &                                                  &                                                  &                                                  &                                                  &                                                    \\
  {n=1:}       & $\bar{\pi}_0 \binom{5}{1} {\theta_0}^0 (1-\theta_0)^4$   & $\bar{\pi}_1 \binom{5}{1} {\theta_1}^1 (1-\theta_1)^4$   & $\bar{\pi}_2 \binom{5}{1} {\theta_2}^1 (1-\theta_2)^4$   & $\bar{\pi}_3 \binom{5}{1} {\theta_3}^1 (1-\theta_3)^4$   & $\bar{\pi}_4 \binom{5}{1} {\theta_4}^1 (1-\theta_4)^4$   & $\bar{\pi}_5 \binom{5}{1} {\theta_5}^1 (1-\theta_5)^4$     \\
               &                                                  &                                                  &                                                  &                                                  &                                                  &                                                    \\
    \hline
               &                                                  &                                                  &                                                  &                                                  &                                                  &                                                    \\
  {n=2:}       & $\bar{\pi}_0 \binom{5}{2} {\theta_0}^2 (1-\theta_0)^3$   & $\bar{\pi}_1 \binom{5}{2} {\theta_1}^2 (1-\theta_1)^3$   & $\bar{\pi}_2 \binom{5}{2} {\theta_2}^2 (1-\theta_2)^3$   & $\bar{\pi}_3 \binom{5}{2} {\theta_3}^2 (1-\theta_3)^3$   & $\bar{\pi}_4 \binom{5}{2} {\theta_4}^2 (1-\theta_4)^3$   & $\bar{\pi}_5 \binom{5}{2} {\theta_5}^2 (1-\theta_5)^3$     \\
               &                                                  &                                                  &                                                  &                                                  &                                                  &                                                    \\
    \hline
               &                                                  &                                                  &                                                  &                                                  &                                                  &                                                    \\
  {n=3:}       & $\bar{\pi}_0 \binom{5}{3} {\theta_0}^3 (1-\theta_0)^2$   & $\bar{\pi}_1 \binom{5}{3} {\theta_1}^3 (1-\theta_1)^2$   & $\bar{\pi}_2 \binom{5}{3} {\theta_2}^3 (1-\theta_2)^2$   & $\bar{\pi}_3 \binom{5}{3} {\theta_3}^3 (1-\theta_3)^2$   & $\bar{\pi}_4 \binom{5}{3} {\theta_4}^3 (1-\theta_4)^2$   & $\bar{\pi}_5 \binom{5}{3} {\theta_5}^3 (1-\theta_5)^2$     \\
               &                                                  &                                                  &                                                  &                                                  &                                                  &                                                    \\
    \hline
               &                                                  &                                                  &                                                  &                                                  &                                                  &                                                    \\
  {n=4:}       & $\bar{\pi}_0 \binom{5}{4} {\theta_0}^4 (1-\theta_0)^1$   & $\bar{\pi}_1 \binom{5}{4} {\theta_1}^4 (1-\theta_1)^1$   & $\bar{\pi}_2 \binom{5}{4} {\theta_2}^4 (1-\theta_2)^1$   & $\bar{\pi}_3 \binom{5}{4} {\theta_3}^4 (1-\theta_3)^1$   & $\bar{\pi}_4 \binom{5}{4} {\theta_4}^4 (1-\theta_4)^1$   & $\bar{\pi}_5 \binom{5}{4} {\theta_5}^4 (1-\theta_5)^1$     \\
               &                                                  &                                                  &                                                  &                                                  &                                                  &                                                    \\
    \hline
               &                                                  &                                                  &                                                  &                                                  &                                                  &                                                    \\
  {n=5:}       & $\bar{\pi}_0 \binom{5}{5} {\theta_0}^5 (1-\theta_0)^0$   & $\bar{\pi}_1 \binom{5}{5} {\theta_1}^5 (1-\theta_1)^0$   & $\bar{\pi}_2 \binom{5}{5} {\theta_2}^5 (1-\theta_2)^0$   & $\bar{\pi}_3 \binom{5}{5} {\theta_3}^5 (1-\theta_3)^0$   & $\bar{\pi}_4 \binom{5}{5} {\theta_4}^5 (1-\theta_4)^0$   & $\bar{\pi}_5 \binom{5}{5} {\theta_5}^5 (1-\theta_5)^0$     \\
               &                                                  &                                                  &                                                  &                                                  &                                                  &                                                    \\
    \hline 
  \end{tabular}
                   }
\caption{ Probability table (probability that the most consistent, admissible value $\theta^*_k$ is being assigned as the estimated value $\hat{p}$, given $n$ observations are observed)
          for a discrete domain limited experiment.
          This table may be used to define numerical values to use as admissible values $\theta_k$.
          Column-to-column weights, $\bar{\pi}_k$, represent the Bayesian prior probability for a particular $\hat{p}\leftarrow\theta_k$ assignment.  
          For \emph{binomial} Bernoulli experiments with equally likely outcomes in the absence of prior information,
          all $\bar{\pi}_k$ may initially be set to $\frac{1}{N+1}$ since there are (N+1) possible outcomes.
          Other choices for $\bar{\pi}_k$, described in the text, 
          may be more appropriate when a success corresponds to an occurrence being assigned (versus not-assigned) to a particular "bin" in a \emph{multinomial} experiment.
 \label{tab:DiscreteProbTable}  }
  \end{table}

These weights $\bar{\pi}_j$ are the \emph{a priori} probability of assigning admissible value $\theta_j$ to $\hat{p}$ in the absence of knowing the observed success count $n$.
This is equivalent to the definition for a Bayes prior probability.
Given the list of values $\Theta_N\in \{\theta_0, \dots \theta_N \}$, and recalling that $P(\hat{p}\leftarrow \theta^*_j|n)$ relates to whether $p_{true}$ is closest to $\theta^*_j$,
if the density of underlying possibilities for $p_{true}$ 
is $\sim \mathcal{U}[0,1;x]$, then $\bar{\pi}_{\diamond}(j,W_j,\theta_j)=1$; $\bar{\pi}_j \propto W_j$, where $W_j$ are the widths of intervals of the possible $p_{true}$ values that are closest to each $\theta_j$.
The $W_j$ may be evaluated as the difference between the half-way points between $\theta_j$ and its neighbors: 
$$ W_j = \phi_{j+1} - \phi_j, ~ j\in\{0,\dots,N\}$$ $$\phi_0 = 0;~~ \phi_j = \frac{1}{2}(\theta_{j-1}+\theta_j),~ j\in\{1,\dots,N\};~~ \phi_{N+1} = 1 $$
However, for a true Bayes prior based on prior knowledge about preferred values of $\theta_j$ in the absence of experimental knowledge about the value $n$, 
one might want to adjust $\bar{\pi}_j$ to account for this additional prior knowledge.

For a multinomial experiment (histogram), if $j$ is the value of $n$ that would give $\theta^*_j$ closest to $p_{i,true}$ for a particular histogram bin $i$, then the additional knowledge that there are $N$ total measurements and a total of $b$ bins, 
means that there are additional statistical prior likelihoods to consider for each value $j$.
These statistical considerations relate to how many ways there are to fill the remainder of the histogram with the remaining $N-j$ observations.
The number of ways to fill $b$ bins with $N$ observations is known to be $\binom{N+b-1}{b-1}$ \citep{brualdi}.  
If $j$ observations are in bin $i$ then the number of ways to fill the remaining $b-1$ bins of the histogram with the remaining $N-j$ observations is $\binom{N-j+b-2}{b-2}$,
so that $\bar{\pi}_{\diamond}() = p_{stat}(j)$, the statistical likelihood of having $j$ observations in bin $i$ of a $b$-bin histogram, is the ratio:
$$ \bar{\pi}_j \propto p_{stat}(j) = \frac{\binom{N-j+b-2}{b-2}}{\binom{N+b-1}{b-1}}= \frac{(N-j+1)_{b-2}\hspace{2pt}(b-1)}{(N+1)_{b-1}} $$
where $(A)_k$ is Pochhammer notation for a repeated product of $k$, sequentially incremented terms starting from $A$.
It is convenient to convert to the recursion formula:
$$ p_{stat}(0) = \frac{b-1}{N+b-1}; ~~ p_{stat}(j+1) = \frac{N-j}{N-j+b-2} \hspace{2pt} p_{stat}(j) $$
which comes from recognizing that the Pochhammer symbol for the next value of $j$ results from including a new multiplicand to one side (numerator) and removing one from the other side (divisor) of the product while keeping the remaining multiplicands in the product unchanged.
For $b$$>$$2$:  $0$$\leq$$\frac{N-j}{N-j+b-2}$$<$$1$, so the underlying statistical likelihood $p_{stat}(j)$ is a continually decreasing function of $j$ that is multiplied by $W_j$ in the prior.
The earlier used priors $\mathcal{B}[\alpha_0,(b-1)\alpha_0;x]$, which had presumed continuity of outcomes $x$, were also usually decreasing functions of $x$, but their functional form differs from that of the discrete prior $\bar{\pi}_j \propto W_j\hspace{2pt}p_{stat}(j)$. 
Thus the multinomial case can be treated by adjusting the prior density $\bar{\pi}_j$ to account for quantifiable differences in statistical likelihood values for particular admissible values $\theta_j$;
and $\bar{\pi}_j$ differs significantly from priors derived by presuming a continuity of possible values for $\hat{p}$.
The prior $\bar{\pi}_j$ is not unique and its derivation presumes $\mathcal{U}[0,1;x]$ as the 
underlying distribution of possibilities for $p_{true}$ 
(Shannon Information);
in principle, $\bar{\pi}_j$ can be further weighted to conform with different local distributions of underlying possibilities for $p_{true}$ based on different preferred information measures.

\renewcommand{\thetable}{B2}
     \begin{table}[t!]
     \centering
    \scalebox{0.36} {
    \centering
   \begin{tabular} { | r | c  c  c  c  c  c  c  c  c  c | c  c  c  c  c  c  c  c  c  c | c  c  c  c  c  c  c  c  c  c | c  c  c  c  c  c  c  c  c  c | c  c  c  c  c  c  c  c  c  c |}
    \hline
                \multicolumn{51}{|c|}{\textbf{Combinatorial Multinomial Prior}:  Expected Number of Bins for $b=100$ Histogram (Range $\mu \pm 3.5 \sigma$) with $\langle n_i\rangle \in \{0,\dots 11\}$  }\\
                \multicolumn{51}{|c|}{Comparison for Varying $N$ with Expected Number of Bins for an Actual Symmetric Generalized Normal Distribution with Varying Excess Kurtosis}\\
    \hline
\multicolumn{1}{|c|}{} & \multicolumn{10}{|c|}{N=40}  &         \multicolumn{10}{|c|}{N=60}                 &         \multicolumn{10}{|c|}{N=80}                 &         \multicolumn{10}{|c|}{N=100}                 &         \multicolumn{10}{|c|}{N=120}           \\
    \hline
 &      \multicolumn{1}{|c}{} & \multicolumn{9}{c|}{Excess Kurtosis:}                  &      \multicolumn{1}{|c}{} & \multicolumn{9}{c|}{Excess Kurtosis:}                  &      \multicolumn{1}{|c}{} &  \multicolumn{9}{c|}{Excess Kurtosis:}                  &      \multicolumn{1}{|c}{} & \multicolumn{9}{c|}{Excess Kurtosis:}         &      \multicolumn{1}{|c}{} & \multicolumn{9}{c|}{Excess Kurtosis:} \\ 
\multicolumn{1}{|c|}{$\langle n_i\rangle$}  
 & C & -1 & $-\frac{1}{2}$ & 0 & $\frac{1}{2}$ & 1  & $1\frac{1}{2}$ & 2 & 3 & 4  & C & -1 & $-\frac{1}{2}$ & 0 & $\frac{1}{2}$ & 1  & $1\frac{1}{2}$ & 2 & 3 & 4  & C & -1 & $-\frac{1}{2}$ & 0 & $\frac{1}{2}$ & 1  & $1\frac{1}{2}$ & 2 & 3 & 4  & C & -1 & $-\frac{1}{2}$ & 0 & $\frac{1}{2}$ & 1  & $1\frac{1}{2}$ & 2 & 3 & 4  & C & -1 & $-\frac{1}{2}$ & 0 & $\frac{1}{2}$ & 1  & $1\frac{1}{2}$ & 2 & 3 & 4  \\ 
    \hline
0  & 71 & 54 & 60 & 64 & 66 & 68 & 70 & 70 & 72 & 74    & 62 & 50 & 52 & 56 & 58 & 60 & 60 & 62 & 64 & 66    & 55 & 48 & 48 & 50 & 52 & 54 & 56 & 56 & 58 & 60    & 50 & 46 & 46 & 46 & 48 & 50 & 50 & 52 & 54 & 56    & 45 & 46 & 44 & 44 & 46 & 46 & 48 & 48 & 50 & 50   \\
1  & 21 & 46 & 40 & 36 & 34 & 32 & 28 & 26 & 22 & 20    & 24 & 50 & 48 & 30 & 28 & 26 & 26 & 24 & 22 & 20    & 25 & 16 & 24 & 24 & 24 & 24 & 22 & 24 & 22 & 22    & 25 & 12 & 18 & 22 & 22 & 22 & 24 & 22 & 22 & 20    & 25 &  8 & 16 & 20 & 20 & 22 & 22 & 22 & 22 & 24   \\
2  &  6 &  0 &  0 &  0 &  0 &  0 &  2 &  4 &  6 &  6    &  9 &  0 &  0 & 14 & 14 & 14 & 14 & 14 & 10 & 10    & 11 & 36 & 28 & 26 & 22 & 16 & 14 & 12 & 10 &  8    & 13 & 42 & 36 & 18 & 16 & 14 & 12 & 12 & 10 & 10    & 14 & 20 & 16 & 14 & 14 & 12 & 12 & 12 & 10 & 10   \\
3  &  2 &  0 &  0 &  0 &  0 &  0 &  0 &  0 &  0 &  0    &  3 &  0 &  0 &  0 &  0 &  0 &  0 &  0 &  4 &  4    &  5 &  0 &  0 &  0 &  2 &  6 &  8 &  8 &  8 &  6    &  6 &  0 &  0 & 14 & 14 & 14 & 10 &  8 &  6 &  6    &  7 & 26 & 24 & 22 & 12 & 10 &  8 &  8 &  8 &  6   \\
4  &  0 &  0 &  0 &  0 &  0 &  0 &  0 &  0 &  0 &  0    &  1 &  0 &  0 &  0 &  0 &  0 &  0 &  0 &  0 &  0    &  2 &  0 &  0 &  0 &  0 &  0 &  0 &  0 &  2 &  4    &  3 &  0 &  0 &  0 &  0 &  0 &  4 &  6 &  6 &  4    &  4 &  0 &  0 &  0 &  8 & 10 &  8 &  6 &  4 &  4   \\
5  &  0 &  0 &  0 &  0 &  0 &  0 &  0 &  0 &  0 &  0    &  0 &  0 &  0 &  0 &  0 &  0 &  0 &  0 &  0 &  0    &  1 &  0 &  0 &  0 &  0 &  0 &  0 &  0 &  0 &  0    &  2 &  0 &  0 &  0 &  0 &  0 &  0 &  0 &  2 &  4    &  2 &  0 &  0 &  0 &  0 &  0 &  2 &  4 &  4 &  2   \\
6  &  0 &  0 &  0 &  0 &  0 &  0 &  0 &  0 &  0 &  0    &  0 &  0 &  0 &  0 &  0 &  0 &  0 &  0 &  0 &  0    &  0 &  0 &  0 &  0 &  0 &  0 &  0 &  0 &  0 &  0    &  1 &  0 &  0 &  0 &  0 &  0 &  0 &  0 &  0 &  0    &  1 &  0 &  0 &  0 &  0 &  0 &  0 &  0 &  2 &  4   \\
7  &  0 &  0 &  0 &  0 &  0 &  0 &  0 &  0 &  0 &  0    &  0 &  0 &  0 &  0 &  0 &  0 &  0 &  0 &  0 &  0    &  0 &  0 &  0 &  0 &  0 &  0 &  0 &  0 &  0 &  0    &  0 &  0 &  0 &  0 &  0 &  0 &  0 &  0 &  0 &  0    &  1 &  0 &  0 &  0 &  0 &  0 &  0 &  0 &  0 &  0   \\
8  &  0 &  0 &  0 &  0 &  0 &  0 &  0 &  0 &  0 &  0    &  0 &  0 &  0 &  0 &  0 &  0 &  0 &  0 &  0 &  0    &  0 &  0 &  0 &  0 &  0 &  0 &  0 &  0 &  0 &  0    &  0 &  0 &  0 &  0 &  0 &  0 &  0 &  0 &  0 &  0    &  0 &  0 &  0 &  0 &  0 &  0 &  0 &  0 &  0 &  0   \\
9  &  0 &  0 &  0 &  0 &  0 &  0 &  0 &  0 &  0 &  0    &  0 &  0 &  0 &  0 &  0 &  0 &  0 &  0 &  0 &  0    &  0 &  0 &  0 &  0 &  0 &  0 &  0 &  0 &  0 &  0    &  0 &  0 &  0 &  0 &  0 &  0 &  0 &  0 &  0 &  0    &  0 &  0 &  0 &  0 &  0 &  0 &  0 &  0 &  0 &  0   \\
10 &  0 &  0 &  0 &  0 &  0 &  0 &  0 &  0 &  0 &  0    &  0 &  0 &  0 &  0 &  0 &  0 &  0 &  0 &  0 &  0    &  0 &  0 &  0 &  0 &  0 &  0 &  0 &  0 &  0 &  0    &  0 &  0 &  0 &  0 &  0 &  0 &  0 &  0 &  0 &  0    &  0 &  0 &  0 &  0 &  0 &  0 &  0 &  0 &  0 &  0   \\
11 &  0 &  0 &  0 &  0 &  0 &  0 &  0 &  0 &  0 &  0    &  0 &  0 &  0 &  0 &  0 &  0 &  0 &  0 &  0 &  0    &  0 &  0 &  0 &  0 &  0 &  0 &  0 &  0 &  0 &  0    &  0 &  0 &  0 &  0 &  0 &  0 &  0 &  0 &  0 &  0    &  0 &  0 &  0 &  0 &  0 &  0 &  0 &  0 &  0 &  0   \\
    \hline
  \end{tabular}
    }
     \centering
    \scalebox{0.36} {
    \centering
   \begin{tabular} { | r | c  c  c  c  c  c  c  c  c  c | c  c  c  c  c  c  c  c  c  c | c  c  c  c  c  c  c  c  c  c | c  c  c  c  c  c  c  c  c  c |}
    \hline
 \multicolumn{1}{|c|}{}  &   \multicolumn{10}{|c|}{N=140}   &         \multicolumn{10}{|c|}{N=160}                 &         \multicolumn{10}{|c|}{N=180}                 &         \multicolumn{10}{|c|}{N=200}   \\
    \hline

   &  & \multicolumn{9}{c|}{Excess Kurtosis:} & \multicolumn{1}{|c}{} & \multicolumn{9}{c|}{Excess Kurtosis:} & \multicolumn{1}{|c}{} & \multicolumn{9}{c|}{Excess Kurtosis:} & \multicolumn{1}{|c}{} & \multicolumn{9}{c|}{Excess Kurtosis:}                \\ 
\multicolumn{1}{|c|}{$\langle n_i\rangle $}
    & C & -1 & $-\frac{1}{2}$ & 0 & $\frac{1}{2}$ & 1  & $1\frac{1}{2}$ & 2 & 3 & 4  
                                         & C & -1 & $-\frac{1}{2}$ & 0 & $\frac{1}{2}$ & 1  & $1\frac{1}{2}$ & 2 & 3 & 4  
                                                                                                     & C & -1 & $-\frac{1}{2}$ & 0 & $\frac{1}{2}$ & 1  & $1\frac{1}{2}$ & 2 & 3 & 4  
                                                                                                                                                & C & -1 & $-\frac{1}{2}$ & 0 & $\frac{1}{2}$ & 1  & $1\frac{1}{2}$ & 2 & 3 & 4  \\ 
    \hline
0    & 41 & 44 & 42 & 42 & 42 & 44 & 44 & 46 & 44 & 46    & 38 & 44 & 40 & 40 & 40 & 42 & 42 & 42 & 42 & 44    & 35 & 42 & 40 & 38 & 38 & 40 & 40 & 38 & 40 & 40    & 33 & 42 & 38 & 38 & 38 & 38 & 38 & 36 & 38 & 38   \\
1    & 24 &  8 & 16 & 18 & 20 & 20 & 22 & 22 & 26 & 24    & 24 &  8 & 14 & 18 & 20 & 20 & 22 & 22 & 24 & 24    & 23 &  8 & 12 & 18 & 20 & 20 & 20 & 24 & 24 & 26    & 22 &  8 & 14 & 16 & 18 & 20 & 20 & 24 & 24 & 26   \\
2    & 14 & 10 & 12 & 12 & 12 & 12 & 12 & 10 & 10 & 10    & 15 &  6 & 12 & 12 & 12 & 12 & 10 & 12 & 10 & 10    & 15 &  6 & 10 & 10 & 10 & 10 & 12 & 10 & 10 & 10    & 15 &  4 &  8 & 10 & 10 & 10 & 12 & 10 & 10 & 10   \\
3    &  8 & 38 & 30 & 14 & 12 & 10 &  8 &  8 &  6 &  6    &  9 & 42 & 14 & 10 &  8 &  8 &  8 &  6 &  8 &  6    & 10 & 10 & 10 & 10 & 10 &  8 &  8 &  8 &  8 &  6    & 10 &  8 & 10 &  8 &  8 &  8 &  6 &  8 &  6 &  6   \\
4    &  5 &  0 &  0 & 14 & 14 &  8 &  6 &  6 &  6 &  6    &  6 &  0 & 20 & 20 & 10 &  8 &  6 &  6 &  4 &  4    &  6 & 34 & 28 & 10 &  8 &  8 &  6 &  6 &  4 &  4    &  7 & 38 & 12 & 10 &  8 &  6 &  6 &  6 &  6 &  4   \\
5    &  3 &  0 &  0 &  0 &  0 &  6 &  8 &  6 &  4 &  2    &  3 &  0 &  0 &  0 & 10 &  8 &  6 &  6 &  4 &  4    &  4 &  0 &  0 & 14 & 10 &  6 &  6 &  4 &  4 &  4    &  4 &  0 & 18 & 14 &  8 &  6 &  6 &  4 &  4 &  4   \\
6    &  2 &  0 &  0 &  0 &  0 &  0 &  0 &  2 &  2 &  4    &  2 &  0 &  0 &  0 &  0 &  2 &  6 &  4 &  4 &  4    &  3 &  0 &  0 &  0 &  4 &  8 &  4 &  6 &  4 &  4    &  3 &  0 &  0 &  4 & 10 &  8 &  6 &  4 &  4 &  4   \\
7    &  1 &  0 &  0 &  0 &  0 &  0 &  0 &  0 &  2 &  2    &  1 &  0 &  0 &  0 &  0 &  0 &  0 &  2 &  2 &  2    &  2 &  0 &  0 &  0 &  0 &  0 &  4 &  4 &  2 &  2    &  2 &  0 &  0 &  0 &  0 &  4 &  4 &  4 &  2 &  2   \\
8    &  1 &  0 &  0 &  0 &  0 &  0 &  0 &  0 &  0 &  0    &  1 &  0 &  0 &  0 &  0 &  0 &  0 &  0 &  2 &  2    &  1 &  0 &  0 &  0 &  0 &  0 &  0 &  0 &  4 &  2    &  1 &  0 &  0 &  0 &  0 &  0 &  2 &  4 &  2 &  2   \\
9    &  0 &  0 &  0 &  0 &  0 &  0 &  0 &  0 &  0 &  0    &  0 &  0 &  0 &  0 &  0 &  0 &  0 &  0 &  0 &  0    &  1 &  0 &  0 &  0 &  0 &  0 &  0 &  0 &  0 &  2    &  1 &  0 &  0 &  0 &  0 &  0 &  0 &  0 &  4 &  2   \\
10   &  0 &  0 &  0 &  0 &  0 &  0 &  0 &  0 &  0 &  0    &  0 &  0 &  0 &  0 &  0 &  0 &  0 &  0 &  0 &  0    &  0 &  0 &  0 &  0 &  0 &  0 &  0 &  0 &  0 &  0    &  1 &  0 &  0 &  0 &  0 &  0 &  0 &  0 &  0 &  0   \\
11   &  0 &  0 &  0 &  0 &  0 &  0 &  0 &  0 &  0 &  0    &  0 &  0 &  0 &  0 &  0 &  0 &  0 &  0 &  0 &  0    &  0 &  0 &  0 &  0 &  0 &  0 &  0 &  0 &  0 &  0    &  0 &  0 &  0 &  0 &  0 &  0 &  0 &  0 &  0 &  2   \\
    \hline
  \end{tabular}
                   }
\caption{ Number of bins, $\bar{b}_e\left(\langle n_i \rangle\right)$, in a $b=100$ histogram (range $\mu \pm 3.5 \sigma$) expected to have $n_i$ occurrences by the combinatorial prior pdf (C)
          based entirely on the statistics of filling a histogram with $N$ observations; 
          comparison to the number with $\bar{b}_e\left( \langle n_i\rangle\right) $ for the particular case of clumped, symmetric Generalized Normal Distributions,
          $ g(x) = \frac{\beta}{2\alpha\Gamma(\frac{1}{\beta})} e^{-(\frac{|x-\mu|}{\alpha})^\beta }$, 
         with different $\alpha$ and $\beta$ values chosen to provide $\sigma=1$ and excess kurtosis -1 through +4.
 \label{tab:DiscretePriorTable}  }
  \end{table}

In the multinomial case, other prior knowledge about the underlying pdf being estimated, such as estimated shape parameters, kurtosis, or skew, can also improve the prior pdf for the discrete estimator. 
The above purely combinatorial prior is inaccurate for small $N$ in the case of commonly occurring underlying pdfs with density clumped about a single value.  
In this case, relative to statistical expectation, there can be increased occurrence of $n_i$ values from the top of the central plateau or from the peripheral baseline of the pdf.
Since the proportion of points in these regions varies with the excess kurtosis value of the underlying pdf, having a prior estimate of the kurtosis value can improve the prior.  
Table \ref{tab:DiscretePriorTable} shows the expected number of bins (nearest integer) for a $b=100$ histogram into which $\langle n_i \rangle$ occurrences have been sorted
for the purely combinatorial prior, based solely on the statistics of histogram filling using the above recursion formula.
These values are compared to the expected number of bins with  $\langle n_i \rangle$ occurrences for histograms derived from
a standard normal distribution (archetypically clumped, excess kurtosis=0.0), or from other Generalized Normal Distributions with known excess kurtosis values from -1 to +4.
As $N$ increases to $\gg 100$ for a $b=100$ histogram (not shown), the excess of the number of possible $n_i$ value outcomes over the number of bins
reduces the probability of a histogram bin having any particular $n_i$ occurrence count, 
but multiple occurrences of the same value $\langle n_i\rangle$ appear in even numbers of bins in the same histogram if the pdf is symmetrical.

The posterior density $P(\hat{p}\leftarrow \theta^*_j|n)$ for each possible outcome $n$ derives from only one row of Table \ref{tab:DiscreteProbTable}
and requires renormalizing the
sum of probabilities in this row to 1.   
When plotting $P(\hat{p}\leftarrow \theta^*_j|n)$, 
(as in Fig \ref{fig:DiscretePosteriorPDFs}), 
the abscissa is dictated by a continuum of values for \emph{underlying} $p_{true}$, 
but probabilities $P(\hat{p}\leftarrow \theta^*_j|n)$ related to the \emph{assigned} estimated value (the ordinate in the plot) 
requires the initially continuous values of $p_{true}$ to be mapped to the discrete list, $\Theta_N$, of admissible values, 
to avoid contributions to the posterior from values for $\hat{p}$ that will never be considered for a given value $N$.  
This mapping $\theta^*_j: p_{true} \rightarrow \Theta_N$ can be done using $\theta^*_j(p_{true}) = \argmin_{\theta_j}(|p_{true}-\theta_j|)$.
The result of this mapping leads to the posterior $P(p_{true}$$\rightarrow$$\Theta_N |n) = P(\hat{p}\leftarrow \theta^*_j|n)$, 
which when plotted against $p_{true}$ is a union of adjacent rectangles of varying widths $W_j$ and heights $C_n\hspace{2pt} p_{stat}(j) \theta^n_j (1-\theta_j)^{N-j}$ (Fig \ref{fig:DiscretePosteriorPDFs}).
The $C_n$ is a normalization constant that adjusts the sum of the rectangular areas to 1.
Similarly to the continuous posterior densities $\mathcal{N}[\mu,\sigma^2;x]$ and $\mathcal{B}[\alpha,\beta; x]$, 
the posterior $P(x \rightarrow\Theta_N |n)$ ---  the discrete union of rectangles --- may be used in the integral in Eq. 6 (Sec. 1.4.1) to define estimated confidence intervals $\hat{\delta}(n,N)$ to a nominal degree of confidence $\xi$.

\renewcommand{\thefigure}{B1}
\begin{figure}
\centering
\scalebox{0.85} {
\centering
\includegraphics[width=15 cm]{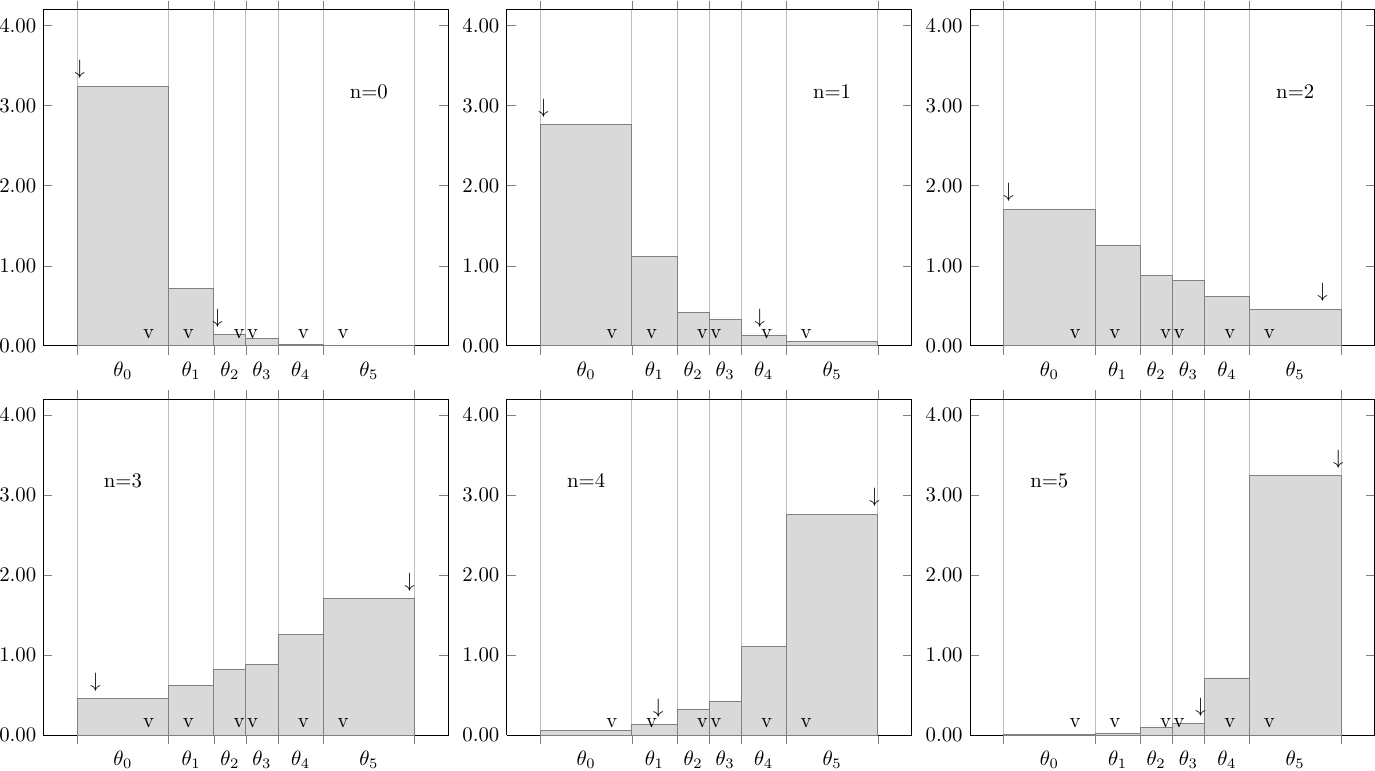}
}
        \caption{ Discrete posterior pdfs (Table \ref{tab:DiscreteProbTable}) for each outcome of a \emph{binomial} experiment with 5 observations. Boundaries of equal tailed $\xi$=0.95 confidence intervals are denoted by downward arrows.
                  Positions of iteratively adjusted median values $\tilde{\theta}_j$ are indicated by the carets.  
                  Boundaries of zones associated with each $\theta_j$  are either at an edge (0 or 1) or half-way between median values.
                  For $j=0\rightarrow 5$: $\theta_j = \{0.21196, 0.32965, 0.48010, 0.51990, 0.67035, 0.78804\}$. 
                        \label{fig:DiscretePosteriorPDFs}}
\end{figure}

The only issue that remains is to define the set of admissible values $\Theta_N = \{\theta_0, \dots, \theta_N\}$.
From the probability table, simultaneous maximization of probability with respect to each $\theta_j$ using partial derivatives (and presuming that a continuum of possible values for $\theta_j$ is allowed \emph{before} assigning a value to $\hat{p}$) gives $(\theta_j)_{mp} = \frac{j}{N}$. 
However, using $(\theta_j)_{mp}$ in the posterior (normalized row of the probability table for the observed value $n$) as weights to calculate a conditional expectation value $\langle \theta | n \rangle $ for each outcome $n$, by using a posterior-weighted average of admissible values,
one finds that $\langle p\rangle = \langle \theta | n \rangle $ and $\hat{p} = (\theta_{j=n})_{mp}$ differ significantly.  
Given the differences, 
this leaves ambiguity about which value, $\langle \theta | n \rangle$ or $(\theta_{j=n})_{mp}$, is better to assign to $\theta_j$ and ultimately to the estimator $\hat{p}$. 
For stable calculations in the Bayes classification problem, the values of $\hat{p}$ needed to differ from exactly $0$ or $1$, so the expected values were tried.
On modifying $\hat{p}$,
by switching  $\theta_{j=n}$ to  $\langle \theta | n \rangle $ for each $j$, $\Theta^{'}_N \leftarrow \langle\Theta_{mp}\rangle  $,
the recalculated conditionally expected value $\langle\Theta^{'}_N\rangle$$=$$\langle \langle\theta_{mp}\rangle_j | n \rangle$ based on the new set of admissible values after the switch again differed from the starting $\Theta^{'}_N$=$\langle\Theta_{N,mp}\rangle$; $\langle"\langle\Theta_{N,mp}\rangle"\rangle \neq \langle\Theta_{N,mp}\rangle$.
Repeated replacement of $\theta_{j=n}$ with recalculated  $\langle \theta | n \rangle $ leads to $\Theta^{{'}\infty}_N$ 
for which the spacing between many adjacent values $\theta_j$ and $\theta_{j+1}$ approach a value of 0; many $\theta_j$ values approach the identical value $\frac{1}{2}$,
making this a poor choice for $\Theta_N$.

One practical possibility was just to use the first set $\langle \Theta_{N,mp} \rangle$, but given the observed instability in these expected values,
it was thought better to find a stably self-consistent set of admissible values $\Theta_N$ that avoided assigning to $\hat{p}$ the values $\theta_0=0$ and $\theta_N=1$ and that avoided arbitrarily selecting one from the many possible sets $\langle\Theta_N\rangle$.
In contrast to the expected values, it was found that replacing each $\theta_{j=n}$ with the median of the confidence interval $\hat{\delta}_{|n}$ defined by Eq. 6:  $\frac{1}{2}(\delta^+_{|n} + \delta^-_{|n})$, led to stability.
Repeated replacement and calculation of new integrals and median values eventually returns a set of updated $\theta^{'}_j$ values that are all practically identical to $\theta_j$ prior to replacement.
The $\theta_j$ values in this self-consistent set $\tilde{\Theta}^{{'}\infty}_N$ are well separated and usually close to the initial $(\theta_j)_{mp}$.
One unusual feature about this approach though is that it causes the estimated values $\hat{p}$ to depend on the confidence levels $\xi$ used to define the interval limits $\delta^{[\xi]}_{|n}$
that define the median values.

For $N < 12800$ in the $b=100$ multinomial case, discrete $\hat{p}_i(n_i,N)$ based on the combinatorial prior leads to values of $\langle(S/N)\rangle_{MC}$ that are comparable to what is found for other estimators (Table S1). 
Systematic differences that occur are usually fairly small
and the larger of these occur for combinations of $N$ and underlying pdf type that 
are expected to be improved by upgrading from the purely combinatorial prior to one based on initial estimates of kurtosis values.
The binomial and multinomial estimators $\hat{\delta}(n,N)$ from the discrete priors and posteriors exhibit comparable $\xi$-to-$C$ matching to that of $\hat{\delta}$ from continuous prior-posterior combinations.

\section*{Supplement A:}
\renewcommand{\thetable}{S1}
\label{app:SupplementA}
Table of S/N values and $N_{eq}$ (equivalent numbers of observations $N$) for different estimator choices.
     \begin{table}[t!]
    \scalebox{0.40} {
    \centering
   \begin{tabular} { | r | c | c   c   c | c   c   c | c   c   c | c   c   c | c   c   c | c   c   c | c   c   c | c   c   c | }
    \hline
                                             &                         & \multicolumn{24}{|c|}{ Estimator Type: }\\ \cline{3-26}  
                 {Underlying}                & {Bayes}                 & \multicolumn{3}{|c|}{Optimized (b=2)}                         & \multicolumn{3}{|c|}{Discrete (b=100)}                      & \multicolumn{3}{|c|}{Uniform (b=100)}                     & \multicolumn{3}{|c|}{Rgbp (b=100)} & \multicolumn{3}{|c|}{Optimized (b=2)}  & \multicolumn{3}{|c|}{Discrete (b=100)} & \multicolumn{3}{|c|}{Uniform (b=100)} & \multicolumn{3}{|c|}{Rgbp (b=100)}   \\
                 {PDF}                       &    Prior:               & \multicolumn{3}{|c|}{$\mathcal{B}[\alpha_0,\alpha_0;x]$}      & \multicolumn{3}{|c|}{combinatorial}            & \multicolumn{3}{|c|}{$\mathcal{B}[1,99;x]$}                     & \multicolumn{3}{|c|}{ }                & \multicolumn{3}{|c|}{$\mathcal{B}[\alpha_0,\alpha_0;x]$} & \multicolumn{3}{|c|}{combinatorial} &  \multicolumn{3}{|c|}{$\mathcal{B}[1,99;x]$}& \multicolumn{3}{|c|}{}       \\
                 {to Identify}               & Posterior:  & \multicolumn{3}{|c|}{$\mathcal{N}[\mu_{p_i},\sigma^2_{p_i};x]$} &   \multicolumn{3}{|c|}{}  & \multicolumn{3}{|c|}{$\mathcal{N}[\mu_{p_i},\sigma^2_{p_i};x]$}& \multicolumn{3}{|c|}{ }                & \multicolumn{3}{|c|}{DE-NOISED}     & \multicolumn{3}{|c|}{DE-NOISED}      & \multicolumn{3}{|c|}{DE-NOISED}        & \multicolumn{3}{|c|}{DE-NOISED}       \\ \cline{3-26}
                                             &    $N$ :                &    S/N    & $N_{eq}$ &  Ratio                                            & S/N    & $N_{eq}$ &   Ratio                             &  S/N    & $N_{eq}$ &   Ratio                               &  S/N    & $N_{eq}$ &   Ratio           &  S/N    & $N_{eq}$ &   Ratio        &  S/N    & $N_{eq}$ & Ratio           &  S/N    & $N_{eq}$ &   Ratio           &  S/N    & $N_{eq}$ & Ratio            \\
        \hline
  \multirow{16}{*}{Gaussian:}                &    40                   & 1.136     &    912   &  22.80                                &  1.193  &    736   & 18.40                              &  1.192   &    749   & 18.73                                 &  1.129  &    767  & 19.18              &   2.81  &     45   & 1.125          &   2.96  &     40  &\textbf{1.000}             &   2.90  &     42   & 1.050             &   2.87  &     43 & 1.075    \\
                                             &    60                   & 1.193     &   1385   &  23.08                                &  1.311  &   1177   & 19.62                              &  1.276   &   1189   & 19.82                                 &  1.229  &   1208  & 20.13              &   3.42  &     68   & 1.133          &   3.65  &     60  &\textbf{1.000}             &   3.50  &     66   & 1.100             &   3.47  &     67 & 1.117    \\
                                             &    80                   & 1.249     &   1673   &  20.91                                &  0.925  &   1451   & 18.14                              &  1.354   &   1461   & 18.26                                 &  1.315  &   1480  & 18.50              &   3.96  &     83   & 1.038          &   3.00  &    110  & 1.375             &   4.02  &     80   &\textbf{1.000}             &   3.99  &     81 & 1.012    \\
                                             &   100                   & 1.304     &   2063   &  20.63                                &  1.082  &   1832   & 18.32                              &  1.428   &   1838   & 18.38                                 &  1.392  &   1857  & 18.57              &   4.42  &    103   & 1.030          &   3.48  &    120  & 1.200             &   4.48  &    100   &\textbf{1.000}             &   4.45  &    101 & 1.010    \\
                                             &   120                   & 1.358     &   2431   &  20.26                                &  1.401  &   2192   & 18.27                              &  1.500   &   2195   & 18.29                                 &  1.465  &   2214  & 18.45              &   4.83  &    123   & 1.025          &   4.51  &    136  & 1.133             &   4.88  &    120   &\textbf{1.000}             &   4.85  &    122 & 1.017    \\
                                             &   140                   & 1.412     &   2781   &  19.86                                &  1.578  &   2537   & 18.12                              &  1.568   &   2537   & 18.12                                 &  1.535  &   2556  & 18.26              &   5.18  &    143   & 1.021          &   4.96  &    161  & 1.150             &   5.23  &    140   &\textbf{1.000}             &   5.20  &    142 & 1.014    \\
                                             &   160                   & 1.462     &   3135   &  19.59                                &  1.635  &   2888   & 18.05                              &  1.630   &   2885   & 18.03                                 &  1.595  &   2903  & 18.14              &   5.52  &    163   & 1.019          &   5.22  &    186  & 1.163             &   5.56  &    160   &\textbf{1.000}             &   5.54  &    162 & 1.012    \\
                                             &   180                   & 1.514     &   3497   &  19.43                                &  1.702  &   3248   & 18.04                              &  1.695   &   3241   & 18.01                                 &  1.661  &   3259  & 18.11              &   5.84  &    183   & 1.017          &   5.47  &    213  & 1.183             &   5.88  &    180   &\textbf{1.000}             &   5.86  &    182 & 1.011    \\
                                             &   200                   & 1.563     &   3831   &  19.16                                &  1.770  &   3584   & 17.92                              &  1.753   &   3574   & 17.87                                 &  1.720  &   3592  & 17.96              &   6.13  &    205   & 1.025          &   5.76  &    242  & 1.210             &   6.17  &    200   &\textbf{1.000}             &   6.14  &    203 & 1.015    \\
                                             &   400                   &  2.02     &   5840   &  14.60                                &   2.27  &   5584   & 13.96                              &   2.26   &   5565   & 13.91                                 &   2.23  &   5582  & 13.96              &   7.61  &    407   & 1.018          &   7.45  &    426  & 1.065             &   7.67  &    400   &\textbf{1.000}             &   7.66  &    402 & 1.005    \\
                                             &   800                   &  2.77     &  10837   &  13.55                                &   3.07  &  10579   & 13.22                              &   3.04   &  10542   & 13.18                                 &   3.01  &  10559  & 13.20              &  10.44  &    814   & 1.018          &  10.36  &    828  & 1.035             &  10.50  &    802   & 1.002             &  10.51  &    800 &\textbf{1.000}    \\
                                             &  1600                   &  3.93     &  19565   &  12.23*                               &   4.21  &  19362   & 12.10*                             &   4.20   &  19319   & 12.07*                                &   4.17  &  19342  & 12.09*             &  14.04  &   1635   & 1.022          &  13.90  &   1681  & 1.051             &  14.10  &   1616   & 1.010             &  14.15  &   1600 &\textbf{1.000}    \\
                                             &  3600                   &  5.97     &  36917   &  10.25*                               &   6.18  &  36927   & 10.26*                             &   6.19   &  36882   & 10.24*                                &   6.17  &  36912  & 10.25*             &  19.31  &   3913   & 1.087          &  18.97  &   4431  & 1.231             &  19.33  &   3901   & 1.084             &  19.47  &   3600 &\textbf{1.000}    \\
                                             &  6400                   &  8.04     &  41186   &   6.44*                               &   8.21  &  41260   &  6.45*                             &   8.22   &  41215   &  6.44*                                &   8.21  &  41248  &  6.45*             &  20.57  &   6400   &\textbf{1.000}          &  20.50  &   6455  & 1.009             &  20.50  &   6457   & 1.009             &  20.51  &   6448 & 1.008    \\
                                             &  9600                   &  9.89     &  56940   &   5.93*                               &  10.02  &  57274   &  5.97*                             &  10.04   &  57233   &  5.96*                                &  10.03  &  57271  &  5.97*             &  24.20  &   9600   &\textbf{1.000}          &  24.10  &   9711  & 1.012             &  24.12  &   9684   & 1.009             &  24.16  &   9641 & 1.004    \\
                                             & 12800                   & 11.43     &  71085   &   5.55*                               &  11.54  &  71675   &  5.60*                             &  11.56   &  71639   &  5.60*                                &  11.55  &  71681  &  5.60*             &  27.05  &  12800   &\textbf{1.000}          &  26.91  &  12972  & 1.013*            &  26.98  &  12893   & 1.007*            &  27.05  &  12802 & 1.000*   \\
        \hline
  \multirow{16}{*}{$\mathcal{B}[3,15;x]$:}   &    40                   & 1.139     &    920   &  23.00                                &  1.224  &    706   & 17.65                              &  1.211   &    721   & 18.02                                 &  1.152  &    739  & 18.48              &   2.89  &     45   & 1.125          &   3.07  &     40  &\textbf{1.000}             &   3.01  &     42   & 1.050             &   2.96  &     43 & 1.075    \\
                                             &    60                   & 1.198     &   1409   &  23.48                                &  1.339  &   1152   & 19.20                              &  1.302   &   1168   & 19.47                                 &  1.259  &   1187  & 19.78              &   3.53  &     72   & 1.200          &   3.83  &     60  &\textbf{1.000}             &   3.65  &     67   & 1.117             &   3.60  &     69 & 1.150    \\
                                             &    80                   & 1.256     &   1624   &  20.30                                &  0.949  &   1352   & 16.90                              &  1.387   &   1367   & 17.09                                 &  1.350  &   1387  & 17.34              &   4.00  &     85   & 1.063          &   2.99  &    112  & 1.400             &   4.12  &     80   &\textbf{1.000}             &   4.08  &     82 & 1.025    \\
                                             &   100                   & 1.313     &   1960   &  19.60                                &  1.112  &   1677   & 16.77                              &  1.470   &   1691   & 16.91                                 &  1.437  &   1712  & 17.12              &   4.42  &    107   & 1.070          &   3.44  &    121  & 1.210             &   4.56  &    100   &\textbf{1.000}             &   4.52  &    102 & 1.020    \\
                                             &   120                   & 1.369     &   2252   &  18.77                                &  1.433  &   1958   & 16.32                              &  1.547   &   1970   & 16.42                                 &  1.513  &   1992  & 16.60              &   4.77  &    127   & 1.058          &   4.52  &    133  & 1.108             &   4.90  &    120   &\textbf{1.000}             &   4.86  &    122 & 1.017    \\
                                             &   140                   & 1.423     &   2587   &  18.48                                &  1.621  &   2283   & 16.31                              &  1.620   &   2293   & 16.38                                 &  1.588  &   2316  & 16.54              &   5.12  &    151   & 1.079          &   5.13  &    152  & 1.086             &   5.28  &    140   &\textbf{1.000}             &   5.24  &    142 & 1.014    \\
                                             &   160                   & 1.477     &   2858   &  17.86                                &  1.686  &   2549   & 15.93                              &  1.690   &   2556   & 15.97                                 &  1.658  &   2580  & 16.13              &   5.39  &    174   & 1.087          &   5.37  &    177  & 1.106             &   5.56  &    160   &\textbf{1.000}             &   5.53  &    163 & 1.019    \\
                                             &   180                   & 1.530     &   3110   &  17.28                                &  1.759  &   2796   & 15.53                              &  1.759   &   2802   & 15.57                                 &  1.727  &   2826  & 15.70              &   5.64  &    193   & 1.072          &   5.60  &    193  & 1.072             &   5.82  &    180   &\textbf{1.000}             &   5.79  &    182 & 1.011    \\
                                             &   200                   & 1.581     &   3418   &  17.09                                &  1.838  &   3099   & 15.49                              &  1.822   &   3103   & 15.52                                 &  1.789  &   3128  & 15.64              &   5.92  &    224   & 1.120          &   5.92  &    220  & 1.100             &   6.11  &    200   &\textbf{1.000}             &   6.08  &    204 & 1.020    \\
                                             &   400                   &  2.06     &   5058   &  12.64                                &   2.40  &   4724   & 11.81                              &   2.38   &   4719   & 11.80                                 &   2.34  &   4746  & 11.87              &   7.30  &    429   & 1.073          &   7.50  &    400  &\textbf{1.000}             &   7.47  &    404   & 1.010             &   7.42  &    412 & 1.030    \\
                                             &   800                   &  2.86     &   8863   &  11.08                                &   3.25  &   8523   & 10.65                              &   3.21   &   8500   & 10.63                                 &   3.18  &   8531  & 10.66              &   9.63  &    894   & 1.117          &  10.02  &    800  &\textbf{1.000}             &   9.79  &    857   & 1.071             &   9.70  &    881 & 1.101    \\
                                             &  1600                   &  4.09     &  14365   &   8.98*                               &   4.46  &  14035   &  8.77*                             &   4.44   &  13999   &  8.75*                                &   4.41  &  14031  &  8.77*             &  12.43  &   1793   & 1.121          &  12.84  &   1600  &\textbf{1.000}             &  12.52  &   1754   & 1.096             &  12.40  &   1820 & 1.137    \\
                                             &  3600                   &  6.28     &  22779   &   6.33*                               &   6.57  &  22511   &  6.25*                             &   6.57   &  22459   &  6.24*                                &   6.54  &  22488  &  6.25*             &  15.94  &   3902   & 1.084          &  16.25  &   3600  &\textbf{1.000}             &  15.96  &   3872   & 1.076             &  15.73  &   4081 & 1.134    \\
                                             &  6400                   &  8.47     &  29857   &   4.67*                               &   8.70  &  29670   &  4.64*                             &   8.71   &  29606   &  4.63*                                &   8.69  &  29629  &  4.63*             &  18.46  &   6612   & 1.033          &  18.65  &   6400  &\textbf{1.000}             &  18.58  &   6473   & 1.011             &  18.41  &   6709 & 1.048    \\
                                             &  9600                   & 10.44     &  38601   &   4.02*                               &  10.63  &  38535   &  4.01*                             &  10.64   &  38457   &  4.01*                                &  10.63  &  38472  &  4.01*             &  21.08  &   9878   & 1.029          &  21.25  &   9600  &\textbf{1.000}             &  21.25  &   9602   &\textbf{1.000}             &  20.71  &  10693 & 1.114    \\
                                             & 12800                   & 12.11     &  45675   &   3.57*                               &  12.27  &  45720   &  3.57*                             &  12.28   &  45631   &  3.56*                                &  12.27  &  45638  &  3.57*             &  22.93  &  13198   & 1.031*         &  23.10  &  12882  & 1.006*            &  23.14  &  12800   &\textbf{1.000}             &  22.21  &  15012 & 1.173*   \\
        \hline
  \multirow{16}{*}{$\mathcal{B}[5,3;x]$:}    &    40                   & 1.134     &    892   &  22.30                                &  1.184  &    714   & 17.85                              &  1.185   &    740   & 18.50                                 &  1.122  &    768  & 19.20              &   2.73  &     45   & 1.125          &   2.89  &     40  &\textbf{1.000}             &   2.83  &     42   & 1.050             &   2.80  &     43 & 1.075    \\
                                             &    60                   & 1.191     &   1417   &  23.62                                &  1.308  &   1201   & 20.02                              &  1.265   &   1233   & 20.55                                 &  1.215  &   1265  & 21.08              &   3.35  &     71   & 1.183          &   3.64  &     60  &\textbf{1.000}             &   3.45  &     67   & 1.117             &   3.42  &     69 & 1.150    \\
                                             &    80                   & 1.246     &   1671   &  20.89                                &  0.916  &   1442   & 18.02                              &  1.340   &   1474   & 18.43                                 &  1.298  &   1508  & 18.85              &   3.87  &     84   & 1.050          &   2.99  &    109  & 1.363             &   3.96  &     80   &\textbf{1.000}             &   3.93  &     81 & 1.012    \\
                                             &   100                   & 1.300     &   2022   &  20.22                                &  1.077  &   1783   & 17.83                              &  1.413   &   1815   & 18.15                                 &  1.374  &   1850  & 18.50              &   4.28  &    104   & 1.040          &   3.46  &    117  & 1.170             &   4.37  &    100   &\textbf{1.000}             &   4.34  &    101 & 1.010    \\
                                             &   120                   & 1.353     &   2399   &  19.99                                &  1.400  &   2151   & 17.93                              &  1.482   &   2183   & 18.19                                 &  1.445  &   2219  & 18.49              &   4.67  &    125   & 1.042          &   4.50  &    130  & 1.083             &   4.77  &    120   &\textbf{1.000}             &   4.74  &    122 & 1.017    \\
                                             &   140                   & 1.405     &   2771   &  19.79                                &  1.581  &   2517   & 17.98                              &  1.546   &   2547   & 18.19                                 &  1.508  &   2584  & 18.46              &   5.03  &    147   & 1.050          &   5.03  &    149  & 1.064             &   5.13  &    140   &\textbf{1.000}             &   5.11  &    142 & 1.014    \\
                                             &   160                   & 1.456     &   3089   &  19.31                                &  1.641  &   2830   & 17.69                              &  1.609   &   2859   & 17.87                                 &  1.570  &   2897  & 18.11              &   5.31  &    168   & 1.050          &   5.25  &    175  & 1.094             &   5.43  &    160   &\textbf{1.000}             &   5.40  &    162 & 1.012    \\
                                             &   180                   & 1.505     &   3423   &  19.02                                &  1.701  &   3162   & 17.57                              &  1.668   &   3189   & 17.72                                 &  1.628  &   3227  & 17.93              &   5.60  &    189   & 1.050          &   5.48  &    198  & 1.100             &   5.72  &    180   &\textbf{1.000}             &   5.69  &    182 & 1.011    \\
                                             &   200                   & 1.555     &   3717   &  18.59                                &  1.767  &   3455   & 17.27                              &  1.729   &   3480   & 17.40                                 &  1.688  &   3518  & 17.59              &   5.85  &    213   & 1.065          &   5.74  &    225  & 1.125             &   5.96  &    200   &\textbf{1.000}             &   5.94  &    203 & 1.015    \\
                                             &   400                   &  2.00     &   5722   &  14.30                                &   2.25  &   5449   & 13.62                              &   2.23   &   5470   & 13.68                                 &   2.18  &   5511  & 13.78              &   7.29  &    419   & 1.048          &   7.30  &    417  & 1.042             &   7.44  &    400   &\textbf{1.000}             &   7.43  &    401 & 1.002    \\
                                             &   800                   &  2.74     &  10233   &  12.79                                &   3.04  &   9959   & 12.45                              &   2.99   &   9966   & 12.46                                 &   2.94  &  10008  & 12.51              &   9.82  &    838   & 1.048          &   9.96  &    807  & 1.009             &   9.95  &    809   & 1.011             &   9.99  &    800 &\textbf{1.000}    \\
                                             &  1600                   &  3.87     &  17526   &  10.95*                               &   4.15  &  17258   & 10.79*                             &   4.11   &  17257   & 10.79*                                &   4.07  &  17301  & 10.81*             &  12.95  &   1690   & 1.056          &  13.01  &   1664  & 1.040             &  13.03  &   1654   & 1.034             &  13.14  &   1600 &\textbf{1.000}    \\
                                             &  3600                   &  5.87     &  28361   &   7.88*                               &   6.08  &  28156   &  7.82*                             &   6.06   &  28138   &  7.82*                                &   6.03  &  28178  &  7.83*             &  16.49  &   3847   & 1.069          &  16.37  &   3915  & 1.087             &  16.55  &   3787   & 1.052             &  16.78  &   3600 &\textbf{1.000}    \\
                                             &  6400                   &  7.87     &  39270   &   6.14*                               &   8.04  &  39161   &  6.12*                             &   8.03   &  39124   &  6.11*                                &   8.01  &  39155  &  6.12*             &  19.40  &   6797   & 1.062          &  19.55  &   6643  & 1.038             &  19.59  &   6595   & 1.030             &  19.79  &   6400 &\textbf{1.000}    \\
                                             &  9600                   &  9.67     &  51699   &   5.39*                               &   9.81  &  51722   &  5.39*                             &   9.81   &  51662   &  5.38*                                &   9.79  &  51680  &  5.38*             &  22.26  &  10277   & 1.071          &  22.35  &  10146  & 1.057             &  22.43  &  10027   & 1.044             &  22.73  &   9600 &\textbf{1.000}    \\
                                             & 12800                   & 11.20     &  61956   &   4.84*                               &  11.32  &  62103   &  4.85*                             &  11.32   &  62023   &  4.85*                                &  11.30  &  62028  &  4.85*             &  24.38  &  13663   & 1.067*         &  24.46  &  13533  & 1.057*            &  24.54  &  13407   & 1.047*            &  24.91  &  12800 &\textbf{1.000}    \\
        \hline
  \multirow{16}{*}{$\mathcal{B}[6,2;x]$:}    &    40                   & 1.139     &    899   &  22.48                                &  1.221  &    689   & 17.23                              &  1.208   &    708   & 17.70                                 &  1.148  &    729  & 18.23              &   2.86  &     45   & 1.125          &   3.02  &     40  &\textbf{1.000}             &   2.96  &     42   & 1.050             &   2.92  &     43 & 1.075    \\
                                             &    60                   & 1.199     &   1346   &  22.43                                &  1.340  &   1096   & 18.27                              &  1.298   &   1118   & 18.63                                 &  1.252  &   1141  & 19.02              &   3.44  &     72   & 1.200          &   3.72  &     60  &\textbf{1.000}             &   3.53  &     68   & 1.133             &   3.49  &     70 & 1.167    \\
                                             &    80                   & 1.256     &   1526   &  19.07                                &  0.948  &   1262   & 15.78                              &  1.383   &   1283   & 16.04                                 &  1.344  &   1308  & 16.35              &   3.87  &     85   & 1.063          &   2.93  &    112  & 1.400             &   3.96  &     80   &\textbf{1.000}             &   3.93  &     82 & 1.025    \\
                                             &   100                   & 1.312     &   1829   &  18.29                                &  1.107  &   1554   & 15.54                              &  1.462   &   1575   & 15.75                                 &  1.426  &   1602  & 16.02              &   4.25  &    106   & 1.060          &   3.37  &    120  & 1.200             &   4.36  &    100   &\textbf{1.000}             &   4.33  &    102 & 1.020    \\
                                             &   120                   & 1.369     &   2119   &  17.66                                &  1.433  &   1833   & 15.28                              &  1.540   &   1852   & 15.43                                 &  1.505  &   1879  & 15.66              &   4.60  &    128   & 1.067          &   4.37  &    132  & 1.100             &   4.72  &    120   &\textbf{1.000}             &   4.69  &    122 & 1.017    \\
                                             &   140                   & 1.424     &   2391   &  17.08                                &  1.628  &   2095   & 14.96                              &  1.615   &   2114   & 15.10                                 &  1.581  &   2141  & 15.29              &   4.90  &    150   & 1.071          &   4.95  &    147  & 1.050             &   5.03  &    140   &\textbf{1.000}             &   5.00  &    142 & 1.014    \\
                                             &   160                   & 1.477     &   2644   &  16.52                                &  1.690  &   2340   & 14.63                              &  1.682   &   2359   & 14.74                                 &  1.646  &   2387  & 14.92              &   5.16  &    172   & 1.075          &   5.18  &    171  & 1.069             &   5.30  &    160   &\textbf{1.000}             &   5.28  &    162 & 1.012    \\
                                             &   180                   & 1.528     &   2881   &  16.01                                &  1.759  &   2571   & 14.28                              &  1.748   &   2589   & 14.38                                 &  1.711  &   2617  & 14.54              &   5.39  &    194   & 1.078          &   5.39  &    191  & 1.061             &   5.55  &    180   &\textbf{1.000}             &   5.52  &    182 & 1.011    \\
                                             &   200                   & 1.581     &   3110   &  15.55                                &  1.838  &   2795   & 13.97                              &  1.813   &   2811   & 14.05                                 &  1.775  &   2840  & 14.20              &   5.61  &    219   & 1.095          &   5.66  &    212  & 1.060             &   5.77  &    200   &\textbf{1.000}             &   5.75  &    203 & 1.015    \\
                                             &   400                   &  2.06     &   4863   &  12.16                                &   2.39  &   4528   & 11.32                              &   2.36   &   4535   & 11.34                                 &   2.32  &   4567  & 11.42              &   7.07  &    433   & 1.083          &   7.29  &    400  &\textbf{1.000}             &   7.26  &    405   & 1.012             &   7.20  &    415 & 1.038    \\
                                             &   800                   &  2.84     &   8488   &  10.61                                &   3.22  &   8144   & 10.18                              &   3.19   &   8138   & 10.17                                 &   3.15  &   8173  & 10.22              &   9.27  &    911   & 1.139          &   9.71  &    800  &\textbf{1.000}             &   9.43  &    873   & 1.091             &   9.32  &    906 & 1.133    \\
                                             &  1600                   &  4.06     &  13679   &   8.55*                               &   4.42  &  13345   &  8.34*                             &   4.39   &  13332   &  8.33*                                &   4.36  &  13370  &  8.36*             &  11.98  &   1811   & 1.132          &  12.43  &   1600  &\textbf{1.000}             &  12.04  &   1787   & 1.117             &  11.85  &   1896 & 1.185    \\
                                             &  3600                   &  6.23     &  21842   &   6.07*                               &   6.52  &  21557   &  5.99*                             &   6.51   &  21535   &  5.98*                                &   6.49  &  21578  &  5.99*             &  15.48  &   3827   & 1.063          &  15.81  &   3600  &\textbf{1.000}             &  15.42  &   3858   & 1.072             &  15.12  &   4056 & 1.127    \\
                                             &  6400                   &  8.41     &  32298   &   5.05*                               &   8.63  &  32125   &  5.02*                             &   8.63   &  32091   &  5.01*                                &   8.61  &  32136  &  5.02*             &  19.10  &   6630   & 1.036          &  19.32  &   6400  &\textbf{1.000}             &  19.13  &   6594   & 1.030             &  18.90  &   6829 & 1.067    \\
                                             &  9600                   & 10.34     &  42171   &   4.39*                               &  10.53  &  42133   &  4.39*                             &  10.53   &  42088   &  4.38*                                &  10.51  &  42133  &  4.39*             &  21.91  &   9941   & 1.036          &  22.14  &   9600  &\textbf{1.000}             &  21.99  &   9812   & 1.022             &  21.77  &  10110 & 1.053    \\
                                             & 12800                   & 12.01     &  50363   &   3.93*                               &  12.17  &  50452   &  3.94*                             &  12.17   &  50397   &  3.94*                                &  12.16  &  50441  &  3.94*             &  23.96  &  13259   & 1.036*         &  24.24  &  12800  &\textbf{1.000}             &  24.09  &  13040   & 1.019*            &  23.94  &  13271 & 1.037*   \\
        \hline
  \multirow{16}{*}{$\mathcal{B}[9,11;x]$:}   &    40                   & 1.134     &    894   &  22.35                                &  1.183  &    723   & 18.07                              &  1.185   &    742   & 18.55                                 &  1.123  &    764  & 19.10              &   2.75  &     44   & 1.100          &   2.90  &     40  &\textbf{1.000}             &   2.85  &     42   & 1.050             &   2.82  &     42 & 1.050    \\
                                             &    60                   & 1.191     &   1401   &  23.35                                &  1.305  &   1195   & 19.92                              &  1.267   &   1217   & 20.28                                 &  1.220  &   1240  & 20.67              &   3.38  &     70   & 1.167          &   3.64  &     60  &\textbf{1.000}             &   3.47  &     66   & 1.100             &   3.44  &     68 & 1.133    \\
                                             &    80                   & 1.247     &   1667   &  20.84                                &  0.919  &   1449   & 18.11                              &  1.344   &   1470   & 18.38                                 &  1.304  &   1494  & 18.68              &   3.89  &     83   & 1.038          &   2.99  &    110  & 1.375             &   3.97  &     80   &\textbf{1.000}             &   3.94  &     81 & 1.012    \\
                                             &   100                   & 1.301     &   2046   &  20.46                                &  1.078  &   1818   & 18.18                              &  1.417   &   1837   & 18.37                                 &  1.380  &   1862  & 18.62              &   4.34  &    103   & 1.030          &   3.48  &    119  & 1.190             &   4.41  &    100   &\textbf{1.000}             &   4.38  &    101 & 1.010    \\
                                             &   120                   & 1.354     &   2436   &  20.30                                &  1.393  &   2200   & 18.33                              &  1.484   &   2216   & 18.47                                 &  1.448  &   2242  & 18.68              &   4.75  &    124   & 1.033          &   4.48  &    136  & 1.133             &   4.82  &    120   &\textbf{1.000}             &   4.80  &    122 & 1.017    \\
                                             &   140                   & 1.407     &   2763   &  19.74                                &  1.573  &   2522   & 18.01                              &  1.551   &   2536   & 18.11                                 &  1.516  &   2562  & 18.30              &   5.07  &    144   & 1.029          &   4.90  &    158  & 1.129             &   5.14  &    140   &\textbf{1.000}             &   5.12  &    141 & 1.007    \\
                                             &   160                   & 1.457     &   3194   &  19.96                                &  1.631  &   2949   & 18.43                              &  1.614   &   2959   & 18.49                                 &  1.578  &   2985  & 18.66              &   5.46  &    166   & 1.038          &   5.18  &    190  & 1.188             &   5.54  &    160   &\textbf{1.000}             &   5.52  &    162 & 1.012    \\
                                             &   180                   & 1.507     &   3481   &  19.34                                &  1.693  &   3234   & 17.97                              &  1.675   &   3242   & 18.01                                 &  1.638  &   3269  & 18.16              &   5.71  &    185   & 1.028          &   5.38  &    211  & 1.172             &   5.79  &    180   &\textbf{1.000}             &   5.77  &    181 & 1.006    \\
                                             &   200                   & 1.557     &   3862   &  19.31                                &  1.760  &   3617   & 18.09                              &  1.734   &   3621   & 18.11                                 &  1.697  &   3647  & 18.23              &   6.03  &    209   & 1.045          &   5.68  &    245  & 1.225             &   6.11  &    200   &\textbf{1.000}             &   6.08  &    202 & 1.010    \\
                                             &   400                   &  2.01     &   5915   &  14.79                                &   2.25  &   5662   & 14.15                              &   2.24   &   5656   & 14.14                                 &   2.20  &   5684  & 14.21              &   7.47  &    415   & 1.038          &   7.34  &    432  & 1.080             &   7.59  &    400   &\textbf{1.000}             &   7.60  &    400 &\textbf{1.000}    \\
                                             &   800                   &  2.74     &  10985   &  13.73                                &   3.03  &  10736   & 13.42                              &   3.00   &  10718   & 13.40                                 &   2.96  &  10745  & 13.43              &  10.17  &    846   & 1.058          &  10.09  &    863  & 1.079             &  10.34  &    812   & 1.015             &  10.40  &    800 &\textbf{1.000}    \\
                                             &  1600                   &  3.89     &  19631   &  12.27*                               &   4.16  &  19408   & 12.13*                             &   4.13   &  19370   & 12.11*                                &   4.10  &  19394  & 12.12*             &  13.59  &   1723   & 1.077          &  13.38  &   1821  & 1.138             &  13.85  &   1638   & 1.024             &  13.96  &   1600 &\textbf{1.000}    \\
                                             &  3600                   &  5.89     &  35133   &   9.76*                               &   6.09  &  35033   &  9.73*                             &   6.09   &  34959   &  9.71*                                &   6.07  &  34967  &  9.71*             &  18.47  &   3952   & 1.098          &  17.71  &   4646  & 1.291             &  18.54  &   3851   & 1.070             &  18.73  &   3600 &\textbf{1.000}    \\
                                             &  6400                   &  7.91     &  42202   &   6.59*                               &   8.07  &  42175   &  6.59*                             &   8.07   &  42084   &  6.58*                                &   8.05  &  42083  &  6.58*             &  20.33  &   6584   & 1.029          &  20.23  &   6680  & 1.044             &  20.46  &   6477   & 1.012             &  20.55  &   6400 &\textbf{1.000}    \\
                                             &  9600                   &  9.71     &  57452   &   5.98*                               &   9.84  &  57605   &  6.00*                             &   9.85   &  57477   &  5.99*                                &   9.83  &  57454  &  5.98*             &  23.83  &   9780   & 1.019          &  23.60  &  10015  & 1.043             &  23.88  &   9731   & 1.014             &  24.01  &   9600 &\textbf{1.000}    \\
                                             & 12800                   & 11.24     &  72470   &   5.66*                               &  11.35  &  72822   &  5.69*                             &  11.36   &  72656   &  5.68*                                &  11.35  &  72609  &  5.67*             &  26.81  &  12996   & 1.015*         &  26.52  &  13354  & 1.043*            &  26.81  &  13003   & 1.016*            &  26.98  &  12800 &\textbf{1.000}    \\
        \hline
  \multirow{16}{*}{Sawtooth:}                &    40                   & 1.143     &    999   &  24.98                                &  1.248  &    760   & 19.00                              &  1.224   &    788   & 19.70                                 &  1.167  &    825  & 20.63              &   3.09  &     46   & 1.150          &   3.32  &     40  &\textbf{1.000}             &   3.23  &     42   & 1.050             &   3.19  &     44 & 1.100    \\
                                             &    60                   & 1.203     &   1519   &  25.32                                &  1.366  &   1233   & 20.55                              &  1.321   &   1268   & 21.13                                 &  1.279  &   1318  & 21.97              &   3.76  &     73   & 1.217          &   4.15  &     60  &\textbf{1.000}             &   3.90  &     68   & 1.133             &   3.86  &     69 & 1.150    \\
                                             &    80                   & 1.263     &   1758   &  21.98                                &  0.966  &   1460   & 18.25                              &  1.413   &   1499   & 18.74                                 &  1.377  &   1554  & 19.43              &   4.35  &     86   & 1.075          &   3.14  &    112  & 1.400             &   4.49  &     80   &\textbf{1.000}             &   4.45  &     82 & 1.025    \\
                                             &   100                   & 1.322     &   2098   &  20.98                                &  1.128  &   1790   & 17.90                              &  1.500   &   1833   & 18.33                                 &  1.465  &   1896  & 18.96              &   4.81  &    106   & 1.060          &   3.62  &    118  & 1.180             &   4.94  &    100   &\textbf{1.000}             &   4.90  &    102 & 1.020    \\
                                             &   120                   & 1.380     &   2429   &  20.24                                &  1.464  &   2114   & 17.62                              &  1.582   &   2161   & 18.01                                 &  1.546  &   2233  & 18.61              &   5.21  &    127   & 1.058          &   5.10  &    127  & 1.058             &   5.35  &    120   &\textbf{1.000}             &   5.31  &    122 & 1.017    \\
                                             &   140                   & 1.436     &   2844   &  20.31                                &  1.674  &   2525   & 18.04                              &  1.658   &   2577   & 18.41                                 &  1.620  &   2659  & 18.99              &   5.57  &    154   & 1.100          &   5.82  &    140  &\textbf{1.000}             &   5.71  &    146   & 1.043             &   5.68  &    148 & 1.057    \\
                                             &   160                   & 1.492     &   3181   &  19.88                                &  1.747  &   2860   & 17.88                              &  1.733   &   2917   & 18.23                                 &  1.692  &   3007  & 18.79              &   5.92  &    176   & 1.100          &   6.18  &    160  &\textbf{1.000}             &   6.07  &    167   & 1.044             &   6.03  &    169 & 1.056    \\
                                             &   180                   & 1.547     &   3536   &  19.64                                &  1.826  &   3216   & 17.87                              &  1.805   &   3277   & 18.21                                 &  1.762  &   3375  & 18.75              &   6.25  &    201   & 1.117          &   6.54  &    180  &\textbf{1.000}             &   6.41  &    189   & 1.050             &   6.38  &    191 & 1.061    \\
                                             &   200                   & 1.600     &   3868   &  19.34                                &  1.913  &   3548   & 17.74                              &  1.871   &   3613   & 18.07                                 &  1.827  &   3720  & 18.60              &   6.53  &    235   & 1.175          &   6.86  &    200  &\textbf{1.000}             &   6.70  &    216   & 1.080             &   6.68  &    219 & 1.095    \\
                                             &   400                   &  2.10     &   5805   &  14.51                                &   2.50  &   5510   & 13.78                              &   2.46   &   5600   & 14.00                                 &   2.40  &   5748  & 14.37              &   8.12  &    443   & 1.107          &   8.48  &    400  &\textbf{1.000}             &   8.29  &    421   & 1.052             &   8.25  &    426 & 1.065    \\
                                             &   800                   &  2.94     &  10411   &  13.01                                &   3.39  &  10210   & 12.76                              &   3.34   &  10368   & 12.96                                 &   3.27  &  10619  & 13.27              &  10.94  &    892   & 1.115          &  11.44  &    800  &\textbf{1.000}             &  11.18  &    847   & 1.059             &  11.15  &    852 & 1.065    \\
                                             &  1600                   &  4.26     &  18780   &  11.74*                               &   4.68  &  18819   & 11.76*                             &   4.63   &  19114   & 11.95*                                &   4.55  &  19529  & 12.21*             &  14.61  &   1832   & 1.145          &  15.34  &   1600  &\textbf{1.000}             &  14.91  &   1735   & 1.084             &  14.94  &   1722 & 1.076    \\
                                             &  3600                   &  6.60     &  34466   &   9.57*                               &   6.91  &  35107   &  9.75*                             &   6.85   &  35661   &  9.91*                                &   6.75  &  36371  & 10.10*             &  19.85  &   4721   & 1.311          &  20.76  &   3600  &\textbf{1.000}             &  20.20  &   4335   & 1.204             &  20.38  &   4164 & 1.157    \\
                                             &  6400                   &  8.92     &  40975   &   6.40*                               &   9.11  &  41896   &  6.55*                             &   9.04   &  42557   &  6.65*                                &   8.93  &  43387  &  6.78*             &  21.95  &   6896   & 1.077          &  22.63  &   6400  &\textbf{1.000}             &  22.14  &   6753   & 1.055             &  22.09  &   6779 & 1.059    \\
                                             &  9600                   & 10.99     &  57112   &   5.95*                               &  11.11  &  58770   &  6.12*                             &  11.03   &  59701   &  6.22*                                &  10.90  &  60822  &  6.34*             &  25.99  &  10315   & 1.074          &  26.71  &   9600  &\textbf{1.000}             &  26.26  &  10035   & 1.045             &  26.27  &  10018 & 1.044    \\
                                             & 12800                   & 12.68     &  71145   &   5.56*                               &  12.75  &  73481   &  5.74*                             &  12.65   &  74646   &  5.83*                                &  12.51  &  76018  &  5.94*             &  29.04  &  13678   & 1.069*         &  29.80  &  12800  &\textbf{1.000}             &  29.39  &  13252   & 1.035*            &  29.43  &  13207 & 1.032*   \\
    \hline 
  \end{tabular}
                   }
\caption{ Histogram-wide average signal-to-error $\langle S/N\rangle_{MC}$ of estimates $\hat{p}_i$ (for a $b$=$100$ histogram) relative to the calculated $p_i$ for known trial pdfs; estimated $N$ ($N_{eq}$) needed for an estimator to match (S/N)$_{max}$ of the most accurate estimator in each row; and the ratio $\frac{N_{eq}}{N}$. Each horizontal triplet of values is from a 10,000 repetition MC trial for a given $N$ and estimator type.  [*: Linear Extrapolation (less accurate) needed to estimate $N_{eq}$.]
 \label{tab:P0Tables}  }
  \end{table}

\vskip 0.2in
\bibliography{BinomMultinomEst}

\end{document}